\documentclass[onecolumn,prd,superscriptaddress,nofootinbib]{revtex4-1}

\usepackage{graphicx}

\usepackage{xcolor}
\usepackage[colorlinks,bookmarks]{hyperref}
\definecolor{linkblue}{rgb}{0,0,0.8}
\definecolor{linkgreen}{rgb}{0,0.5,0}

\hypersetup{pdfpagemode=UseNone, pdfstartview=FitH, linkcolor=linkblue, %
            citecolor=linkgreen, urlcolor=linkblue}

\usepackage{mathtools}  
\usepackage[utf8]{inputenc}
\usepackage[english]{babel}

\usepackage{makecell}
\usepackage{subfigure}
\usepackage{xspace}
\newcommand{\be}{\begin{eqnarray}}

\newcommand{\ee}{\end{eqnarray}}
\usepackage{ulem}

\newcommand{\comment}[1]{{}}
\def\beq{\begin{equation}}
\def\eeq{\end{equation}}
\def\beqn{\begin{eqnarray}}
\def\eeqn{\end{eqnarray}}

\def\d{\rm{d}}

\def\l{\left}
\def\r{\right}

\def\2gcm{\textrm{g cm$^{-2}$}}

\def\modu#1{\l |{#1}\r |}

\def\H0{\ensuremath{\mathrm{H}_0}}

\def\nn{\nonumber}

\newcommand{\cmbav}[1]{\left\langle #1 \right\rangle}

\newcommand{\dontshow}[1]{}

\newcommand{\bsimple}{b_{1}}

\graphicspath{{./}{figures/}}

\renewcommand{\em}{\emph}
\renewcommand{\citealt}{\cite}

\newcommand{\fnl}{f_\text{NL}}

 \color{black}
 \color{black}
 \color{black}
 \color{black}

\newcommand{\ba}{\begin{aligned}}
\newcommand{\ea}{\end{aligned}}

\newcommand{\tcm}{21$\,$cm\xspace}  %

\newcommand{\dirac}{\delta_{\rm D}}

\newcommand{\deltag}{\delta_{\rm g}}

\newcommand{\invMpc}{h\, {\rm Mpc}^{-1}\,}

\def\beq{\begin{equation}}
\def\eeq{\end{equation}}
\def\d{\partial}
\newcommand{\lp}{\left(}
\newcommand{\rp}{\right)}
\newcommand{\lb}{\left[}
\newcommand{\rb}{\right]}
\newcommand{\vx}{\boldsymbol{x}}
\newcommand{\vk}{\boldsymbol{k}}
\newcommand{\vK}{\boldsymbol{K}}
\newcommand{\vq}{\boldsymbol{q}}
\newcommand{\vl}{\boldsymbol{\ell}}
\newcommand{\vL}{\boldsymbol{L}}
\def\Omm{\Omega_{\rm m}}
\newcommand\numberthis{\addtocounter{equation}{1}\tag{\theequation}}

\usepackage{bm}
\renewcommand{\vec}{\bm}

\begin{document}

\title{Density reconstruction from biased tracers and its application to primordial non-Gaussianity}

\author{Omar Darwish}
\affiliation{Department of Applied Mathematics and Theoretical Physics, \\
University of Cambridge, Wilberforce Road, \\
Cambridge CB3 0WA, United Kingdom}

\author{Simon Foreman}
\affiliation{Perimeter Institute for Theoretical Physics, \\
31 Caroline Street North, Waterloo, ON N2L 2Y5, Canada}
\affiliation{Dominion Radio Astrophysical Observatory, 
Herzberg Astronomy \& Astrophysics Research Centre,  \\
National Research Council Canada, P.O.\ Box 248, Penticton, BC V2A 6J9, Canada}

\author{Muntazir M.~Abidi}
\affiliation{Department of Applied Mathematics and Theoretical Physics, \\
University of Cambridge, Wilberforce Road, \\
Cambridge CB3 0WA, United Kingdom}

\author{Tobias Baldauf}
\affiliation{Department of Applied Mathematics and Theoretical Physics, \\
University of Cambridge, Wilberforce Road, \\
Cambridge CB3 0WA, United Kingdom}

\author{Blake D.~Sherwin}
\affiliation{Department of Applied Mathematics and Theoretical Physics, \\
University of Cambridge, Wilberforce Road, \\
Cambridge CB3 0WA, United Kingdom}
\affiliation{Kavli Institute for Cosmology, \\
University of Cambridge, \\
Cambridge CB3 0HA, United Kingdom}

\author{P.~Daniel Meerburg}
\affiliation{Van Swinderen Institute for Particle Physics and Gravity,\\ University of Groningen,
Nijenborgh 4, 9747 AG Groningen, The Netherlands}

\begin{abstract}
Large-scale Fourier modes of the cosmic density field are of great value for learning about cosmology because of their well-understood relationship to fluctuations in the early universe. However, cosmic variance generally limits the statistical precision that can be achieved when constraining model parameters using these modes as measured in galaxy surveys, and moreover, these modes are sometimes inaccessible due to observational systematics or foregrounds. For some applications, both limitations can be circumvented by reconstructing large-scale modes using the correlations they induce between smaller-scale modes of an observed tracer (such as galaxy positions). In this paper, we further develop a formalism for this reconstruction, using a quadratic estimator similar to the one used for lensing of the cosmic microwave background. We incorporate nonlinearities from gravity, nonlinear biasing, and local-type primordial non-Gaussianity, and verify that the estimator gives the expected results when applied to $N$-body simulations. We then carry out forecasts for several upcoming surveys, demonstrating that, when reconstructed modes are included alongside directly-observed tracer density modes, constraints on local primordial non-Gaussianity are generically tightened by tens of percents compared to standard single-tracer analyses. In certain cases, these improvements arise from cosmic variance cancellation, with reconstructed modes taking the place of modes of a separate tracer, thus enabling an effective ``multitracer" approach with single-tracer observations.
\end{abstract}
\keywords{cosmology, primordial non gaussianity --- 
quadratic estimators --- forecasting --- galaxy surveys}

\maketitle

\tableofcontents

\section{Introduction}

Our understanding of the Universe has benefited tremendously from measurements of the cosmic microwave background (CMB), primarily because of the linear relationship between fluctuations in the CMB and fluctuations generated in the very early universe. This relationship allows us to connect CMB measurements to the statistics of the initial fluctuations and their time evolution, and has led to the establishment of the current cosmological model. Extraction of similar information from the large-scale structure (LSS) of the universe is limited by nonlinear clustering at smaller distances and lower redshifts, requiring more elaborate modelling to interpret observations. This modelling burden is greatly reduced at the largest distances we can resolve with galaxy surveys, but this regime is in turn obscured by both statistical and systematic errors. In this paper, we explore a method to access these large scales while bypassing both types of errors: quadratic density reconstruction.

This idea of density reconstruction relies on the fact that a fixed long-wavelength density fluctuation correlates two different small-scale modes due to non-linear evolution and higher-order biasing, with the amount of correlation proportional to the long-wavelength mode. This can be understood as arising from a violation of statistical homogeneity if the long-wavelength mode is considered as fixed and the shorter-wavelength modes are averaged over in an ensemble. Writing down a quadratic estimator that probes this induced correlation between two different modes, we can estimate the long-wavelength modes from the statistical properties of the smaller-scale modes.\footnote{In fact, these statements are independent of the relative wavelengths of the modes, and the formalism we present in this paper is not restricted to the so-called ``squeezed limit" of the three modes involved. However, for our applications, the modes we are seeking to reconstruct have longer wavelengths than the two modes whose correlations are used for the reconstruction, so we focus on that situation in this paper.}

There is a close analogy between this procedure and the common method of CMB lensing reconstruction, in which a quadratic estimator, making use of the lensing-induced correlation between two different CMB temperature modes, is used to reconstruct the lensing field (e.g.\ \citealt{Hu:2001kj}). It is using this analogy that many of the methods for density reconstruction were derived. The idea of using a standard quadratic estimator in the CMB lensing form to perform this reconstruction was first proposed by \cite{Foreman:2018gnv}, building on earlier work (\citealt{Pen:2012ft,Zhu:2015zlh,Zhu:2016esh}, albeit with a somewhat modified estimator). Significant further work in this area has been presented by  \cite{Li:2018izh,Modi:2019hnu,Karacayli:2019iyd,Li:2020uug,Li:2020luq}; see further discussion in Section~\ref{sec:discussion}.

The work in this paper broadly divides in two parts. In the first part, we present a considerable expansion of current technology for density reconstruction. We discuss the application of density reconstruction to biased tracers, including, for the first time, a full non-linear bias model in such a formalism. We further validate our method on a suite of realistic $N$-body simulations, demonstrating that our methods perform just as expected from theoretical calculations for both the reconstruction and its noise level.

We note that this reconstruction has a wealth of applications. One simple application is the following: LSS surveys are often plagued by observational systematics that manifest at large scales, impeding the direct observation of low-$k$ modes. Galaxy and quasar surveys are affected, for example, by variations in the density of foreground stars, seeing, and galactic dust extinction (e.g.\ \citealt{Ross:2012sx,Ho:2013lda,Kalus:2018qsy}), while \tcm surveys cannot access modes with low line-of-sight wavenumbers that are dominated by galactic foregrounds, and imperfect knowledge of the instrument can spread this contamination throughout a wider region of Fourier space (e.g.\ \citealt{Parsons:2012qh,Liu:2014bba,Liu:2014yxa}). A method of reconstructing these inaccessible modes using correlations between smaller-scale modes will improve the constraining power of a given survey for large-scale signals such as local non-Gaussianity, and allow cross-correlations involving \tcm surveys that would otherwise be impossible (e.g.\ \citealt{Li:2018izh}). In this paper, we parameterize large-scale systematics with a wavenumber $K_{\rm min}$ below which the tracer modes are assumed to be inaccessible, and explore the precision with which modes with $K<K_{\rm min}$ can be recovered by our estimator. We note that this assumes that the relevant systematics can be parameterized as a large-scale additive component, rather than a possible modulation that might also significantly affect small scales; while there is evidence that this is a reasonable assumption for some of the currently known systematics (e.g.~\cite{Kalus:2018qsy}), it may not hold in all cases.

In contrast to this general application, the second goal of our paper is to explore, in detail, a much more subtle application of density reconstruction: improving constraints on local-type primordial non-Gaussianity. We will briefly motivate the measurement of primordial non-Gaussianity and the utility of density reconstruction for improving these constraints in the following paragraphs.

The CMB has taught us that the statistics of the primordial fluctuations can be accurately described by a red-tilted power law. If the initial conditions are completely described by this power law, they have to be Gaussian distributed, with statistics determined by only two degrees of freedom: the amplitude ($A_{\rm s}$) and tilt ($n_{\rm s}$) of the power law.  If this is the case, however, it will be difficult to reach beyond our current understanding of the early Universe. The most widely accepted theory is known as cosmic inflation, which postulates a short early period of accelerated cosmic expansion. Effectively, the constraints we derive from the CMB tell us that inflation can be very well described by a scalar field slowly rolling down a potential (``single-field slow roll", or SFSR), with only (weak) gravitational interactions. While such a model is certainly possible (it was the first to be considered \citep{Gut81,Lin82,Lin82b}), it will not provide us with simple opportunities to understand the physics of inflation. If a proposed model of the early Universe has to comply with Gaussian initial conditions, effectively the model will observationally resemble SFSR. Any further distinction could be extracted from the details of the scale dependence of the primordial power spectrum \citep{Slosar:2019gvt}, but so far, observations do not reveal any obvious deviations from a single parameter power law \citep{Akrami:2018odb,Planck2019IX}.

A much more powerful model discriminator would be available if the initial conditions showed a (small) deviation from Gaussianity. In the presence of non-Gaussianity, all moments beyond the power spectrum will generically  be excited (starting with the 3-point function or bispectrum). Technically, these higher-point spectra probe the dynamics of the field(s) driving inflation. As a result, a measurement of non-Gaussianity would reveal details of inflation that can be directly related to the underlying fundamental physics. For example, non-Gaussianity could reveal the presence of more fields relevant during inflation, or could provide clues to how strongly coupled the inflation field is (see e.g.~\cite{Meerburg2019} and references therein). These powerful constraints cannot be exposed through any other measurement, making non-Gaussianities a unique probe of the early Universe.

To lowest order, primordial non-Gaussianities modulate the gravitational potential  $\Phi$ via
\be
\Phi(\vk) = \varphi_{\rm G} (\vk)+ f_{\rm NL}^X \int \frac{d^3 q}{(2\pi)^3} G_{\rm NL}^X(\vq,\vk-\vq)\varphi_{\rm G} (\vq)\varphi_{\rm G}(\vk-\vq)\ ,
\ee
where $\varphi_{\rm G}$ is the Gaussian potential and $G_{\rm NL}^X$ is a kernel that describes how the potential is modulated. In this paper, we are interested in local non-Gaussianities for which $G_{\rm NL}^{\rm local} = 1$, i.e. 
\beq
\Phi(\vx)=\varphi_{\rm G}(\vx)+\fnl (\varphi_{\rm G}^2(\vx)-\langle \varphi_{\rm G}^2\rangle),
\label{eq:localNGs}
\eeq
where we have subtracted the mean to yield zero expectation value for the fluctuations and have renamed $f_{\rm NL}^{\rm local}$ to $f_{\rm NL}$.  Current constraints set $\sigma(f_{\rm NL}) \sim \mathcal{O}(5)$ \citep{Planck2019IX}, while $f_{\rm NL} \sim 1$ has been identified as a compelling theoretical threshold~\citep{Alvarez:2014vva} which provides a strong motivation to go beyond current limits: if a measurement is made showing $f_{\rm NL}$ above this limit, it would effectively rule out SFSR inflation as a viable scenario. 
 Future ground-based CMB experiments \citep{SO2019,Abazajian:2016yjj} may be able reach $\sigma(f_{\rm NL}) \sim \mathcal{O}(1)$, but poor scaling and galactic and cosmological foregrounds will likely prevent the CMB from reaching (far) beyond this limit. Fortunately, the large scale structure (LSS) in the universe provides access to many more modes, for which $\sigma(f_{\rm NL}) \propto ( k_{\rm max}^3 \log k_{\rm max}/k_{\rm min}  )^{-1/2}$ \citep{Scoccimarro_2004}. While increased dimensionality will help to improve constraints, the use of LSS will introduce many complications. For one, the scaling argument breaks down when $k_{\rm max}$ exceeds the nonlinear scale $k_{\rm NL}$, which is of order $0.2\invMpc$ for current galaxy surveys \citep{DAmico:2019fhj,Ivanov:2019pdj}. Furthermore, line-of-sight information, which will be crucial in obtaining a sufficient number of modes, will require a careful treatment mainly due to redshift space effects \citep{Gil_Mar_n_2014}. Obtaining cosmological constraints from a measurement of the full LSS bispectrum will therefore be challenging, not least because of non-Gaussian covariance \citep{Scoccimarro_2004,Sefusatti_2006,kayo2013cosmological} which will likely require (a large number of) simulations to estimate \citep{Chan_2017}. Some of these difficulties can be overcome by simplifying the full bispectrum into more compressed statistics \citep{Schmittfull:2014tca,Fergusson:2010ia,Byun:2017fkz,Dai:2020adm,MoradinezhadDizgah:2019xun,Chiang:2014oga,dePutter:2018jqk,Gualdi:2018pyw}. The advantage of these statistics is that they should capture nearly all the information \citep{MoradinezhadDizgah:2019xun}, but are computationally and observationally less challenging.
 
Unlike in the CMB, in LSS local primordial non-Gaussianity can also significantly affect the power spectrum of biased tracers, such as galaxies. Specifically, it has been shown \citep{Dalal:2007cu,Matarrese_2008,Slosar_2008,Desjacques:2010jw,Schmidt_2010} that tracer bias will be affected by the primordial non-Gaussianity, with the bias acquiring a unique $1/k^2$ contribution, which is hard to produce otherwise. This signature has been used to place constraints on $\fnl$ with current surveys \citep{Giannantonio:2013uqa,Leistedt:2014zqa,Castorina:2019wmr}.
 Unfortunately, although the signal should be distinguishable from other effects, on large scales, the precision with which we can measure the power spectrum is ultimately limited by cosmic variance from the number of available modes. However, it was shown that this cosmic variance can be mitigated \citep{Seljak:2008xr,McDonald:2008sh,Hamaus_2011,Schmittfull:2017ffw,Liu:2020izx} by using multiple tracers of the same underlying density field (with different biases), which essentially allows a measurement of scale-dependent bias via a mode-by-mode comparison of the different tracers. A combination of two (or more, e.g.\ \citealt{Schmittfull:2017ffw,Ballardini:2019wxj}) tracers will allow for cosmic variance cancellation, limiting a measurement of the scale-dependent bias from local primordial non-Gaussianity only by the number density of these tracers. Forecasts show that these techniques enable constraints to reach  $\sigma(f_{\rm NL})\sim 1$ this decade \citep{Schmittfull:2017ffw,Ballardini:2019wxj,Munchmeyer:2018eey,SO2019}.

In this paper, we show that this cosmic variance cancellation can also be achieved, to some extent, using only a single tracer. In order to do this, we compare our reconstructed density field (which provides information from higher-point functions) with a directly-measured tracer field. In the end, the constraints on $f_{\rm NL}$ will depend on the auto-correlation of the tracer field $P_{\rm gg}$\footnote{Since we will focus on the use of galaxies as tracers in this paper, we will use the subscript g to refer to these tracers, although the method we describe is equally applicable to quasars, line intensity maps, or other tracers.}, the cross correlation of the tracer and the reconstructed field $P_{\rm gr}$, and the auto-correlation of the reconstructed field $P_{\rm rr}$. This idea is related to~\cite{dePutter:2018jqk}, where similar ideas are used to simplify a forecast of the combined information in the power spectrum, bispectrum and trispectrum. However, unlike in \cite{dePutter:2018jqk}, we examine the reconstruction approach as a possible analysis tool rather than a method for more easily computing complex forecasts. In addition, whereas \cite{dePutter:2018jqk} relies on an extension of position-dependent power spectra \citep{Chiang_2014,Chiang_2015,Chiang_2017,Adhikari_2016}, which draw information only from the squeezed limit, here we use a quadratic estimator formalism for the reconstructed field without imposing a squeezed-limit constraint.

Let us briefly summarize our most important results:
\begin{itemize}
    \item The modes of the tracer overdensity will be coupled due to nonlinearities from gravity, nonlinear bias, and primordial non-Gaussianity. The amplitudes (parameterized with bias coefficients) of several of these mode-couplings are unknown a priori.
     We incorporate this in our characterization of the quadratic estimator for long-wavelength modes, and marginalize over the unknown coefficients in our forecasts. We also highlight the important contribution of tracer shot noise to the noise on the reconstructed modes.
    \item We demonstrate density reconstruction using dark matter halos in $N$-body simulations, verifying that the performance agrees well with that predicted from analytical formulas. Though additional work using simulations will be required for a practical analysis, our results indicate that our forecasts are realistic.
    \item We show that the quadratic estimator is able to reconstruct long-wavelength modes at high signal-to-noise for a wide range of upcoming surveys (see Fig.~\ref{fig:prr-ebars}).
    \item The addition of the reconstructed field to forecasts using the large-scale biased tracer field can improve constraints on $\fnl$ by tens of percents depending on the survey configuration. The improvement arises from a combination of two sources: sample variance cancellation of signal in the large-scale tracer field, and additional scale-dependent signal in the reconstructed field on scales where the tracer field may be obscured by observational systematics. The additional information in the reconstructed modes can be viewed as a signature of non-Gaussian signal in the three and four-point functions, and our approach can be viewed as a simple method to obtain combined information from the three- and four-point functions and the power spectrum.
    \item The performance of this approach to constraining $\fnl$ is limited by a combination of tracer number density and maximum wavenumber of modes that can be used for reconstruction, with the details again depending on the survey configuration. Potential improvements using response function approaches \citep{Barreira:2017sqa, Barreira:2017kxd} could be explored to extend the reconstruction wavenumber and gain signal-to-noise.
\end{itemize}

The outline of our paper is as follows. In Section~\ref{sec:formalism}, we describe our methodology for density reconstruction, including the quadratic estimator formalism and bias expansion we use. In Section~\ref{sec:sims}, we apply this method to halos in $N$-body simulations. In Section~\ref{sec:forecasts}, we present our forecasts for the expected precision on reconstructed modes, as well as constraints on local non-Gaussianity. We compare this reconstruction formalism to other work involving higher-point statistics in Section~\ref{sec:discussion}. Finally, we conclude in Section~\ref{sec:conclusions}. Several derivations and technical details are included in the appendices, and a summary of our notation can be found in Table~\ref{tab:notation}. Except for in Sec.~\ref{sec:sims}, we use cosmological parameters from the Planck 2015 results, given in the ``TT,TE,EE+lowP+lensing+ext" column of Table~4 of \citealt{Ade:2015xua}.

\begin{table}
\begin{centering}
\begin{tabular}{ |l|l|l| }
\hline
Quantity	&	Symbol	& Defined in \\ \hline\hline

Dirac delta function in 3d & $\dirac(\vec{k})$ & --- \\

Wavenumbers of modes used in reconstruction & $\vk$, $\vq$, etc. & --- \\

Wavenumbers of modes used for $\fnl$ constraints & $\vK$, $\vK'$, etc. & --- \\

\hline

Amplitude of local primordial non-Gaussianity & $\fnl$ & Eq.~\eqref{eq:localNGs} \\

Factor relating primordial potential and $\delta_1$  & $M(k,z)$ & Eqs.~\eqref{eq:phidef}-\eqref{eq:M} \\

\hline

Linear matter overdensity & $\delta_1(\vk,z)$ & --- \\

Linear matter power spectrum & $P_\text{lin}(k,z)$ & --- \\

Tracer overdensity & $\delta_{\rm g}(\vk,z)$ & Eq.~\eqref{eq:deltag-generic} [generic]; \\
 & & Eq.~\eqref{eq:deltag-condensed} [second-order bias model] \\

Second-order mode-coupling & $F_{\alpha}(\vec{k}_1, \vec{k}_2)$ & Eq.~\eqref{eq:deltag-generic} [generic]; \\
 & & Eq.~\eqref{eq:deltag-condensed} [second-order bias model] \\
 
Second-order response of small-scale power spectrum to long mode & $f_{\alpha}(\vec{k}_1, \vec{k}_2,z)$ & Eq.~\eqref{eq:falpha} \\

Coefficient of $F_\alpha$ in second-order bias model for $\delta_{\rm g}$ & $c_\alpha$ & Eq.~\eqref{eq:deltag-condensed} \\

Linear tracer bias & $b_1\equiv b_{10}^\text{E}$ & Eq.~\eqref{eq:deltag-condensed} \\

Quadratic tracer bias & $b_2\equiv b_{20}^\text{E}$ & Eq.~\eqref{eq:deltag-condensed} \\

Other second-order bias parameters & $b_{s^2}^{\rm E}$, $b_{01}^{\rm E}$, $\cdots$ & Sec.~\ref{sec:bias} \\

\hline

Quadratic estimator for mode with wavenumber $\vec{K}$ & $\hat{\Delta}_{\alpha}(\vec{K})$ & Eqs.~\eqref{eq:quad-def}, \eqref{eq:quadest} \\

Weight function in $\hat{\Delta}_{\alpha}(\vec{K})$ & $g_{\alpha}(\vk_1,\vk_2)$ & Eq.~\eqref{eq:galpha} \\

Normalization and Gaussian noise of $\hat{\Delta}_{\alpha}(\vec{K})$ 
	& $N_{\alpha\beta}(\vec{K})$ & Eq.~\eqref{eq:nab} \\
	
Mode reconstructed with growth-coupling estimator $\hat{\Delta}_{\rm G}(\vec{K},z)$
	& $\delta_{\rm r}(\vk,z)$ & --- \\

Power spectrum of $\delta_{\rm g}$, ignoring shot noise contribution & $P_{\rm gg}$ & --- \\

Sum of $P_{\rm gg}$ and shot noise contribution & $P_{\rm tot}$ & Sec.~\ref{sec:qe} \\

Cross power spectrum between $\delta_{\rm g}$ and $\delta_{\rm r}$, ignoring shot noise contribution & $P_{\rm gr}$ & --- \\

Power spectrum of $\delta_{\rm r}$, ignoring shot noise contribution  & $P_{\rm rr}$ & --- \\

Shot noise contribution to $\delta_{\rm g}$ power spectrum & $P_{\rm gg,shot}$ & Eq.~\eqref{eq:pggshot} \\

Shot noise contribution to $\delta_{\rm r}$ power spectrum & $P_{\rm rr,shot}$ 
	& Eqs.~\eqref{eq:nrrshot}-\eqref{eq:prrshot-appendix} \\

Shot noise contribution to $\delta_{\rm g}$-$\delta_{\rm r}$ cross power spectrum & $P_{\rm gr,shot}$ 
	& Eqs.~\eqref{eq:nrtshot}-\eqref{eq:pgrshot-appendix} \\

\hline

Lowest wavenumber within survey volume & $K_{\rm f}$ & Sec.~\ref{sec:scales} \\

Wavenumber below which we assume $\delta_{\rm g}$ cannot be measured & $K_{\rm min}$ & Sec.~\ref{sec:scales} \\

Maximum wavenumber used for $\fnl$ constraints & $K_{\rm max}$ & Sec.~\ref{sec:scales} \\

Maximum wavenumber used in quadratic estimator for reconstructed modes & $k_{\rm max}$ & Sec.~\ref{sec:scales} \\

 \hline
\end{tabular}
\caption{
\label{tab:notation}
Notation used for important quantities in this paper
}
\end{centering}
\end{table}

\section{Density reconstruction}
\label{sec:formalism}

\subsection{Quadratic estimator: general case}
\label{sec:qe}

In this section, we will develop the general formalism for reconstructing large-scale\footnote{We again remind the reader that the our formalism is generally applicable, without any strong assumptions on the wavelenghts of the modes.} density modes using observations of a biased tracer. This is largely based on the treatment in \cite{Foreman:2018gnv}, but we have adapted their expressions to 3D wavenumbers rather than a separate treatment of line-of-sight and transverse components of~$\vk$.

Suppose that the overdensity field of the tracer, $\deltag$, is well-described by a linear bias with respect to the linear matter overdensity $\delta_1$, plus a set of quadratic terms that couple modes of $\delta_1$ with kernels $F_\alpha$ and amplitudes $c_\alpha$:
\begin{equation}
\deltag(\vec{k}, z) \approx \bsimple(z)\delta_1(\vec{k}, z)
	+\sum_{\alpha}c_{\alpha}(z) \int_{\vec{q}}F_\alpha(\vec{q}, \vec{k}-\vec{q}; z)
	\delta_1(\vec{q}, z)\delta_1(\vec{k}-\vec{q}, z)\ ,
\label{eq:deltag-generic}
\end{equation}
where $\int_{\vq} \equiv (2\pi)^{-3} \int d^3\vq$.
For example, if we took $\deltag$ to be the matter overdensity rather than a biased tracer, we would have $b_{1}=1$ and the sum would run over the second-order mode-couplings induced by gravitational evolution, which take the form (e.g.\ \citealt{Sherwin:2012nh})
\beq
F_{\rm G}(\vk_1,\vk_2;z) \equiv \frac{17}{21}\ , \quad
	F_{\rm S}(\vk_1,\vk_2;z) 
	\equiv \frac{1}{2} \lp \frac{1}{k_1^2}+\frac{1}{k_2^2} \rp \vk_1\cdot\vk_2\ , \quad
	F_{\rm T}(\vk_1,\vk_2;z) 
	\equiv \frac{2}{7} \lb \frac{\lp \vk_1\cdot\vk_2 \rp^2}{k_1^2 k_2^2} -\frac{1}{3} \rb\ ,
	\label{eq:f2kernels}
\eeq
with $c_{\rm G} = c_{\rm S} = c_{\rm T} = 1$ and the subscripts indicating that these functions arise from isotropic {\bf G}rowth, a large-scale coordinate {\bf S}hift, and a {\bf T}idal coupling. For a biased tracer, nonlinear biasing will lead to $c_\alpha\neq 1$ for the above couplings, and primordial non-Gaussianity will introduce additional mode-couplings. In Sec.~\ref{sec:bias}, we will introduce the full set of mode-couplings that must be considered, but we note here that many of the corresponding $c_\alpha$ coefficients  will not be known a priori, and this must be accounted for in the density reconstruction procedure. Henceforth, we will drop the $z$-dependence from the quantities defined above.

Now, we would like to use the mode-couplings in Eq.~\eqref{eq:deltag-generic} to construct a quadratic estimator for a given mode of~$\delta_1$. We will present the logic in some detail, for readers who may not be familiar with the relevant arguments, but a reader who is comfortable with peak-background-split arguments or the CMB lensing formalism may wish to skip to the final result in Eqs.~\eqref{eq:covoffdiag}-\eqref{eq:falpha}.

The analogous procedure for CMB lensing is to first consider an ensemble average over CMB fluctuations while keeping fluctuations in the lower-redshift matter density fixed. In this case, the fixed modes of the lensing potential~$\phi$ (which is a line-of-sight projection of the lower-redshift density field -- see e.g.\ \citealt{Hu:2001kj}) break the statistical isotropy of the CMB fluctuations, inducing correlations between CMB fluctuation modes with different wavenumbers: for temperature modes on the flat sky, the specific effect is given by
\beq
\left\langle T(\vl) T(\vL-\vl) \right\rangle_{\phi\text{ fixed}} 
	= (2\pi)^2 \dirac(\vL) C_L + f_\phi(\vl,\vL-\vl) \phi(\vL)\ .
\label{eq:cmblensing}
\eeq
When analyzing CMB simulations or data, the temperature two-point function is estimated by a (weighted) sum over~$\vl$ within a given CMB realization, and this in fact approximates the ensemble average above, with $\phi$ modes effectively fixed because they do not explicitly enter the sum. Eq.~\eqref{eq:cmblensing} is an efficient starting point for deriving quadratic estimators for a specific mode of $\phi$, and we would like to find the analogous starting point for density reconstruction.

To proceed, we consider an ensemble average over all modes of $\delta_1$ except those with wavenumbers in a small neighborhood around $\vK$, with $\delta_1(\vK)$ being the mode we will eventually want to reconstruct. (We must consider a neighborhood around $\vK$ because we are working in the continuum limit, where we have integrals instead of discrete sums over wavenumbers; we will return to this point below.) In this ensemble average, which we denote by ``$\sim$$\vK$ fixed", and using Eq.~\eqref{eq:deltag-generic}, the two-point function of $\deltag$ is at next-to-leading-order in $\delta_1$ is
\begin{align*}
\left\langle \deltag(\vec{k}) \deltag(\vec{K}-\vec{k})\right\rangle_{\sim\vK\text{ fixed}} 
	&= \bsimple^2 \langle \delta_1(\vec{k})\delta_1(\vec{K}-\vec{k}) \rangle \\
&\quad + \bsimple \int_{\vq} \sum_\alpha c_\alpha
	F_{\alpha}(\vec{q},\vec{k}-\vec{q}) \,
	\langle \delta_1(\vec{q}) \delta_1(\vec{k}-\vec{q})\delta_1(\vec{K}-\vec{k})\rangle_{\sim\vK\text{ fixed}}
	+ [\vec{k}\leftrightarrow\vec{K}-\vec{k}]\ .
	\numberthis
\end{align*}
In the first line, we have assumed that $\vk$ is not within the chosen neighborhood of $\vK$ or the equivalent neighborhood of $0$, so there is no difference between our special ensemble average and the standard one. In the second line, the integrand evaluates to zero if $\vq$ and $\vK-\vq$ are not within the neighborhood of $\vK$, since in that case, all three $\delta_1$ modes are averaged over, and the three-point function is zero for $\vK\neq 0$. When $\vq \sim \vK$ or $\vK-\vq \sim \vK$, where we use ``$\sim\vK$" to indicate a vector falling within the neighborhood of $\vK$, then $\delta_1(\vq)$ or $\delta_1(\vK-\vq)$ factor out of the ensemble average because they are held fixed, and the remaining two modes are averaged over:
\begin{align*}
\left\langle \deltag(\vec{k}) \deltag(\vec{K}-\vec{k})\right\rangle_{\sim\vK\text{ fixed}}
	&= \bsimple^2 \langle \delta_1(\vec{k})\delta_1(\vec{K}-\vec{k}) \rangle \\
&\quad + \bsimple \int_{\vec{q}\,\sim\, \vec{K}}  \sum_\alpha c_\alpha
	F_{\alpha}(\vec{q},\vec{k}-\vec{q})\delta_1(\vec{q}) \,
	\langle  \delta_1(\vec{k}-\vec{q})\delta_1(\vec{K}-\vec{k})\rangle
	+ [\vec{k}\leftrightarrow\vec{K}-\vec{k}] \\
&\quad + \bsimple \int_{\vec{K}-\vec{q}\,\sim\, \vec{K}} \sum_{\alpha} c_{\alpha} 
	F_{\alpha}(\vec{q},\vec{k}-\vec{q})\delta_1(\vec{K}-\vec{q}) \,
	\langle \delta_1(\vec{q})\delta_1(\vec{K}-\vec{k})\rangle
	+ [\vec{k}\leftrightarrow\vec{K}-\vec{k}] \ .
	\numberthis
\end{align*}
From here, we simply evaluate the two-point correlators and use the resulting Dirac delta functions to collapse the $\vq$ integrals.\footnote{One must integrate in a neighborhood around the argument of a Dirac delta function for this collapse to take place, and this is why we considered a neighborhood around $\vK$ in the first place. In the discrete case, where we have sums instead of integrals over wavenumbers, we could define our ensemble average to keep a single mode $\delta_1(\vK)$ fixed, since we would then have Kronecker deltas instead of Dirac delta functions.} The final result is
\beq
\left\langle \deltag(\vec{k}) \deltag(\vec{K}-\vec{k})\right\rangle_{\sim\vK\text{ fixed}}
	= (2\pi)^3 \dirac(\vK) \, \bsimple^2 P_{\rm lin}(k)
	+ \bsimple \sum_{\alpha} c_{\alpha} f_\alpha(\vec{k}, \vec{K}-\vec{k}) \delta_1(\vec{K})\ ,
	\label{eq:covoffdiag}
\eeq
where
\begin{equation}
    f_{\alpha}(\vec{k}_1, \vec{k}_2) \equiv 
    	2 [F_{\alpha}(\vec{k}_1+\vec{k}_2, -\vec{k}_1) P_\text{lin}(\vec{k}_1,z) + 1\leftrightarrow 2]\ .
    \label{eq:falpha}
\end{equation}
In Eq.~\eqref{eq:covoffdiag}, we find the same structure as in the CMB lensing case in Eq.~\eqref{eq:cmblensing}: the standard power spectrum term, plus a term from off-diagonal correlations induced by the fixed background mode.

Eq.~\eqref{eq:covoffdiag} suggests we can multiply two different modes of the measured tracer field and then simply ``divide" by the coupling strength $\bsimple \sum_{\alpha}c_{\alpha}f_\alpha(\vec{k}, \vec{K}-\vec{k})$  to obtain an estimate of the linear field $\delta_1$ at large scales. Unfortunately, in general we do not know the bias coefficients $b_1$ or $c_{\alpha}$ a priori, so the best we can do is to use the galaxy mode couplings to estimate the product $b_1 c_{\alpha}\delta_1$ for a chosen $\alpha$. To reduce variance on the estimate, we will sum over all the mode couplings that involve the same large-scale mode. This can be achieved by writing the following general quadratic estimator
\begin{equation}
   \hat{\Delta}_{\alpha}(\vec{K}) \equiv \widehat{b_1 c_{\alpha}\delta_1}(\vec{K})=\int_{\vec{q}}g_{\alpha}(\vec{q}, \vec{K}-\vec{q}) \delta_{\text{g}}(\vec{q})\delta_{\text{g}}(\vec{K}-\vec{q})\ ,
   \label{eq:quad-def}
\end{equation}
with weights $g_{\alpha}$, similar to what is done for CMB lensing \citep{Hu:2001kj} or ``clustering fossils" from primordial gravitational waves \citep{Masui:2010cz,Jeong:2012df,Masui:2017fzw}. For an alternative derivation of this estimator, based on optimizing the cross-correlation of a quadratic combination of measured modes with the true linear mode to be reconstructed, see Appendix~\ref{app:rec-from-b}.

The covariance between two such estimators $\alpha$ and $\beta$ of the biased matter density field on large scales can be split into a Gaussian part, coming from all disconnected contributions, and a non-Gaussian part that includes all connected contributions:
\beq
    \langle \hat{\Delta}_{\alpha}(\vec{K}) \hat{\Delta}^{*}_{\beta}(\vec{K'})   \rangle
    - \langle\hat{\Delta}_{\alpha}(\vec{K})\rangle\langle \hat{\Delta}^{*}_{\beta}(\vec{K'}) \rangle 
    = (2\pi)^3\dirac(\vec{K}-\vec{K'})
    \left[ {\rm Cov_G}(\hat{\Delta}_{\alpha}(\vec{K}), \hat{\Delta}^{*}_{\beta}(\vec{K'}) )+{\rm Cov_{NG}}(\hat{\Delta}_{\alpha}(\vec{K}), \hat{\Delta}^{*}_{\beta}(\vec{K'})) \right]\ .
\eeq
We constrain the weights to provide an estimator that is optimal in the sense of minimizing the Gaussian contributions to its variance,
\begin{equation}
    \mathrm{Var}_{\rm{G}}[\hat{\Delta}_{\alpha}](\vec{K})\equiv \mathrm{Cov}_{\rm{G}}(\hat{\Delta}_{\alpha}(\vec{K}), \hat{\Delta}^{*}_{\alpha}(\vec{K'}) )\ ,
\end{equation}
while requiring that it be unbiased if there were only a single mode-coupling, i.e.
\begin{equation}
    \int_{\vec{q}}g_{\alpha}(\vec{q}, \vec{K}-\vec{q}) f_{\alpha}(\vec{q}, \vec{K}-\vec{q})  = 1\ .
    \label{eq:gab}
\end{equation}
These criteria lead to the familiar quadratic estimator weights:
\beq
g_\alpha(\vk_1,\vk_2) = N_{\alpha\alpha}(\vk_1+\vk_2) 
	\frac{f_\alpha(\vk_1,\vk_2)}{2P_{\rm tot}(k_1) P_{\rm tot}(k_2)}\ ,
\label{eq:galpha}
\eeq
where $P_{\rm tot}$ is the sum of the clustering and shot noise contributions to the tracer power spectrum. The normalization is given by
\begin{equation}
\label{eq:nab}
    N_{\alpha\beta}(\vec{K})=\bigg( \int_{\vec{q}}\frac{f_{\alpha}(\vec{q}, \vec{K}-\vec{q})f_{\beta}(\vec{q}, \vec{K}-\vec{q})}{2P_{\rm tot}(q)P_{\rm tot}(|\vec{K}-\vec{q}|)} \bigg)^{-1}
\end{equation}
which guarantees that $N_{\alpha\alpha}$ is equal to the Gaussian part of the variance of $\hat{\Delta}_{\alpha}$. We will refer to  $N_{\alpha\alpha}$ as the reconstruction noise, which incorporates cosmic variance in the reconstruction and the disconnected contribution from shot noise of the tracer field. (It should be noted that $N_{\alpha \beta}$ is not equal to the noise when $\alpha \neq \beta$.) With the weights in Eq.~\eqref{eq:galpha}, the estimator in Eq.~\eqref{eq:quad-def} becomes
\begin{equation}
\label{eq:quadest}
   \hat{\Delta}_{\alpha}(\vec{K}) = N_{\alpha \alpha}(\vec{K})\int_{\vec{q}}\frac{f_{\alpha}(\vec{q}, \vec{K}-\vec{q})}{2P_{\text{tot}}(q)P_{\text{tot}}(|\vec{K}-\vec{q}|)} \delta_{\text{g}}(\vec{q})\delta_{\text{g}}(\vec{K}-\vec{q})\ .
\end{equation}

The non-Gaussian part of the variance includes a trispectrum contribution from clustering of the tracers, and further contributions from tracer shot noise. We neglect the former, because it is subdominant to the latter; our comparisons with simulations in Sec.~\ref{sec:sims} show that this is a valid approximation.  Importantly, the shot noise contributions can dominate over the Gaussian reconstruction noise in many cases, because these contributions couple to large-scale modes with large variance, while the Gaussian contribution only involves small-scale modes, which have smaller variance due to the shape of the matter power spectrum. We derive the full expressions for these contributions and discuss their hierarchy further in Appendix~\ref{app:shot}.

The expectation value of the estimator in Eq.~\eqref{eq:quadest}, for a given realization of the linear field at wavevector~$\vec{K}$, is
\begin{equation}
    \left\langle \hat{\Delta}_{\alpha}(\vec{K}) \right\rangle_{\delta_1(\vec{K})\text{ fixed}}  
    = b_1\left[c_\alpha +\sum_{\beta \neq \alpha} c_{\beta}\frac{N_{\alpha\alpha}(K)}{N_{\alpha \beta}(K)} \right] \delta_1(\vec{K})\ .\label{eq:meanfield}
\end{equation}
We clearly see that there is a contamination of the estimator with respect to the case of only a single mode-coupling, given by the product of the (Gaussian) noise for the estimator $\alpha$ and a sum of bias terms divided by the cross normalization between estimators $\alpha$ and $\beta$.\footnote{It can be seen from Eq.~\eqref{eq:nab} that as the overlap integral of the two mode-couplings goes to zero, $N_{\alpha \beta}$ becomes very large and the contamination vanishes.} If the goal is to just reconstruct the linear mode of interest, then it is important to account for this contribution. One can attempt to construct a so-called ``bias-hardened" estimator by forming a linear combination of the original estimators that is free of this contamination at leading order (e.g.\ \citealt{Namikawa:2012pe,Osborne:2013nna,Foreman:2018gnv}). However, for the specific mode-couplings relevant in this situation, the high degree of correlation between the original estimators implies that the noise on the new estimator will be so high that it is no longer useful; see Appendix~\ref{app:biashardening} for details.

We claim that, for extracting non-Gaussianity, this contamination can actually be useful. As we will see later, some of these contaminating terms induce scale-dependence that reproduces the $1/K^2$ scaling created by primordial non-Gaussianity. Depending of the signs of these terms, they can either raise or lower the signal to noise on $\fnl$ from the reconstructed field. We will discuss this further in Sec.~\ref{sec:noisecont}.

\subsection{Non-Gaussianity and bias expansion}
\label{sec:bias}

As we discussed in the introduction, primordial non-Gaussianity of the local type introduces a quadratic contribution to the metric perturbation. The metric perturbation (gravitational potential) $\varphi$ is related to the linear matter overdensity through the usual Poisson equation (dropping the subscript $_{\rm G}$)
\beq
\varphi(\vk,z) = \frac{\delta_1(\vk,z)}{M(k,z)}\, ,
\label{eq:phidef}
\eeq
where the Poisson factor $M(k,z)$ is given by
\beq
M(k,z) = \frac{2c^2}{3H_0^2\Omm} D(z) k^2 T(k)\, .
\label{eq:M}
\eeq
Here, the growth factor $D(z)$ is normalized to agree with the scale factor $1/(1+z)$ during matter domination.

Galaxies and \tcm fluctuations of the density field are biased tracers of the underlying, dynamically dominant matter distribution. In the presence of local primordial non-Gaussianity, the coupling of the long and short modes leads to an additional modulation of the abundance of collapsed objects by the long wavelength potential fluctuations~$\varphi$.
To describe biased tracers we thus follow \cite{Giannantonio:2009ak,Baldauf:2010vn} in performing a double expansion of the Eulerian galaxy (or tracer) density field in the non-linear density and linear potential\footnote{In the peak-background split formalism, the abundance of collapsed objects is given by $e^{-\nu^2/2}$ where $\nu=\delta_\text{c}/\sigma$ with $\delta_\text{c}$ the collapse threshold and $\sigma$ the variance. The long wavelength density modulates the collapse threshold as $\delta_c\to \delta_c-\delta$, whereas the metric perturbation $\varphi$ modulates the variance $\sigma\to\sigma(1+2\fnl \varphi)$.}
\begin{equation}
\begin{split}
\delta_{\rm g}^{\rm E}(\vx) =& b_{10}^{\rm E} \delta(\vx) + b_{01}^{\rm E} \varphi(\vx_{\rm L}[\vx])
	+ b_{20}^{\rm E} \delta^2(\vx) + b_{11}^{\rm E} \delta(\vx) \varphi(\vx_{\rm L}[\vx])
	+ b_{02}^{\rm E} \varphi^2(\vx)\\
	&+ b_{s^2}^{\rm E} s_{ij}(\vx) s^{ij}(\vx)
	+ \varepsilon(\vx) + \varepsilon_\delta(\vx) \delta(\vx) + \varepsilon_\varphi(\vx) \varphi(\vx_{\rm L}[\vx]) 
	+ \cdots
	\label{eq:nongausbiasexpansion}
\end{split}
\end{equation}
Here the $b_{ij}^{\rm E}$ are the Eulerian bias parameters\footnote{In the introductory discussion in Sec.~\ref{sec:qe}, we employed the notation $b_1\equiv b_{10}^\text{E}$ for the sake of simplicity.}, $s_{ij}$ is the tidal tensor
\beq
s_{ij}(\vx) = \lb \frac{\nabla_i \nabla_j}{\nabla^2} - \frac{1}{3} \delta_{ij}^{\rm (K)} \rb \delta(\vx)\, ,
\eeq
and $\varepsilon$ is the stochasticity, which correlates with itself but not with the linear density field. In the simplest case where galaxies are a Poisson sample of the underlying matter field, the stochasticity leads to the fiducial $1/\bar n$ power spectrum. The higher order stochasticity contributions $\varepsilon_\delta \delta$ and $\varepsilon_\varphi \varphi$ lead to stochasticity contributions in the bispectrum \citep{Desjacques:2016bnm}, as we review in App.~\ref{app:shot}.
 In simple local-Lagrangian bias models, the tidal tensor bias $b_{s^2}^\text{E}$ can be related to the linear density bias as $b_{s^2}^\text{E}=-2/7\ \left(b_{10}^\text{E}-1\right)$ \citep{Baldauf:2012ev}. Employing realistic simplifying assumptions, we will see that all of the bias parameters $b_{ij}^\text{E}$ can be expressed in terms of $b_{10}^\text{E}$ and $b_{20}^\text{E}$. We are truncating the above expansion at second order, since we will only consider tree level power spectra and bispectra as well as the Gaussian disconnected trispectrum in our derivations.
 We can thus also neglect higher derivative contributions, such as $k^2 \delta_1(\vec k)$, as they are equivalent to cubic contributions to the matter and galaxy density fields. Note that all of the $\delta$ terms in Eq.~\eqref{eq:nongausbiasexpansion} refer to the underlying non-linear matter density field including its quadratic couplings. The potential $\varphi$, in turn, is linear, as the dependence of the halo abundance on long wavelength potential fluctuations is set up in the early Universe.

There is, however, a non-linearity in the potential terms that arises from the fact that the abundance of galaxies in the peak-background split is set up in Lagrangian space with coordinates $\vx_{\rm L}$. These Lagrangian positions are related to the Eulerian coordinates by $\vx_{\rm L}[\vx] = \vx - \boldsymbol{\Psi}(\vx)$ at leading order.  %
The potential is thus advected by long wavelength displacements as \citep{Tellarini:2015faa}
\beq
\varphi(\vx_{\rm L}[\vx]) = \varphi(\vx) - \boldsymbol{\Psi}(\vx) \cdot \boldsymbol{\nabla}\varphi(\vx) + \cdots\ .
\eeq
The Fourier transform of the linear displacement field $\boldsymbol{\Psi}(\vx)$ is related to the linear matter overdensity by $\boldsymbol{\Psi}(\vk)=i(\vk/k^2)\delta(\vk)$. 

At second order, the matter density field picks up a new quadratic contribution from primordial non-Gaussianity according to Eq.~\eqref{eq:localNGs}:
\beq
\delta(\vk) = \delta_1(\vk) + \int_{\vq} \lb \sum_{\alpha=\text{G,S,T}} F_\alpha(\vq,\vk-\vq) \rb  
\delta_1(\vq)\delta_1(\vk-\vq)
	+ \fnl M(k) \int_{\vq}\varphi(\vq) \varphi(\vk-\vq) + \cdots,
\eeq
where the growth, shift and tidal components of the gravitational coupling kernel are given by Eq.~\eqref{eq:f2kernels}. For biased tracers, this expression gets multiplied by $b_{10}^\text{E}$. 
We can rewrite the last term in terms of the density field using the Poisson equation, resulting in a new quadratic coupling
\beq
F_{\varphi\varphi}(\vec k_1,\vec k_2)=\frac{M(|\vk_1+\vk_2|)}{M(k_1)M(k_2)},
\eeq
such that
\beq
\delta(\vk) = \delta_1(\vk) + \int_{\vq} \lb \sum_{\alpha=\text{G,S,T},\varphi\varphi} c_\alpha \fnl^{p_\alpha} F_\alpha(\vq,\vk-\vq) \rb  
\delta_1(\vq)\delta_1(\vk-\vq) + \cdots,
\eeq
where now $c_\alpha=\left\{1,1,1,1\right\}$ and $p_\alpha=\left\{0,0,0,1\right\}$.

Combining this result with the Fourier transform of the other second order bias terms in Eq.~\eqref{eq:nongausbiasexpansion} yields
\begin{align*}
\delta_\text{g}^{\rm E}(\vk) &= \lb b_{10}^{\rm E} + \frac{b_{01}^{\rm E}}{M(k)} \rb \delta_1(\vk)
	+ b_{01}^{\rm E} \int_{\vq} \frac{1}{2} 
		\lb \frac{\vq\cdot(\vk-\vq)}{q^2 M(|\vk-\vq|)} + \frac{\vq\cdot(\vk-\vq)}{|\vk-\vq|^2 M(q)} \rb
		\delta_1(\vq)\delta_1(\vk-\vq) \\
&\quad
	+ b_{10}^{\rm E}\int_{\vq} \lb \sum_{\alpha=\text{G,S,T,}\varphi\varphi} F_\alpha(\vq,\vk-\vq) \rb 
		\delta_1(\vq)\delta_1(\vk-\vq) \\
&\quad
	+ \fnl b_{10}^{\rm E} \int_{\vq} \frac{M(k)}{M(q)M(|\vk-\vq|)} \delta_1(\vq)\delta_1(\vk-\vq)
	+ b_{20}^{\rm E}  \int_{\vq}  \delta_1(\vq)\delta_1(\vk-\vq) \\
&\quad
	+ b_{11}^{\rm E} \int_{\vq} \frac{1}{2} \lp \frac{1}{M(q)} + \frac{1}{M(|\vk-\vq|)} \rp
		\delta_1(\vq)\delta_1(\vk-\vq)
	+ b_{02}^{\rm E} \int_{\vq} \frac{1}{M(q)M(|\vk-\vq|)} \delta_1(\vq)\delta_1(\vk-\vq) \\
&\quad
	+ b_{s^2}^{\rm E} \int_{\vq} \lb \frac{[\vq\cdot(\vk-\vq)]^2}{q^2|\vk-\vq|^2} - \frac{1}{3} \rb
		\delta_1(\vq)\delta_1(\vk-\vq).
	\numberthis
\end{align*}

The additional terms arising from the non-Gaussian bias can be encoded by the new quadratic coupling kernels
\beq
\begin{split}
F_{01}=\frac{1}{2} \vk_1\cdot \vk_2 \lp \frac{1}{k_2^2 M(k_1)}+\frac{1}{k_1^2 M(k_2)} \rp\, , \ \
F_{11}=\frac{1}{2} \lp\frac{1}{M(k_1)}+\frac{1}{M(k_2)} \rp\, , \ \ 
F_{02}=\frac{1}{M(k_1)M(k_2)}\, .
\end{split}
\eeq

The Eulerian bias parameters can be related to their Lagrangian counterparts through a spherical collapse calculation \citep{Giannantonio:2009ak,Baldauf:2010vn}:
\begin{align}
b_{10}^{\rm E} &= b_{10}^{\rm L}+1\ , \\
b_{20}^{\rm E} &= 2(a_1+a_2)b_{10}^{\rm L} + a_1^2 b_{20}^{\rm L}\ , \\
b_{01}^{\rm E} &= b_{01}^{\rm L}\ , \\
\label{eq:b11E}
b_{11}^{\rm E} &= a_1 b_{11}^{\rm L} + b_{01}^{\rm L}\ , \\
\label{eq:b02E}
b_{02}^{\rm E} &= b_{02}^{\rm L}\, ,
\end{align}
where $a_1=1$ and $a_2=-17/21$ are spherical collapse expansion factors. The non-Gaussian Lagrangian bias parameters can be obtained using the peak background split. They are given as the the derivatives of the mass function with respect to the long wavelength potential fluctuations. Assuming a universal mass function, the derivatives with respect to the potential can be related to the derivatives with respect to the long wavelength density, and consequently the bias parameters of the potential terms can be related to the bias parameters of the density terms:
\begin{align}
\label{eq:b01Lfinal}
b_{01}^{\rm L} &= 2\fnl \delta_{\rm c} \lp b_{10}^{\rm E} - 1 \rp\ , \\
\label{eq:b11Lfinal}
b_{11}^{\rm L} &= 2\fnl \lp \delta_{\rm c} 
	\lb \frac{b_{20}^{\rm E} - 2(a_1+a_2) \lp b_{10}^{\rm E}-1 \rp}{a_1^2} \rb 
	- \lb b_{10}^{\rm E} - 1 \rb \rp\ , \\
\label{eq:b02Lfinal}
b_{02}^{\rm L} &= 4 \fnl^2 \delta_{\rm c} \lp
	\delta_{\rm c} \lb \frac{b_{20}^{\rm E} - 2(a_1+a_2) \lp b_{10}^{\rm E}-1 \rp}{a_1^2} \rb 
	- 2 \lb b_{10}^{\rm E} - 1 \rb \rp\, ,
\end{align}
where $\delta_\text{c}$ is the spherical collapse threshold.
Note that small deviations from this simple scaling of non-Gaussian bias $b_{01}^{\rm L}$ with Gaussian bias $b_{10}^{\rm E}$ have been found in simulations \citep{Biagetti:2016ywx} and seem to depend on the way halos are identified.

\begin{table}
\begin{centering}
\renewcommand{\arraystretch}{1.7}
\begin{tabular}{ c|c|c|c }
 Mode Coupling ($\alpha$) & $p_\alpha$ &  $c_{\alpha}$	
 									& $F_{\alpha}(\vk_1, \vk_2)$ 
\\ 
\hline
\hline
G		& 0	& $b_1+\frac{21}{17} b_2$ 
									& $\frac{17}{21}$
\\ 
S		& 0	& $b_1$					& $\frac{1}{2}\lb\frac{1}{k_1^2}+\frac{1}{k_2^2}\rb(\vk_1\cdot\vk_2)$
 \\  
T		& 0	& $b_1 + \frac{7}{2}b_{s^2}$ 	& $\frac{2}{7} \lb \frac{(\vk_1\cdot \vk_2)^2}{k_1^2 k_2^2} 
										-\frac{1}{3} \rb$
 \\ 
$\varphi\varphi$ & 1	& $b_1$					& $\frac{M(|\vk_1+\vk_2|, z)}{M(k_1)M(k_2)}$
 \\  
$01$		& 1	& $2\delta_c(b_{1}-1)$ 		& $\frac{1}{2} \vk_1\cdot \vk_2 \lp \frac{1}{k_2^2 M(k_1)}
    +\frac{1}{k_1^2 M(k_2)} \rp$
 \\ 
$11$		& 1	& $2 \lp \delta_{\rm c} 
				\lb \frac{b_{2} - 2(a_1+a_2) \lp b_{1} -1 \rp}{a_1} \rb 
				- a_1 \lb b_{1} - 1 \rb \rp
				+ 2\delta_{\rm c} \lp b_{1} - 1 \rp$			
									& $\frac{1}{2} \lp\frac{1}{M(k_1)}+\frac{1}{M(k_2)} \rp$
 \\ 
$02$		& 2	& $4 \delta_{\rm c} \lp
				\delta_{\rm c} \lb \frac{b_2 - 2(a_1+a_2) \lp b_1-1 \rp}{a_1^2} \rb 
				- 2 \lb b_1 - 1 \rb \rp$			
									& $\frac{1}{M(k_1)M(k_2)}$
 \\ 
\end{tabular}
\caption{
\label{tab:modecouplings}
Mode couplings, $\fnl$ exponents, bias parameters and coupling kernels of the quadratic interactions for Eq.~\eqref{eq:deltag-condensed}.
}
\end{centering}
\end{table}

In summary, we can write for the galaxy density field up to second order in the presence of local type primordial non-Gaussianity:
\beq
\delta_{\rm g}(\vk) = \lb b_{10}^\text{E} + \fnl \frac{c_{01}}{M(k)} \rb \delta_1(\vk)
	+ \int_{\vq} \lb \sum_{\alpha} c_\alpha \fnl^{p_\alpha} F_\alpha(\vq,\vk-\vq) \rb 
	\delta_1(\vq) \delta_1(\vk-\vq)\ ,
	\label{eq:deltag-condensed}
\eeq
where $\alpha$ now runs over $\{ \text{G}, \text{S}, \text{T}, \varphi\varphi, 01, 11, 02 \}$ with the couplings given in Table~\ref{tab:modecouplings}. In this table, Eq.~\eqref{eq:deltag-condensed}, and throughout the rest of the paper, we have simplified the notation to $b_1 \equiv b_{10}^{\text{E}}$, $b_2 \equiv b_{20}^{\text{E}}$, and $b_{s^2} \equiv b_{s^2}^{\text{E}}$. Note that we have not included mode-couplings due to lensing, which are expected to be a subdominant contribution that is somewhat degenerate with the S term~\cite{Foreman:2018gnv}, nor have we incorporated redshift space distortions or anisotropic selection effects (see Sec.~\ref{sec:rsd} for discussion).

\subsection{Reconstruction noise and contamination}
\label{sec:noisecont}

With this formalism in place, we can now examine the noise of the reconstructed modes, and the contamination arising from the presence of multiple mode-couplings in the tracer field used for reconstruction.\footnote{For producing matter power spectra for forecasts, we relied on the nbodykit code (\url{https://github.com/bccp/nbodykit}). } We will show these quantities for a DESI-like survey (with specifications given in Sec.~\ref{sec:config}), but we have checked that the conclusions we draw from this case also apply to the other surveys we consider.

\begin{figure}[t]
\centering
\includegraphics[width=0.5\textwidth, trim = 10 10 10 10 ]{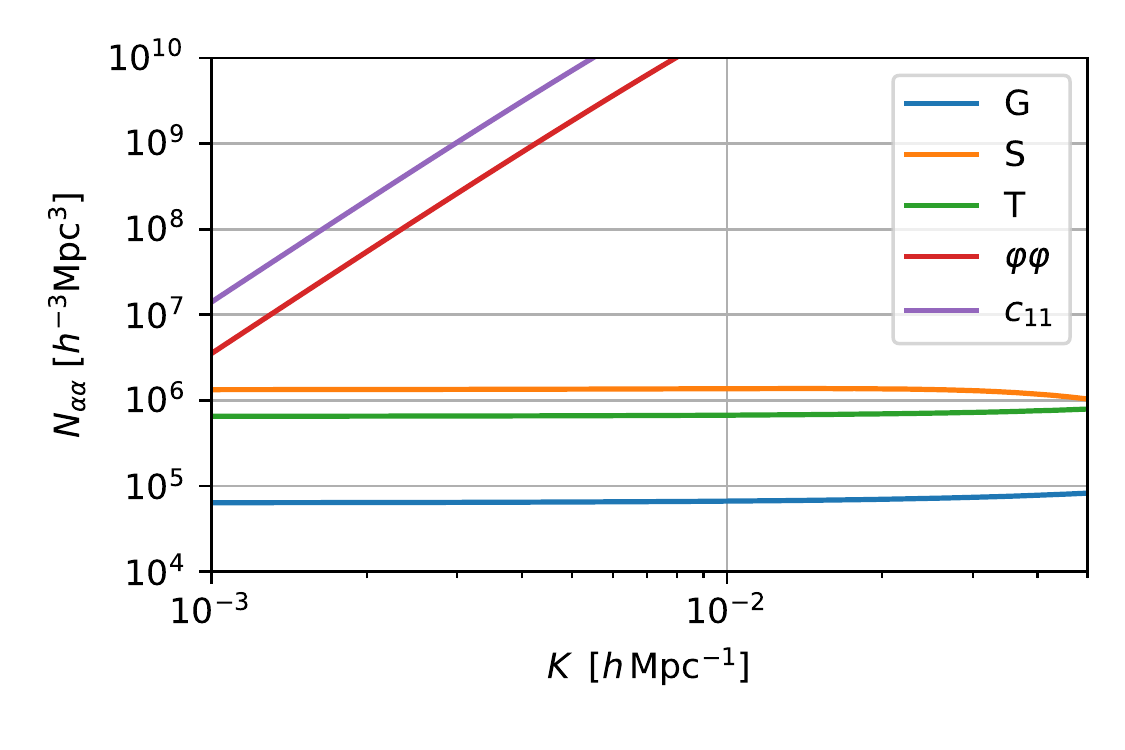} 
\caption{
\label{fig:noises}
Reconstruction noise power spectra for estimators that use each of the quadratic mode-couplings discussed in Sec.~\ref{sec:bias}. We omit curves for the $c_{01}$ and $c_{02}$ estimators, which are greater than the upper limit of the plot. The G (``growth") estimator has the lowest noise by far. These curves are computed for a DESI-like survey, but the hierarchy between them is unchanged for the other surveys we consider. The signal to noise on reconstructed modes (not shown) is likewise much higher for the G estimator than for S or T, justifying our use of the G estimator for our main results.
}
\end{figure}

Fig.~\ref{fig:noises} shows the reconstruction noise power spectrum corresponding to estimators that use each of the quadratic mode-couplings discussed in Sec.~\ref{sec:bias}. We see that the ``growth" estimator has the lowest noise by far. We compare the predicted noise for the G, S, and T estimators with results from $N$-body simulations in Sec.~\ref{sec:sims} (among other tests), finding good agreement. Thus, we use the growth estimator in our forecasts for reconstruction\footnote{
Out of the G, S, and T estimators, the G estimator yields both the lowest noise and the highest signal to noise on reconstructed modes. However, some of the other estimators (e.g.\ $\alpha=\varphi\varphi$) also have signal to noise approaching that of the G estimator, since the contaminating terms in Eq.~\eqref{eq:meanfield} act as ``signal" in a signal to noise computation. This indicates that a more optimal choice of estimator weights may be possible, although we leave this to future work.
}, henceforth referring to reconstructed modes as $\delta_{\rm r}(\vK)$ instead of $\hat{\Delta}_{\rm G}(\vK)$. However, as we discussed in Sec.~\ref{sec:qe}, the output of the G estimator (or any other single estimator) will be contaminated by the other mode-couplings, with the specific contamination given by Eq.~\eqref{eq:meanfield}, and we must incorporate this contamination into our forecasts.

\begin{figure}[t]
\centering
\includegraphics[width=\textwidth, trim = 10 10 10 10 ]{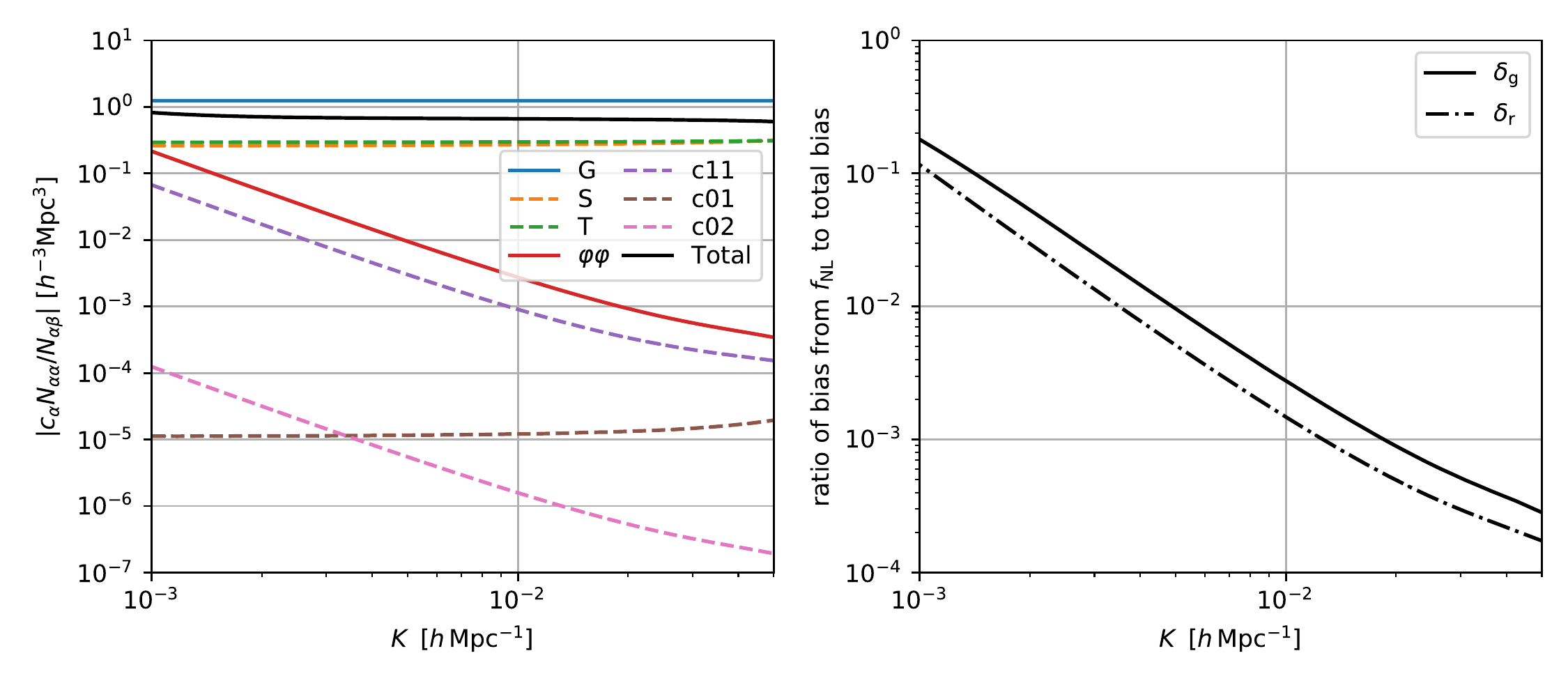} 
\caption{
\label{fig:contamination}
{\it Left:} Contamination in the expectation value of G estimator, corresponding to separate multiplicative biases on the amplitude of a reconstructed mode, computed for a DESI-like survey. Blue solid line is the estimator growth bias shown for comparison. Dashed lines indicate negative values. Several of these curves inherit the $k^{-2}$ scaling of the scale-dependent bias in $\deltag$ arising from nonzero $\fnl$, implying that reconstructed modes can be used to constrain $\fnl$ in the same way. {\it Right:} Ratio of scale-dependent bias from $\fnl$ (for a fiducial value of $\fnl=1$) to total bias for $\deltag$ (solid) and $\delta_{\rm r}$ (dot-dashed). Local primordial non-Gaussianity has roughly the same relative contribution to the bias of $\deltag$ or reconstructed modes.
}
\end{figure}

We show this contamination in the left panel of Fig.~\ref{fig:contamination}, in the form of each term $c_\beta N_{\rm GG}/N_{\rm G\beta}$ in the square brackets of Eq.~\eqref{eq:meanfield}. These curves each represent separate multiplicative biases on the amplitude of a reconstructed mode. Those arising from late-time gravitational evolution (S, T) or from advection of the primordial potential ($c_{01}$) are white in~$K$.
In contrast, those arising from couplings between $\delta$ and $\varphi$ ($c_{11}$) or $\varphi$ and itself ($\varphi\varphi$, $c_{02}$) scale like $M(K)^{-1} \propto K^{-2}$. We derive these scalings analytically in the large-scale limit in Appendix~\ref{app:meanfield}. Importantly, all terms that scale like~$K^{-2}$ involve $\fnl$, such that, as for $\deltag$, low-$K$ scale-dependent bias in the reconstructed modes can be used as a probe of local primordial non-Gaussianity. The right panel of Fig.~\ref{fig:contamination} shows that the relative size of this scale-dependent bias is comparable for $\delta_{\rm g}$ and $\delta_{\rm r}$, reaching $\mathcal{O}(10\%)$ at $K\sim 0.001\invMpc$, assuming $\fnl=1$. 

Fig.~\ref{fig:contamination} also shows that the contamination from other mode-couplings is subdominant to the intrinsic bias on the reconstructed field (i.e.\ the $c_{\rm G}$ term in Eq.~\ref{eq:meanfield}). Thus, using $c_{\rm G}=b_1+(21/17)b_2$ from Table~\ref{tab:modecouplings}, we can derive the rough dependence of $P_{\rm rr}$ and $P_{\rm gr}$ on $b_1$ and $b_2$:
\beq
P_{\rm rr} \propto b_1^2 (b_1+b_2)^2\ , 
	\quad P_{\rm gr} \propto b_1^2 (b_1+b_2)\ .
\eeq
If galaxy shot noise is negligible compared to $P_{\rm gg}$, then the reconstruction noise $N_{\rm GG}$ satisfies $N_{\rm GG} \propto b_1^4$, implying that $P_{\rm rr} / N_{\rm GG} \propto (1+b_2/b_1)^2$ in this regime. This scaling will be useful to help understand the behavior of our forecasts when we change the fiducial value of $b_2$.

\section{Simulations}
\label{sec:sims}

To validate the quadratic estimator framework presented in Sec.~\ref{sec:formalism}, we use a suite of 15 realisations of a cosmological $N$-body simulation. The initial conditions are generated with the second order Lagrangian Perturbation Theory (\textbf{2-LPT}) code \citep{Scoccimarro:2011pz} at the initial redshift $z_\text{i}=99$ and are subsequently evolved using \textbf{Gadget-2} \citep{Springel:2005mi}. The simulations are performed with $N_\text{p} = 1024^{3}$  dark matter particles in a cubic box of length $L=1500 h^{-1}$ Mpc with periodic boundary conditions. We assume a flat $\Lambda$CDM cosmology with the cosmological parameters $\Omega_\text{m}=0.272$, $\Omega_\Lambda=0.728$, $h=0.704$, $n_\text{s}=0.967$, $\sigma_8 =0.81$.

Dark matter halos in the final $z=0$ density field are identified using a Friends-of-Friends (FoF) algorithm with linking length $l=0.2$ times the mean interparticle distance. The halos are binned in mass, with each bin spanning a factor of three in mass. We have checked the viability of our reconstruction method for a range of masses, finding qualitatively similar results in all cases; however, for simplicity, we present only the results for the lowest mass bin, the properties of which are given in Table~\ref{halotable}. Particles and halos are assigned to a regular grid using the Cloud-in-Cell (CIC) scheme. We Fourier transform the matter and halo density fields using the publicly available \textbf{FFTW} library\footnote{\url{http://www.fftw.org}}.

\begin{table}[t]
\small
\begin{center}
\begin{tabular}{|c|c|c|c|c|c|c|}
\hline
  Mass Bin & Mean Halo Mass $[10^{13} h^{-1}M_{\odot}]$ & $\bar{n}$ $[10^{-6}h^{3}$ Mpc$^{-3}$] & $b_1$ &$c_g$&$c_t$&$c_s$\\  [0.2ex] 
\hline
I  & 0.77 & 627 & 1.07&0.62&1.14&1.07 \\ 
\hline
\end{tabular}
\end{center}
\caption{Properties of the halo mass bin employed in this study: the mean mass of the sample, the number density of halos~$\bar{n}$, the linear bias $b_1$, and the three relevant $c_\alpha$ parameters defined in Table~\ref{tab:modecouplings}. The measured bias parameters are taken from~\cite{Abidi2018}, which is based on the same simulations and mass bin we use here.} 
\label{halotable}
\end{table}

\subsection{Generation of quadratic estimators}
We generate quadratic estimators from the halo density field $\delta_\text{g}$ in $N$-body simulations using the convolution theorem. This means that we use a sequence of multiplications with powers of wavenumbers in Fourier space, Fourier transforms, and subsequent multiplication of the weighted fields in configuration space. We generate three quadratic estimators corresponding to the growth term $\delta^2$, shift term $\Psi\cdot\nabla\delta$, and the tidal term $s^2$, with associated Fourier-space kernels given in Eq.~\eqref{eq:f2kernels}. The first step in our procedure is to remove very small scale modes by applying a cut-off $k_{\text{max}}$ in Fourier space through multiplication of the Fourier space density field with a filtering function. While the exact form of the cutoff is not important, we adopt a Gaussian filter $W(R \vk) = \exp\left(-{k^2R^2}/{2}\right)$ for numerical stability.
We define the smoothed density field by $\delta_{\text{g}}^{R}(\vk)$. We choose three external smoothing scales: $R=20 h^{-1}$ Mpc, $R=10 h^{-1}$ Mpc, and $R=4 h^{-1}$ Mpc, corresponding to maximum wavenumbers $k_{\text{max}}\approx 0.05 \invMpc$, $k_{\text{max}}\approx0.1 \invMpc$, and $k_{\text{max}}\approx 0.25\invMpc$ respectively. The smoothing scale removes all wavenumbers $k>k_{\text{max}}$, such that we reconstruct long wavelength modes using modes  $k<k_{\text{max}}$ for three different cases. 

The mode coupling functions $g_{\alpha}(\vq,\vk-\vq)$ defined by Eq.~\eqref{eq:galpha} contain a Wiener filter, which we implement by first generating the linear power spectrum on the simulation grid, and then defining two fields:
\begin{equation}
    \delta_{A}(\vk) = \frac{\delta_{\text{g}}^{R}(\vk)}{b_1^2 P_{\text{lin}}(\vk) + \bar{n}^{-1}}\qquad \text{and} \qquad
    \delta_{B}(\vk) = \frac{\delta_{\text{g}}^{R}(\vk)P_{\text{lin}}(\vk)}{b_1^2 P_{\text{lin}}(\vk) + \bar{n}^{-1}},
    \label{eq:dB}
\end{equation}
where $b_1$ and $\bar{n}$ are the linear bias and halo number density corresponding to the halo mass bin defined in Table~\ref{halotable}. Using $\delta_{A}$ and $\delta_B$ we generate growth, shift and tidal estimators using  multiplications of powers of wavenumbers in Fourier space, Fourier transforms, and multiplication of fields in configurations space. For example, we generate the growth estimator as follows. First, we inverse Fourier transform both fields defined in Eq.~\eqref{eq:dB} to obtain $\delta_{A}(\vx)$ and $\delta_{B}(\vx)$. Next, in configuration space, we multiply the product of both fields by $17/21$ (Table~\ref{tab:modecouplings}) and finally Fourier transform back to obtain the growth estimator in Fourier space. We generate shift and tidal estimators with a similar procedure. 

Note that the main computational cost in generating the quadratic estimators comes from performing the Fourier transforms. The auto- and cross-spectrum analysis of quadratic estimators only requires the computational cost of a power spectrum analysis, which is quite efficient. In all our figures in this section, we estimate the errorbars of our measurements using the standard deviation of 15 simulation realisations.

\begin{figure}[t]
    \centering
    \includegraphics[width=0.95\textwidth]{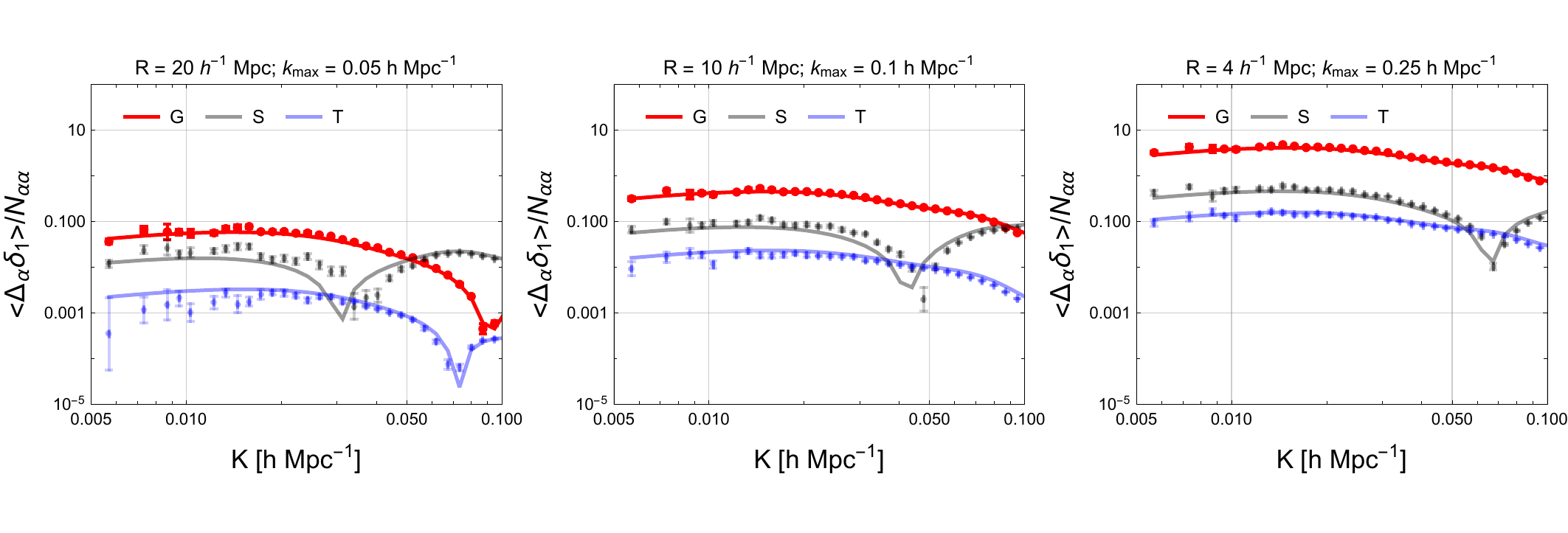}
    \caption{Cross correlations of estimators $\hat{\Delta}_{\alpha}$ corresponding to the growth, shift, and tidal mode-couplings with the linear density field $\delta_1$. We compare theory predictions (lines) with simulations (points) for three different smoothing scales, $R=20 h^{-1}$ Mpc, $R=10 h^{-1}$ Mpc and $R=4 h^{-1}$ Mpc, corresponding to maximum wavenumbers $k_{\text{max}} = 0.05 \invMpc$, $0.1\invMpc$, and $0.25 \invMpc$ respectively.
    In this figure, we plot $\langle\hat{\Delta}_{\alpha} \delta_1 \rangle/N_{\alpha \alpha}$ (in contrast to what is defined in Eq.~\eqref{eq:quadest}, in simulations we define the estimators $\hat{\Delta}_{\alpha}$ without a prefactor $N_{\alpha\alpha}$). We find very good agreement for the growth estimator for all smoothing scales, and also reasonably good agreement for the other estimators.}
    \label{fig:simAidelta}
\end{figure}

\subsection{Cross-correlation of quadratic estimators with the initial linear field}
In this section, we describe our results for the cross-correlations of three quadratic estimators $\hat{\Delta}_{\alpha}(\vk)$ with the initial linear field $\delta_{1}(\vk)$, and compare the theory predictions with simulations. The prediction is given by
\begin{align*}
    \langle\hat{\Delta}_{\alpha}(\vk) \delta_1(\vk')\rangle' 
    &= b_1 N_{\alpha\alpha}(\vk)\sum_{\beta\in\{G,S,T\}}c_{\beta}P_{\text{lin}}(\vk)
    \int_{\vq} \frac{f_{\alpha}(\vq,\vk-\vq)f_{\beta}(\vq,\vk-\vq)}{2P_{\text{tot}}(\vq)P_{\text{tot}}(\vk-\vq)}
    W(R\vq)W(R(\vk-\vq))+ P_{\alpha,\text{shot}}(\vk)\\
    &= b_1P_{\text{lin}}(\vk)\sum_{\beta\in\{G,S,T\}}c_{\beta}\frac{N_{\alpha\alpha}(\vk)}{N_{\alpha\beta}(\vk)} + P_{\alpha,\text{shot}}(\vk)\ ,
    \numberthis
    \label{eq:Aidelta}
\end{align*}
where the prime on the left-hand side denotes that the factor of $(2\pi)^3 \dirac(\vk+\vk')$ has been omitted, and
 $c_{\beta}$ are bias parameters corresponding to the growth, shift and tidal terms and can be measured from either simulations or data. In our analysis we use the bias parameters from Table~\ref{halotable}, measured in simulations in \cite{Abidi2018}. In Eq.~\eqref{eq:Aidelta}, $P_{\alpha,\text{shot}}$ is the bispectrum shot noise term. Since one field is the linear field, all contribution to this shot noise comes from the stochastic bias terms in the two galaxy fields $\deltag$ in the quadratic estimator, such as $\varepsilon$ and $\varepsilon_{\delta} \delta$ (see App.~\ref{app:shot} or \citealt{Desjacques:2016bnm} for more discussion about stochastic bias terms). The expression for this shot noise contribution in this case can also be derived from Eq.~\eqref{eq:nrtshot} and it takes the form
\begin{equation}
    P_{\alpha,\text{shot}}(\vk) = \frac{b_1}{\bar{n}}P_{\text{lin}}(\vk)N_{\alpha\alpha}(\vk)\int_{\vq}
    \frac{f_{\alpha}(\vq,\vk-\vq)}{2P_{\text{tot}}(\vq)P_{\text{tot}}(\vk-\vq)}
    W(R\vq)W(R(\vk-\vq)) .
    \label{eq:shotalpha}
\end{equation}

\begin{figure*}[t]
\centering
	 \includegraphics[width=0.95\textwidth]{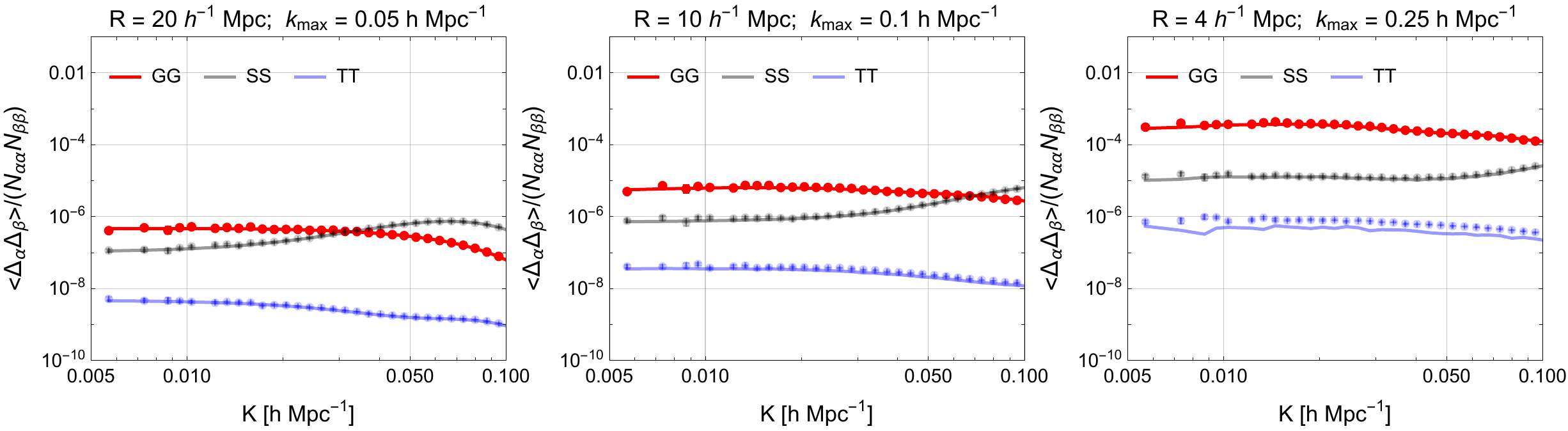}
        \caption{Auto-correlations of the quadratic estimators $\hat{\Delta}_{\alpha}$, for the same smoothing scales shown in Fig.~\ref{fig:simAidelta}. The predictions for the growth estimator agree with simulations for all smoothing scales. However, for other estimators predictions agree with simulations for large smoothing scales but for the low smoothing scales, the predictions slightly disagree with simulation results as higher-order terms become more important.}
        \label{fig:AiAjsim}
\end{figure*}
In Fig.~\ref{fig:simAidelta}, we compare theory with simulations for three different values of $k_{\text{max}}$. Although for the Fisher analysis in this work, we only use the growth estimator, here we also compare results in simulations for the shift and the tidal estimators. For the growth estimator, we find that the theory predictions agree very well with simulation results for up to $k_{\text{max}}=0.25\invMpc$ at redshift $z=0$. For the other estimators, we also find reasonably good agreement; however, upon close inspection we can see small disagreements which might arise from higher-order terms ignored in our theory predictions.

Interestingly, for $k_{\text{max}}=0.25\invMpc$, we can see in Fig.~\ref{fig:simAidelta} that the shape of the cross-correlation of growth estimators with the density field is very similar to the linear power spectrum on large scales. The scale-dependent bias factor in Eq.~\eqref{eq:Aidelta} is flat on large scales, indicating that the reconstruction works very well for large $k_{\text{max}}$.

\subsection{Auto- and cross-correlations of quadratic estimators}
In this section we discuss our results for the auto- and cross-correlations of three quadratic estimators from simulations and compare the results with our linear order theoretical prediction, given by
\begin{align*}
    \langle\hat{\Delta}_{\alpha}(\vk) \hat{\Delta}_{\beta}(\vk')\rangle'
    & = b_1^4N_{\alpha \alpha}(\vk)N_{\beta \beta}(\vk)
    \int_{\vq} \frac{f_{\alpha}(\vq,\vk-\vq)f_{\beta}(\vq,\vk-\vq)}{\left[ 2P_{\text{tot}}(\vq)P_{\text{tot}}(\vk-\vq)\right]^2}
    W(R\vq)^2W(R(\vk-\vq))^2
    P_{\text{lin}}(\vq)
    P_{\text{lin}}(\vk-\vq) \\
    &\quad+ b_1^2 P_{\text{lin}}(\vk)\sum_{i,j}c_ic_j\frac{N_{\alpha \alpha}(\vk)}{N_{\alpha i}(\vk)}\frac{N_{\beta \beta}(\vk)}{N_{\beta j}(\vk)} + P_{\alpha\beta,\text{shot}}(\vk)\ .
    \numberthis
\label{eq:simAiAj}
\end{align*}
The first term is of order $\mathcal{O}(\delta_1^4)$, while the second and third are of order $\mathcal{O}(\delta_1^6)$. The third term, $P_{\alpha\beta,\text{shot}}$, is the contribution arising from halo shot noise, and is given in App.~\ref{app:shot}.

In Fig.~\ref{fig:AiAjsim} we compare cross-correlation results from simulations with theory, for the growth, shift, and tidal estimators, using the same three smoothing scales as above. The simulations and theory agree very well up to $k_{\text{max}}=0.1\invMpc$ at $z=0$. For larger $k_{\text{max}}$ we see good agreement for the growth estimator and reasonable agreement for the tidal and shift estimators. The small disagreement of linear predictions for the tidal and shift estimators with simulations for the higher $k_{\text{max}}$ show that higher-order terms become important for these estimators. The detailed impact of these higher order corrections from biasing or scale dependent stochasticity will be subject of future inquiry. Although we appear to have excellent agreement for the growth term at higher $k_{\text{max}}$, to be conservative, we still set the scale $k_{\text{max}}=0.1 h$ Mpc$^{-1}$ at redshift $z=0$ in our forecasts in Sec.~\ref{sec:forecasts}. We scale this to other redshifts by making use of the fact that perturbation theory and the bias expansion at a given order will be valid at higher $k$ for higher redshifts.

\begin{figure*}[t]
\centering
	 \includegraphics[height=4.8in,width=0.9\textwidth]{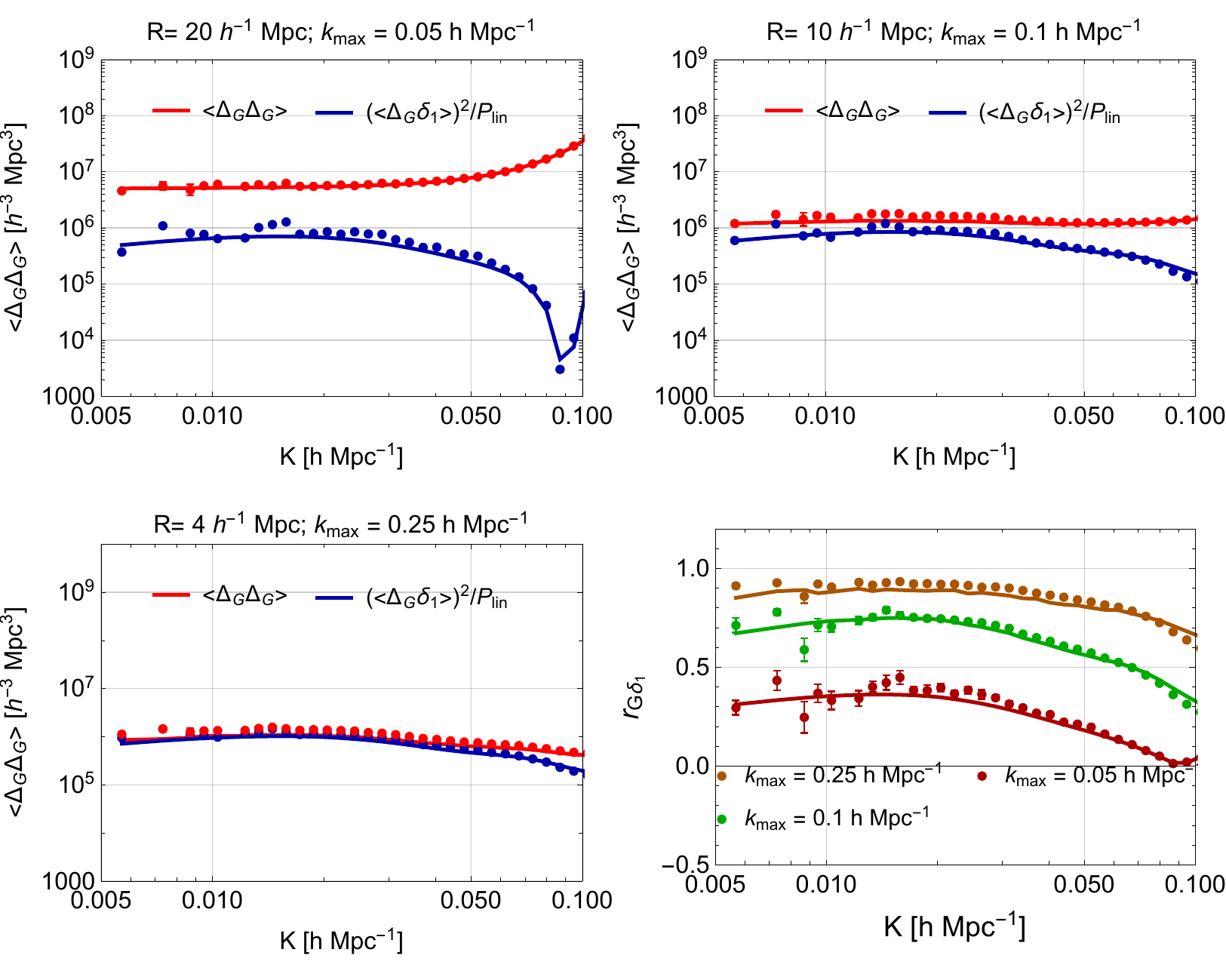}
        \caption{Comparison of the auto power spectrum of the growth estimator $\hat{\Delta}_{\text{G}}$, normalised by $N_{\text{GG}}$ computed from theory (in red), with $(\langle\hat{\Delta}_{\text{G}}\delta_1\rangle)^2/P_{\text{lin}}$ (in blue). We compare simulation results (points) with theory predictions (lines) for the same smoothing scales as Figs.~\ref{fig:simAidelta} and~\ref{fig:AiAjsim}. We again find excellent agreement between simulations and theory. In the bottom right panel, we plot the cross-correlation coefficient $r_{\text{G}\delta_1}$ between the growth estimator and the linear density field for three smoothing scales. We see that $r_{\text{G}\delta_1} >0.9$ for $R=4 h^{-1}$ Mpc which is why in the bottom left panel, $\langle\hat{\Delta}_{\text{G}} \hat{\Delta}_{\text{G}}\rangle$ is signal dominated.} 
        \label{fig:growthsim}
\end{figure*}

In Fig.~\ref{fig:growthsim}, we plot the auto spectra of the growth estimator, normalized with $N_{\rm GG}$ (unlike in the previous plots), in order to compare them to an approximation of the signal power spectrum, given by the second term in Eq.~\eqref{eq:simAiAj} (the first and third terms represent noise).
Since the contribution of the cross-shot noise is small, the signal part can be approximated by cross-correlating the growth estimator with the linear density field and dividing it by the linear power spectrum to ensure the correct normalization, i.e. $(\langle\hat{\Delta}_{\text{G}}\delta_1\rangle)^2/P_{\text{lin}}$; we show this in blue in Fig.~\ref{fig:growthsim}. For the two larger smoothing scales, the spectra of the estimator are dominated by the noise contribution (which is white at low $k$). The excellent agreement between theory (red solid lines) and simulations (red points) for all smoothing scales serves as an additional verification that the reconstruction procedure is working as expected for reasonable values of $k_{\rm max}$. In addition to the auto spectra, to check how well the reconstruction is working, we plot the cross-correlation coefficients between the growth estimator and the linear density field in the bottom right panel of Fig.~\ref{fig:growthsim} for three different $k_{\text{max}}$. The cross-correlation coefficient for low $k_{\text{max}}$ is very low, $r_{\text{G}\delta_1} < 0.4$, explaining why the auto spectra in the top left panel are noise dominated. However, for highest $k_{\rm max}$ we consider,  $0.25\invMpc$, the cross-correlation coefficient is $r_{\text{G}\delta_1} > 0.9$, which explains why the reconstruction works very well and the auto spectra for high $k_{\text{max}}$ are signal dominated. 

\subsection{Visualization of reconstructed field}
\begin{figure}[t]
    \centering
    \includegraphics[width=0.42\textwidth]{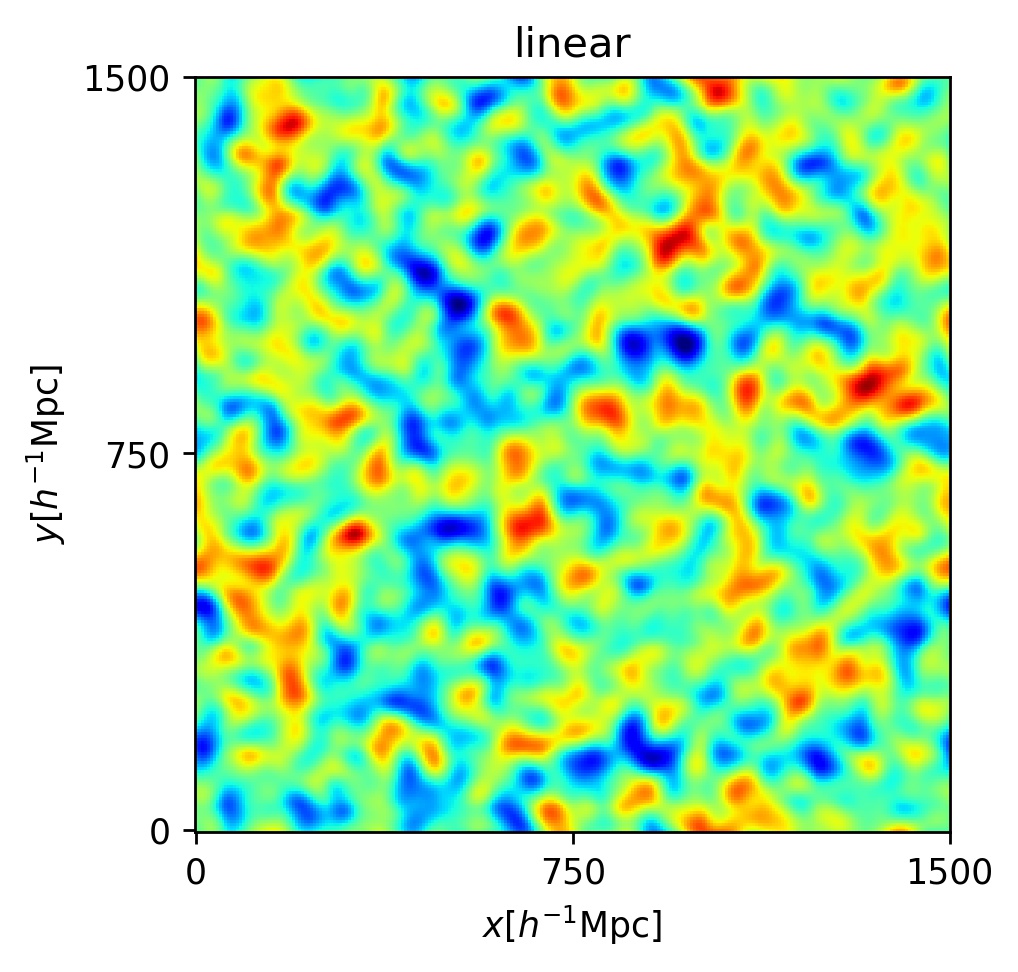}
    \includegraphics[width=0.485\textwidth]{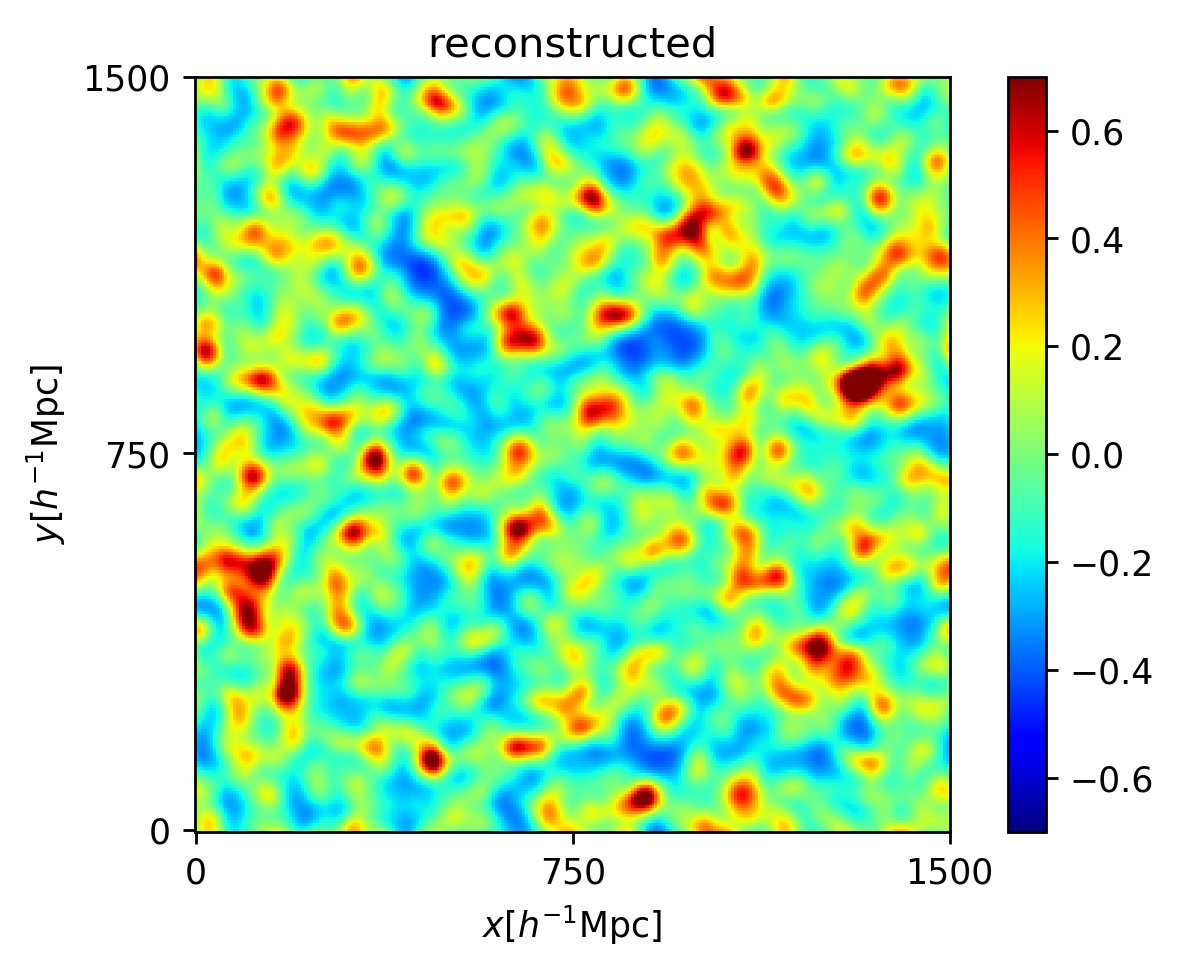}
    \caption{2D slices of the 3D linear density field (left panel) and the growth estimator $\hat{\Delta}_{\text{G}}$ (right panel). For the growth estimator we used $R=4 h^{-1}$ Mpc smoothing which corresponds to $k_{\text{max}}=0.25 h$ Mpc$^{-1}$. We apply an external smoothing of $R=20 h^{-1}$ Mpc to both the linear and reconstructed fields. As expected, we find that the reconstruction reproduces many of the large-scale features in the linear density field.}
    \label{fig:densplot}
\end{figure}

\begin{figure}
    \centering
    \includegraphics[width=0.55\textwidth]{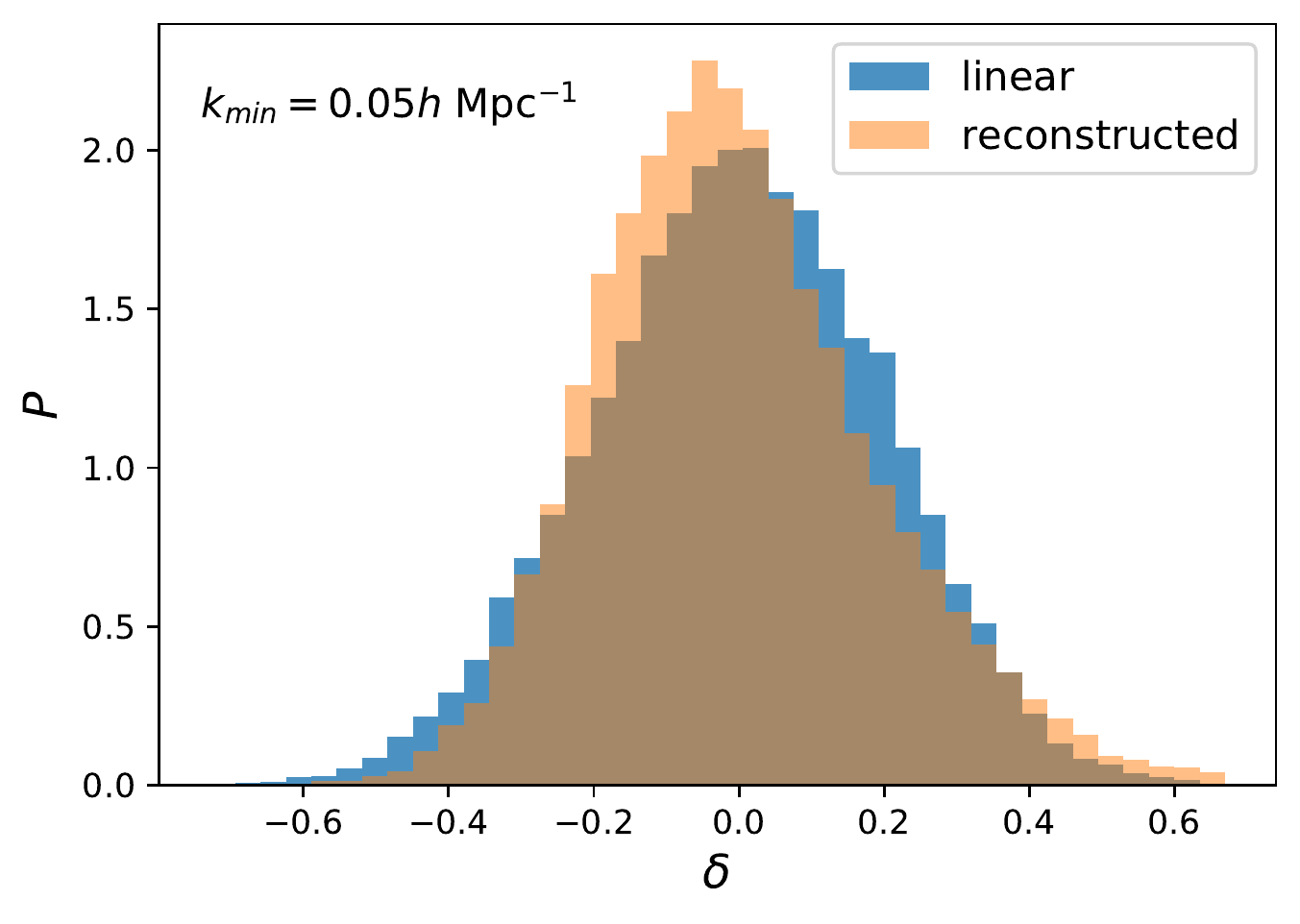}
    \caption{Probability distribution functions (histograms) of the linear density field and the reconstructed field from the halo density field of mass bin I. As in Fig.~\ref{fig:densplot}, we use $k_{\text{max}} = 0.25 \invMpc$ for the reconstruction and apply an external smoothing scale of $R=20h^{-1}$ Mpc to both the linear field and the reconstructed field. The PDFs of the reconstructed field are scaled to have the same variance as the linear field, and shifted to have mean 0. We find that the PDF of the reconstructed field is very close to Gaussian. Note that here we have applied a low-$k$ cutoff to the modes used for reconstruction of $k_\text{min}=0.05 h\ \text{Mpc}^{-1}$ in order to match the approach in our forecast section below.}
    \label{fig:pdfsdh}
\end{figure}

To visualize how well we are reconstructing the linear density field on large scales in simulations, we compare 2D slices of thickness $6 h^{-1}\text{Mpc}$ of the linear density field and the reconstructed field in Fig.~\ref{fig:densplot}. We perform the reconstruction using $k_{\text{max}} = 0.25 h$ Mpc$^{-1}$, i.e.,  smoothing at a scale of $R= 4 h^{-1}$ Mpc. In the visualization, we apply an external smoothing of $R=20 h^{-1}$ Mpc to both the linear field and the reconstructed field, which removes all modes with $k > 0.05 \invMpc$. Our comparison of the linear and reconstructed fields in Fig.~\ref{fig:densplot} shows that the reconstruction indeed recovers most of the large scale features in the linear density field.
In Fig.~\ref{fig:pdfsdh} we show histograms, probing the one-point probability distribution functions, of the linear density field and the reconstructed field. We see that the reconstructed field is nearly Gaussian, partially justifying our approximation of a Gaussian likelihood in the next section.

\section{Forecasts}
\label{sec:forecasts}

\subsection{Fisher matrix setup }

To perform a Fisher forecast, we make the usual assumption that the measured tracer overdensity $\deltag$ and the reconstructed field $\delta_{\rm r}$ obey a Gaussian likelihood. For the matter and galaxy field, this approximation is partially justified by the fact that we are analyzing very large scales; for the reconstruction noise, this is partially justified by the fact that the reconstruction sums over a large number of mode pairs, so that to some extent the central limit theorem applies (although the pairs may not all be independent). Figure~\ref{fig:pdfsdh} supplies additional evidence that a simple Fisher forecast is sufficient, in that the PDFs of the density field (smoothed to correspond with our analysis range) do not greatly deviate from a Gaussian. This indicates that the influence of higher moments of the density field and noise is comparatively small for the purposes of a forecast.

Making this approximation and including the fact that $\deltag$ has zero statistical mean, the Fisher matrix per mode $\vec{K}$ and redshift $z$ is given by (e.g.\ \citealt{Tegmark:1996bz})
\beq
\tilde{F}_{\rm{a}\rm{b}}(\vec{K}, z) = \frac{1}{2}\text{Tr}
	\Big[ \d_{\rm{a}} C(\vec{K}, z) C^{-1}(\vec{K}, z) \d_{\rm{b}} C(\vec{K}, z) C^{-1}(\vec{K}, z)\Big],
	\label{eq:fisherpermodegeneral}
\eeq
where $C$ is the total (signal plus noise) covariance matrix for our data vector $\vec{d}(\vec{K})=\left(\delta_{\text{g}}(\vec{K}), \delta_{\text{r}}(\vec{K})\right)^{\text{T}}$, $\text{Tr}$ is the trace matrix operator, $\partial_{\rm{b}}C(\vec{K}, z)\equiv \frac{\partial}{\partial{\rm{b}}}C(\vec{K}, z)$, and $\rm{a}, \rm{b}$ are the parameters on which our quantities depend (in this case, $f_{\text{NL}}$ and bias parameters). If the data vector is drawn from a Gaussian distribution and nothing is known about the parameters, then the inverse of the Fisher matrix gives the covariance matrix of the parameters, and the square root of the diagonal elements of $F^{-1}$ give the errorbars on the parameters and represent the minimum error achievable. Our goal is to calculate this minimum error, as it will determine our best ability to constrain parameters.

In reality, we do not just measure a single mode, but we measure several modes whose information can combined together in an integrated Fisher matrix for a specific redshift bin, i.e. 
\begin{equation}
\label{eq:integratedfishermatrix}
    F_{\rm{a}\rm{b}}(z) = \frac{V}{(2\pi)^2} \int_{K_{\text{min}}}^{K_{\text{max}}}
    \text{d}K \int_{-1}^{1} \text{d}\mu \,  K^2 \tilde{F}_{\rm{a}\rm{b}}(K, \mu, z).  
\end{equation}
Here $V$ is the survey volume, $K_{\text{min}}$ and $K_{\text{max}}$ are the minimum and maximum moduli of the modes probed, and we already integrated over the azimuthal direction, supposing no dependence from it in the integrand.

For our specific case, the original field $\deltag$ and the reconstructed field $\delta_{\rm r}$ will give the total covariance matrix (which only depends on the magnitude of $\vK$)
\begin{equation}
C(K, z)=\left[ \begin{array}{ccc} 
	C^{\text{gg}}(K, z) & C^{\text{gr}}(K, z)\\
	C^{\text{gr}}(K, z) & C^{\text{rr}}(K, z)\ 
	\end{array} \right], \label{eq:covmatrix}
\end{equation}
with elements
\begin{align}
\label{eq:cgg}
C^{\text{gg}}(K, z) 
	&= \lp b_1(z) + \frac{c_{01} \fnl}{M(K,z)} D(z) \rp^2
	P_\text{lin}(K, z) + P_{\text{gg,shot}}(K,z)\ , \\
\label{eq:cgr}
C^{\text{gr}}(K, z) 
	&= \lp b_1(z) + \frac{c_{01} \fnl}{M(K,z)} D(z) \rp b_{\rm r}(K,z)
	P_\text{lin}(K, z) + P_{\text{gr,shot}}(K,z)\ , \\
\label{eq:crr}
C^{\text{rr}}(K, z) 
	&= b_{\rm r}(K,z)^2
	P_\text{lin}(K, z) + N_{\rm GG}(K,z) + P_{\text{rr,shot}}(K,z)\ ,
\end{align}
where
\beq
b_{\rm r} \equiv b_1 \left( c_{\rm G} + \sum_{\beta \neq {\rm G}} c_{\beta}
	\frac{N_{\rm GG}}{N_{\text{G} \beta}} \right),
\eeq
and the sum runs over the mode-couplings found in Table~\ref{tab:modecouplings}. We do not include redshift space distortions in these expressions; see Sec.~\ref{sec:rsd} for discussion.

The tracer shot noise is simply 
\beq
P_\text{gg,shot}(K,z) = \frac{1}{\bar{n}(z)}\, ,
\label{eq:pggshot}
\eeq
where $\bar{n}$ is the comoving number density of observed tracers, while $P_{\text{rr,shot}}$ and $P_{\text{gr,shot}}$ are given in Eqs.~\eqref{eq:nrrshot}-\eqref{eq:prrshot-appendix} and \eqref{eq:nrtshot}-\eqref{eq:pgrshot-appendix} respectively. We will neglect the dependence of the reconstruction shot noise on $\fnl$. This is because in general these shot noise terms include the small scale  tracer power spectrum, whose response to a change of $\fnl$ is negligible compared to the response experienced by the large scale power spectrum. Moreover, even when the large scale tracer power spectrum enters the reconstruction shot noise, as in $P_{\rm rr,shot}$ where there is a coupling between large and small scales as we explain in Appendix \ref{app:shot}, a small change from $f_{\rm{NL}}=0$, our fiducial value, is barely detectable. In principle, it may be possible extract additional information from the $\fnl$-dependence of the shot noise contributions, but this will likely be difficult in practice, and therefore we conservatively choose not to consider these contributions as observables.

Substituting Eq.~\eqref{eq:covmatrix} into Eq.~\eqref{eq:fisherpermodegeneral}, we can derive an explicit formula for the Fisher matrix per mode for our case, which can then be inserted into Eq.~\eqref{eq:integratedfishermatrix}:
\begin{align*}
\tilde{F}_{\rm{a}\rm{b}} &= 
	\frac{1}{2} \lp \frac{1}{C^{\text{rr}}C^{\text{gg}}(1-r_{cc}^2)} \rp^2 
	\lb C^{\text{gg}} \left\{ \partial_{\rm{b}}C^{\text{gr}}\Big(-C^{\text{gr}}\partial_{\rm{a}}C^{\text{rr}}+C^{\text{rr}}\partial_{\rm{a}}C^{\text{gr}}\Big)+\partial_{\rm{b}}C^{\text{rr}}\Big(C^{\text{gg}}\partial_{\rm{a}}C^{\text{rr}}-C^{\text{gr}}\partial_{\rm{a}}C^{\text{gr}}\Big) \right\} \right. \\
  &\qquad\qquad\qquad\qquad\qquad\qquad
  -C^{\text{gr}} \left\{ \partial_{\rm{b}}C^{\text{gg}}\Big(-C^{\text{gr}}\partial_{\rm{a}}C^{\text{rr}}+C^{\text{rr}}\partial_{\rm{a}}C^{\text{gr}}\Big)+\partial_{\rm{b}}C^{\text{gr}}\Big(C^{\text{gg}}\partial_{\rm{a}}C^{\text{rr}}-C^{\text{gr}}\partial_{\rm{a}}C^{\text{gr}}\Big) \right\} \\
 &\qquad\qquad\qquad\qquad\qquad\qquad
 -C^{\text{gr}}\left\{ \partial_{\rm{b}}C^{\text{gr}}\Big(-C^{\text{gr}}\partial_{\rm{a}}C^{\text{gr}}+C^{\text{rr}}\partial_{\rm{a}}C^{\text{gg}}\Big)+\partial_{\rm{b}}C^{\text{rr}}\Big(C^{\text{gg}}\partial_{\rm{a}}C^{\text{gr}}-C^{\text{gr}}\partial_{\rm{a}}C^{\text{gg}}\Big) \right\} \\
&\qquad\qquad\qquad\qquad\qquad\qquad
+ \left. C^{\text{rr}} \left\{ \partial_{\rm{b}}C^{\text{gg}}\Big(-C^{\text{gr}}\partial_{\rm{a}}C^{\text{gr}}+C^{\text{rr}}\partial_{\rm{b}}C^{\text{gg}}\Big)+\partial_{\rm{b}}C^{\text{gr}}\Big(C^{\text{gg}}\partial_{\rm{a}}C^{\text{gr}}-C^{\text{gr}}\partial_{\rm{a}}C^{\text{gg}}\Big) \right\} \rb
\numberthis
    \label{eq:FisherMat},
\end{align*}
where $r_{cc}$ is the g-r cross correlation coefficient: %
\beq
r_{cc} \equiv \frac{C^{\text{gr}}}{\sqrt{C^{\text{gg}}C^{\text{rr}}}} \ .
\eeq
For $\alpha=\beta$, we obtain
\begin{align*}
\tilde{F}_{\rm{a}\rm{a}} &= \frac{1}{2(1-r_{cc}^2)^2}
	\left[\left(\frac{\partial_{\rm{a}}C^{\text{gg}}}{C^{\text{gg}}}-2r_{cc}^2\frac{\partial_{\rm{a}}C^{\text{gr}}}{C^{\text{gr}}}\right)^2
	+2r_{cc}^2\left(1-r_{cc}^2\right)\left(\frac{\partial_{\rm{a}}C^{\text{gr}}}{C^{\text{gr}}}\right)^2  \right. \\
&\qquad\qquad\qquad\quad\left.
	+\, 2r_{cc}^2\frac{\partial_{\rm{a}} C^{\text{rr}}}{C^{\text{rr}}}\left(\frac{\partial_{\rm{a}}C^{\text{gg}}}{C^{\text{gg}}}-2\frac{\partial_{\rm{a}}C^{\text{gr}}}{C^{\text{gr}}}\right)+\left(\frac{\partial_{\rm{a}}C^{\text{rr}}}{C^{\text{rr}}}\right)^2\right] \ .
	\numberthis
	 \label{eq:fisheronepar}
\end{align*}
On the other hand, if we only use $\deltag$, we get
\beq
\tilde{F}_{\rm{a}\rm{a}}^\text{(g only)} = \frac{1}{2} \lp \frac{\d_{\rm{a}} C^{\rm gg}}{C^{\rm gg}} \rp^2\ .
\eeq

\subsection{Analytical derivation of cosmic variance cancellation}
\label{sec:lowshotlimit}

Cosmic variance cancellation will occur in the limit of low noise on the measured fields -- that is, low reconstruction noise on the quadratic estimator, and low galaxy shot noise. To investigate this case analytically, let us work in the limit of very low shot noise, so that
\begin{align*}
C^{\rm gg}(K) &= b_{\rm g}(K)^2 P_{\rm lin}(K)\ , \\
C^{\rm gr}(K) &= b_{\rm g}(K) b_{\rm r}(K) P_{\rm lin}(K)\ , \\
C^{\rm rr}(K) &= b_{\rm r}(K)^2 P_{\rm lin}(K) + N_{\rm GG}(K)\ .
\numberthis
\end{align*}
Further, let us assume that $\fnl$ is the only unknown parameter. If we define
\beq
x(K) \equiv \frac{N_{\rm GG}(K)}{b_{\rm r}(K)^2 P_{\rm lin}(K)}\ ,
\qquad
R_p(K) \equiv \lp \frac{\d_{\fnl} b_{\rm r}(K)}{b_{\rm r}(K)} \rp
	\lp \frac{\d_{\fnl} b_{\rm g}(K)}{b_{\rm g}(K)} \rp^{-1}\ ,
\eeq
where $x(K)$ is the inverse of the signal to noise per mode of the reconstructed field and $R_p(K)$ is a measure of similarity between the response of the bias of the reconstructed field and the one of the original tracer field, then a short calculation gives the unmarginalized errorbar on $\fnl$ per $K$-mode:
\beq
  \label{eq:lownoiseerror}
   \sigma^2_{\fnl}(K)=\sigma_{\fnl,\text{ g only}}^2(K)
   \frac{2x(K)}{\lp R_p(K)-1 \rp^2}
   \frac{1}{1+ 2\lp R_p(K)-1 \rp^{-2} x(K)}\ ,
\eeq
where $\sigma_{\fnl,\text{ g only}} = [F_{\rm{a}\rm{a}}^\text{(g only)}]^{-1/2}$.

Let us investigate the general behavior of this equation in some limiting cases. If $\lp R_p-1 \rp^{-2}x$ is small, the $R_p<0$ case (when $\d_{\fnl}b_{\rm r}$ and $b_{\rm r}$ have opposite signs) will result in smaller errorbars than the $R_p>0$ case, because the signatures of $\fnl$ in $b_{\rm r}$ and $b_{\rm g}$ will be more distinguishable in that case. 
Expanding Eq.~\eqref{eq:lownoiseerror} in limit of small $\lp R_p-1 \rp^{-2}x$  gives
\beq
   \sigma^2_{\fnl}(K)=\sigma_{\fnl,\text{ g only}}^2(K)
   \frac{2x(K)}{\lp R_p(K)-1 \rp^2} 
   \sum_{n=0}^\infty \lb - 2\lp R_p(K)-1 \rp^{-2} x(K) \rb^n\ .
\eeq
As we will see in Appendix~\ref{app:meanfield}, $N_{\rm GG} \propto k_{\rm max}^{-3}$ in the low-$K$ limit, so that we arrive at
\beq
    \lim_{x\rightarrow 0} \sigma^2_{\fnl}(K)
    \propto 2 \sigma_{\fnl,\text{ g only}}^2(K)
    \lb k_{\rm max}^{-3} + \mathcal{O}(k_{\rm max}^{-6}) \rb\ ,
    \label{eq:lownoiseapproximated}
\eeq
where we assume that $R_p-1$ varies slowly with $K$. This demonstrates that constraints on $\fnl$ that use both reconstructed modes and modes of the original tracer will improve on a tracer-only analysis in a way that is only limited by the noise on the reconstructed modes (if shot noise is negligible).

Cosmic variance cancellation clearly requires that $\delta_{\rm r}$ and $\deltag$ are measured at the same wavenumber and in the same volume. To verify this, we can repeat the derivation above with $C^{\rm gr}=0$, corresponding to $\delta_{\rm r}$ and $\deltag$ being measured in different volumes. In this case, Eq.~\eqref{eq:lownoiseerror} becomes
\beq
   \sigma^2_{\fnl}(K)=\sigma_{\fnl,\text{ g-only}}^2(K)
   \frac{\lb 1+x(K) \rb^2}{ R_p(K)^2 + \lb 1+x(K) \rb^2}\ ,
\eeq
which approaches a finite limit as $x\to 0$; thus, the improvement realized in Eq.~\eqref{eq:lownoiseapproximated} is only possible if $\delta_{\rm r}$ and $\deltag$ can be compared mode-by-mode in the same volume.

\subsection{Assumptions and experimental configurations}
\label{sec:config}

\subsubsection{Scales}
\label{sec:scales}

In each forecast, for measuring $\fnl$, we use $\deltag$ modes and reconstructed modes with wavenumber $K$ satisfying $K_{\rm min} < K < K_{\rm max}$, and we also use reconstructed modes with $K_{\rm f} < K < K_{\rm min}$, where $K_{\rm f} \approx 0.002\invMpc$ is the lowest measurable wavenumber within each survey volume. In this way, $K_{\rm min}$ accounts for possible systematic effects that can prevent direct measurements of $\deltag$ on large scales, but that do not impede reconstruction of these large-scale modes using smaller-scale correlations; an example is foreground contamination for intensity mapping experiments, which as been a primary motivator for other work on reconstruction methods \cite{Zhu:2015zlh,Zhu:2016esh,Foreman:2018gnv,Li:2018izh,Karacayli:2019iyd,Modi:2019hnu}.  As input to the density-field reconstruction, we use modes with wavenumber $k$ satisfying $K_{\rm max} <k<k_{\rm max}$. We consider a range of possible $K_{\rm min}$ values in our forecasts, while $k_{\rm max}$ and $K_{\rm max}$ are fixed for each survey, as described below.

\subsubsection{Surveys}

\begin{table}
\begin{centering}
\begin{tabular}{ l | c | c c | c c }
& DESI-like & \multicolumn{2}{c|}{MegaMapper-like} & \multicolumn{2}{c}{PUMA-like}   \\
 & $0.6<z<1.6$ & $2<z<2.5$ & $4.5<z<5$ & $2<z<3$ & $5<z<6$ \\
 \hline
 {\bf Survey parameters} & & & & \\
 Survey volume (Gpc$^3$)
 	& $100$ & $80$ & $66$ & $266$ & $203$ \\
 Mean galaxy density $\Bar{n}$ (Mpc$^{-3}$) 
 	& $10^{-4}$ & $6\times10^{-4}$ & $2\times10^{-5}$ 
	& $2\times10^{-3}$ ($6\times10^{-3}$)  
	& $1\times10^{-3}$ ($2\times10^{-2}$)  \\
 $K_\text{max}$ for $\fnl$ constraint ($h \,\text{Mpc}^{-1}$)
 	& $0.05$ & $0.08$ & $0.14$ & $0.09$ & $0.15$   \\ 
$k_\text{max}$ for reconstruction ($h \,\text{Mpc}^{-1}$)
	& $0.15$ & $0.24$ & $0.4$ & $0.26$ & $0.47$ \\
 \hline
 {\bf Fiducial bias parameters} & & & & \\
$b_1$ 			& $1.6$ 	& $2.9$	& $7.0$	& $2.1$	& $3.7$ \\
$b_2$ 			& $-0.30$	& $1.1$	& $17$	& $0.041$	& $2.8$\\
$b_{s^2}$			& $-0.17$	& $-0.54$	& $-1.7$	& $-0.31$	& $-0.77$\\
$b_{11}^{\rm E}$	& $-3.0$ 	& $-2.5$	& $37$	& $-3.5$	& $0.58$\\
$b_{02}^{\rm E}$ 	& $-14$ 	& $-21$	& $85$	& $-19$	& $-16$\\ \hline
\end{tabular}
\caption{\label{tab:surveys}
Survey characteristics used for our main forecasts. The DESI-like survey is based on the expected DESI emission-line galaxy sample, the MegaMapper-like survey is a next-generation survey targeting high-redshift ``dropout" galaxies, and the PUMA-like survey represents a future \tcm intensity mapping effort over half the sky.  We marginalize over $b_1$, $b_2$, and $b_{s^2}$ in our forecasts, and determine $b_{11}^{\rm E}$ and $b_{02}^{\rm E}$ using the relationships in Sec.~\ref{sec:bias}. For the PUMA-like forecast, the main $\bar{n}$ values represent effective number densities that reproduce the same noise level as the sum of shot and instrumental noise power at $k=k_{\rm max}$, while the expected physical number densities are shown in parentheses. For this forecast, we also consider the effects of the so-called ``foreground wedge" that will prevent direct measurement of certain modes. See main text for details.
}
\end{centering}
\end{table}

In our main forecasts, we consider three galaxy surveys, with properties summarized in Table~\ref{tab:surveys}. The first is similar to the emission-line galaxy sample expected from DESI \citep{Aghamousa:2016zmz}. For this survey, following \cite{Munchmeyer:2018eey}, we consider $14000\,{\rm deg}^2$ of sky area over $0.6<z<1.6$, which translates into a total comoving volume of roughly $100\,{\rm Gpc}^3$ and a mean redshift of $\bar{z} \approx 1$. We use a mean galaxy number density of $\bar{n}=10^{-4}\,{\rm Mpc}^{-3}$, obtained by dividing the expected total number of redshifts in the DESI ELG sample ($1.7\times10^7$, from \citealt{Aghamousa:2016zmz}) by the survey volume, and assume a mean linear galaxy bias of $b_1=1.6$. We take $K_{\rm max}=0.05\invMpc$, since linear bias is expected to be an acceptable approximation for $K<K_{\rm max}$ at $z=1$, and $k_{\rm max}=0.15\invMpc$, since our quadratic bias expansion is valid for $k<k_{\rm max}$ at $z=1$ (see Sec.~\ref{sec:sims} for justification based on simulations).

The second survey, which we call ``MegaMapper-like", is modelled on proposals for a next-generation spectroscopic survey targeting high-redshift ``dropout" galaxies in the southern hemisphere \citep{Wilson:2019brt,Ferraro:2019uce,Schlegel:2019eqc}. For this, we assume a $14000\,{\rm deg}^2$ survey, and separately consider two redshift bins, at $2<z<2.5$ and $4.5<z<5$, which have volumes of $80\,{\rm Gpc}^3$ and $66\,{\rm Gpc}^3$ respectively. The mean number density and linear bias in each bin are obtained from averages of the values at the bin edges, taken from Table 1 of \cite{Ferraro:2019uce}; this yields $\bar{n}=6\times10^{-4}\,{\rm Mpc}^{-3}$ and $b_1=2.9$ for the lower-redshift bin, and $\bar{n}=2\times10^{-5}\,{\rm Mpc}^{-3}$ and $b_1=7.0$ for the higher-redshift bin. For $K_{\rm max}$ and $k_{\rm max}$, we scale the DESI values using the ratio of linear growth factors between the mean redshifts of each redshift bin, to account for the increased range of validity of our perturbative expressions at higher redshift.\footnote{In reality, the scaling of $k_{\rm max}$ with redshift is more complicated, involving the power spectrum tilt at the relevant wavenumbers (e.g.~\citealt{Carrasco:2013mua}), but the simple growth factor scaling we use here should at least be roughly indicative of the useful scales for our forecasts.}

The third survey is based on specifications for PUMA, an envisioned radio interferometer designed for \tcm intensity mapping \citep{Ansari:2018ury,Bandura:2019uvb}. We assume a survey over half the sky, and again consider two redshift bins, this time at $2<z<3$ and $5<z<6$, with volumes $266\,{\rm Gpc}^3$ and $203\,{\rm Gpc}^3$ respectively. For simplicity, we treat this survey as observing galaxy positions directly, rather than brightness temperature (which is just a rescaled biased tracer of the matter density). To do so, we set the noise contribution to the tracer power spectrum $P_{\rm gg}$ to equal the sum of the shot noise and instrumental noise power spectra computed using the PUMA noise calculator\footnote{\url{https://github.com/slosar/PUMANoise}}, evaluated at $k=k_{\rm max}$ in each redshift bin. We quote an effective number density that would result in the same noise level in Table~\ref{tab:surveys}. When computing the shot noise contributions to $P_{\rm gr}$ and $P_{\rm rr}$, we use the expected number densities of \tcm emitters, also taken from the PUMA noise calculator and quoted in parentheses in Table~\ref{tab:surveys}. For the linear bias in each bin, we use values from Fig.~33 of \cite{Ansari:2018ury}, evaluated at the mean redshifts. As for the MegaMapper-like survey, we scale~$K_{\rm max}$ and~$k_{\rm max}$ from DESI by the appropriate ratios of linear growth factors.

In our derivation of stochastic contributions to the noise of the estimator and the cross-correlation between estimator and galaxy fields in App.~\ref{app:shot}, we assume that the noise is Poissonian, i.e., that $\left \langle \varepsilon \varepsilon\right \rangle =(2\pi)^3\dirac(\vec K+\vec K')/\bar n$. There is evidence for halo stochasticity being sub-Poissonian for high mass haloes and super-Poissonian for low mass haloes \citep{Hamaus:2010min,Baldauf:2013hka}. Since the stochasticity corrections arise from small-scale exclusion and higher order biases, the actual shot noise levels cannot be theoretically predicted, implying that it may be advisable to marginalize over the stochasticity parameter(s). This approach is indeed adopted by some for the $\fnl$ forecasting literature (e.g.~\citep{Castorina:2020blr}) but certainly not all of it (e.g.~\citep{Schmittfull:2017ffw,Munchmeyer:2018eey}).
Here we decide to fix the stochasticity parameters to their fiducial Poissonian values and defer a more detailed investigation of the impact of noise corrections on the reconstructed fields to future work. We do note however, that we expect the impact of shot noise marginalization to be rather small, since we do not include the additional non-Gaussian signal arising in combination with stochastic terms in Eqs.~(\ref{eq:cgg}-\ref{eq:crr}).

 \subsubsection{\tcm foregrounds}
 
An additional consideration for \tcm intensity mapping is the presence of foreground radiation, dominantly synchrotron from our own galaxy, that is brighter than the cosmological signal by several orders of magnitude. These foregrounds are extremely smooth in frequency, which implies that they mainly populate Fourier modes with low line-of-sight wavenumber $k_\parallel$; these modes will therefore likely not be usable for cosmology. Furthermore, the chromatic properties of interferometers generically spread foreground power from the low-$k_\parallel$ modes into a wedge-shaped region in the $k_\parallel-k_\perp$ plane (e.g.~\citealt{Parsons:2012qh,Liu:2014bba,Liu:2014yxa}), although this contamination can be removed with sufficiently precise instrumental calibration (e.g.~\citealt{Shaw:2014khi,Ghosh:2017woo}).
 
For constraining $\fnl$, the wedge will have two effects: it will reduce the number of short-wavelength modes available for the quadratic estimator, therefore increasing the noise $N_{\rm GG}$ on the reconstructed modes, and it will also reduce the number of long-wavelength $\deltag$ modes available for measuring the scale-dependent bias induced by primordial non-Gaussianity. 

We account for both effects in our forecasts for the PUMA-like survey, assuming a foreground wedge defined by 3~times the primary beam width, following \cite{Ansari:2018ury}; see Appendix~\ref{app:fg-implementation} for details of how this is implemented in our computations. In addition, we perform forecasts that ignore the wedge, to represent the case when it can be completely eliminated via calibration. We account for lost low-$K_\parallel$ modes in two ways: either by restricting $\delta_{\rm g}$ to have $K>K_{\parallel,{\rm min}}$, or by approximating $K_{\parallel,{\rm min}}$ as an isotropic $K_{\rm min}$, matching our procedure for DESI and MegaMapper. The former approach is more realistic, while the latter is easier to compare with the other surveys, so we present the latter in the main text, and the former in Appendix~\ref{app:fg-pumakparmin}.

\subsubsection{Bias parameters}

For every survey, to perform forecasts, we assume a fiducial value of the quadratic bias parameter $b_2$ derived from the fitting formula of \cite{Lazeyras:2015lgp}, which was fit to halo bias in separate-universe simulations over the range $1\lesssim b_1\lesssim 10$:
\beq
b_2(b_1) = 2\lp 0.412 - 2.143b_1 + 0.929b_1^2 + 0.008b_1^3 \rp\ ,
\eeq
where the extra factor of 2 arises from our different definition of $b_2$ compared to \cite{Lazeyras:2015lgp}. The fiducial value of the tidal bias $b_{s^2}$ is found from
\beq
b_{s^2} = -\frac{2}{7} \lp b_1 -1 \rp\ ,
\eeq
which assumes that the tidal bias in Lagrangian space is zero. In our forecasts, $b_1$, $b_2$, and $b_{s^2}$ are allowed to vary independently (i.e.\ are marginalized over when we estimate uncertainties on $\fnl$), while $b_{11}$ and $b_{02}$ are assumed to obey the relationships in Eqs.~\eqref{eq:b11E}-\eqref{eq:b02E} and~\eqref{eq:b01Lfinal}-\eqref{eq:b02Lfinal}. We take wide, flat priors on $b_1$, $b_2$, and $b_{s^2}$; we have also implemented 10\% Gaussian priors on $b_2$ and $b_{s^2}$, but these have a negligible effect on our baseline results.

\subsubsection{Redshift space distortions}
\label{sec:rsd}

The line-of-sight component of a galaxy's position is observationally inferred from the galaxy's redshift, and the associated ``redshift-space distortions" of $\deltag$ should be included in a full treatment of the observed galaxy clustering. The leading-order effect is to add a $f\mu^2$ term to the linear bias of $\deltag$, such that Eq.~\eqref{eq:deltag-condensed} is modified to
\beq
\deltag(\vk) = \lb b_1 + \fnl \frac{c_{01}}{M(k)} + f\mu^2 \rb \delta_1(\vk) + \cdots \ ,
\eeq
where $f\equiv d\log D / d\log a$, $\mu \equiv k_\parallel/k$, and $D$ is the linear growth factor \citep{Kaiser:1987qv}. Higher-order effects will create additional mode-couplings that can be described in perturbation theory (e.g.~\citealt{Perko:2016puo,delaBella:2017qjy}). In a real tracer catalogue, there will also be line-of-sight-dependent selection effects that can be treated perturbatively \citep{Desjacques:2018pfv}.

We do not include any of these effects in our baseline forecasts, leaving them for future work. However, as a first step in this direction, we have checked the impact of including the Kaiser term. This raises the reconstruction noise $N_{\rm GG}$ by increasing $P_{\rm gg,tot}$ in the denominator of Eq.~\eqref{eq:quadest}, while also increasing the amplitude of $P_{\rm gg}$ and $P_{\rm gr}$, thereby increasing the signal to noise on those quantities. For all surveys we consider, the former effect overcomes the latter, with the result that $\sigma(\fnl)$ increases by roughly 10\%, and the improvement in $\sigma(\fnl)$ from including reconstructed modes decreases by no more than the same amount. Additional mode-couplings from nonlinear redshift-space effects will likely dominate over this change, and a detailed analysis will be worthwhile to pursue, especially since some of these mode-couplings could potentially carry additional information about $\fnl$ \citep{Castorina:2020blr}.

\subsection{Expected precision on reconstructed modes}
\label{sec:prec_on_r}

\begin{figure}[t]
\includegraphics[width=\textwidth, trim = 10 10 10 10 ]{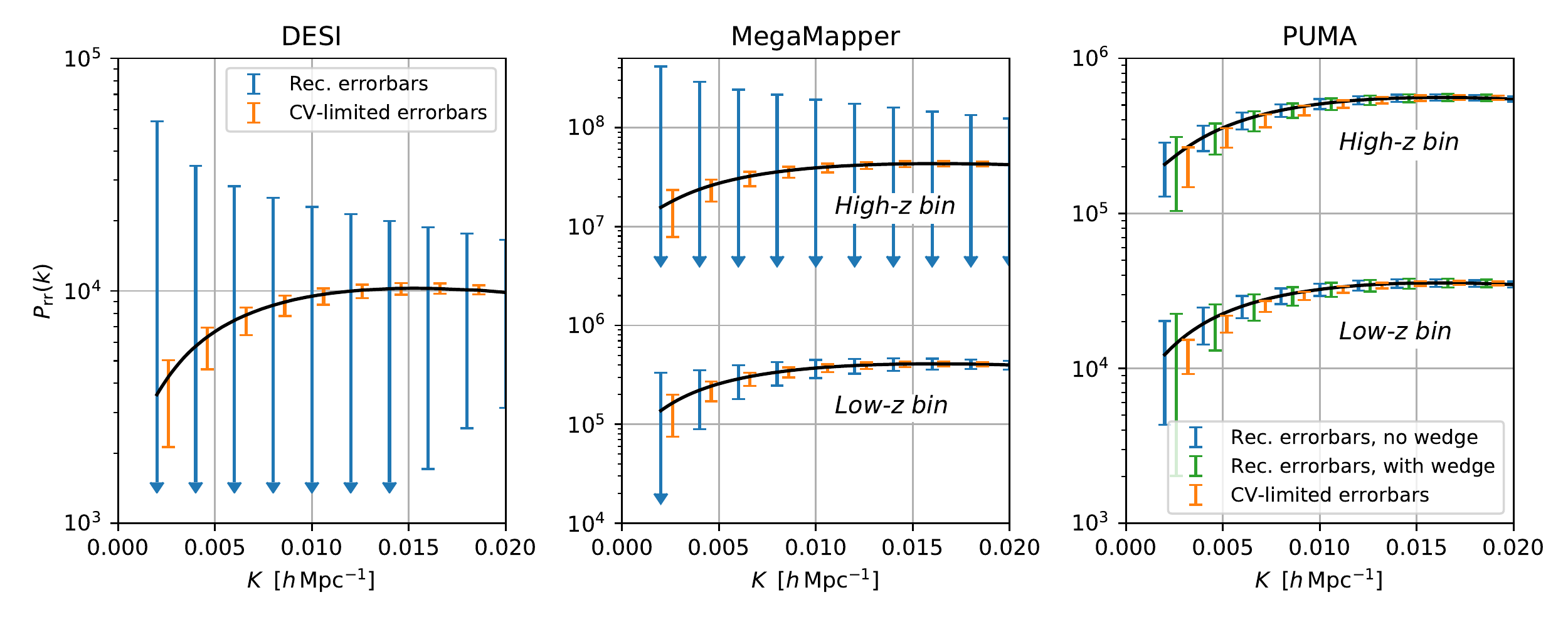} 
\caption{
\label{fig:prr-ebars}
Expected errorbars on the reconstructed power spectrum $P_{\rm rr}$ for the surveys and redshift bins we consider {\it (blue)}, along with cosmic-variance-limited errorbars {\it (orange)}, computed for bandpowers with $\Delta K = 0.002\invMpc$. Downward arrows indicate errorbars whose lower limits fall outside of the $y$ axis range. High-precision measurements of the power spectrum of reconstructed modes will be possible in several cases, even in the presence of a \tcm foreground wedge for PUMA.
}
\end{figure}

Aside from primordial local non-Gaussianity, there are many other applications of reconstructing large-scale modes, including more general constraints on cosmology, tests of predictions for the power spectrum on the largest scales, calibration of photometric redshifts \citep{Modi:2019hnu}, cross-correlations with other tracers (such as kSZ fluctuations in the CMB, e.g.\ \citealt{Li:2018izh}), and removing contamination from measurements of lensing of \tcm fluctuations \citep{Foreman:2018gnv}. To represent the general utility of reconstructed modes from different surveys, in Fig.~\ref{fig:prr-ebars} we show the expected precision on the auto power spectrum of the reconstructed modes (plotted using the fiducial bias parameters from Table~\ref{tab:surveys}), computed in wavenumber bins with $\Delta K = 0.002\invMpc$. While these errorbars are substantial for $K\lesssim 0.01\invMpc$ in DESI and the high-$z$ bin of MegaMapper, the precision is expected to be much better for MegaMapper at low $z$ and across the entire redshift range of PUMA, with most errorbars approaching the cosmic variance limit. This will enhance many scientific applications of these surveys, particularly for PUMA, where large-scale modes can be reconstructed at high precision even in the presence of the foreground wedge.

\cite{Ansari:2018ury} also estimates the total signal to noise in reconstructed modes from PUMA over $1<z<6$, following the methodology of \cite{Foreman:2018gnv}, finding $\mathcal{O}(1300)$ in the no-wedge case and $\mathcal{O}(500)$ for the same wedge model we use here. For comparison, we find a total S/N of 135 (108) for $2<z<3$ and 161 (134) for $5<z<6$ in the no-wedge (wedge) case. A direct comparison between the two sets of forecasts is difficult, because they use several distinct approximations: \cite{Ansari:2018ury} treats the \tcm brightness temperature as a linearly biased tracer of the matter density, while we have incorporated second-order biasing; \cite{Ansari:2018ury} neglects the shot noise contribution to the reconstructed mode power spectrum, while we include it; \cite{Ansari:2018ury} bias-harden their results against mode-couplings from gravitational lensing, while we do not; and, most importantly, \cite{Ansari:2018ury} only consider reconstruction of modes that are purely transverse to the line of sight ($k_\parallel=0$), while we use a 3d reconstruction formalism. Nevertheless, both forecasts reach the same broad conclusion that PUMA will be able to reconstruct long-wavelength density modes with total signal to noise of several hundred, which is strong motivation for continued studies of the density reconstruction method we have presented in this paper.

\subsection{Results: constraints on non-Gaussianity}
\label{sec:results}

\subsubsection{DESI}

\begin{figure}[t]
\includegraphics[width=\textwidth, trim = 10 10 0 10 ]{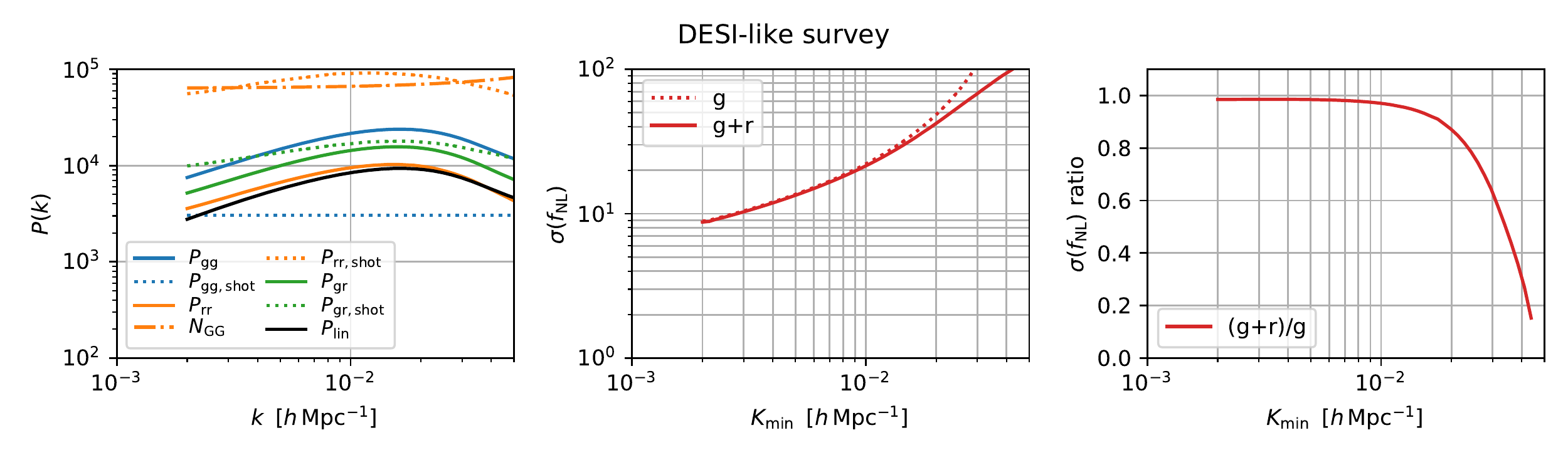} 
\caption{
\label{fig:desi}
Forecasts for a DESI-like survey. 
{\it Left:} Signal and noise power spectra involved in the forecast. The galaxy auto spectrum is well above the shot noise, while the auto spectrum of reconstructed modes ($P_{\rm rr}$) is roughly an order of magnitude below both the reconstruction noise ($N_{\rm GG}$) in the quadratic estimator and the shot noise contribution to the estimator variance.
{\it Center:} Expected constraints on $\fnl$ when only $\delta_{\rm g}$ is used {\it (solid)}, or when $\delta_{\rm r}$ is also used {\it (dotted)}. We assume that $\delta_{\rm g}(\vK)$ cannot be directly measured for $K<K_{\rm min}$,  and marginalize over the $b_1$, $b_2$, and $b_{s^2}$ bias parameters.
{\it Right:} Ratio of $\delta_{\rm g}+\delta_{\rm r}$ and $\delta_{\rm g}$-only cases from the center panel. We only notice an improvement for higher values of $K_{\rm min}$, corresponding to using $\delta_{\rm r}$ but not $\delta_{\rm g}$ at $K<K_{\rm min}$.
}
\end{figure}

Fig.~\ref{fig:desi} shows the results of our forecasts for the DESI-like survey. The left panel shows the various power spectra of interest, of linear matter density, galaxy number density, and reconstructed matter density modes, along with the cross spectrum between galaxies and reconstructed modes. This panel also shows the shot noise on $P_{\rm gg}$, $P_{\rm gr}$, and $P_{\rm rr}$, as well as the statistical noise ($N_{\rm GG}$) on reconstructed modes. For DESI, the galaxy power spectrum is well above the shot noise, while the reconstructed power spectrum is about an order of magnitude lower than the reconstruction noise. Despite the fact that galaxy shot noise is below $P_{\rm gg}$, the shot noise contributions for both $P_{\rm gr}$, and $P_{\rm rr}$ are above the signal power spectra. As explained in Appendix~\ref{app:shot}, this is due to coupling between galaxy shot noise and clustering at large scales, where the variance is larger than at small scales and therefore these shot noise spectra are significantly boosted compared to the $\bar{n}^{-1}$ contribution.

The middle panel of Fig.~\ref{fig:desi} shows the expected constraints on $\fnl$ when only $\delta_{\rm g}$ is used, or when reconstructed modes are also incorporated. The right panel shows the ratio of $\sigma(\fnl)$ in these two cases. The improvement in $\sigma(\fnl)$ is negligible at the lowest $K_{\rm min}$ we consider, which corresponds to $\delta_{\rm g}$ being measured on all scales resolvable within the survey volume ($K_{\rm min}=K_{\rm f}$). However, a larger improvement is seen when $K_{\rm min}$ is assumed to be higher: for $K_{\rm min}=0.02\invMpc$, for example, $\sigma(\fnl)$ improves by around 15\% when reconstructed modes are used.

\begin{figure}[t]
\centering
\includegraphics[width=0.5\textwidth, trim = 10 10 0 10 ]{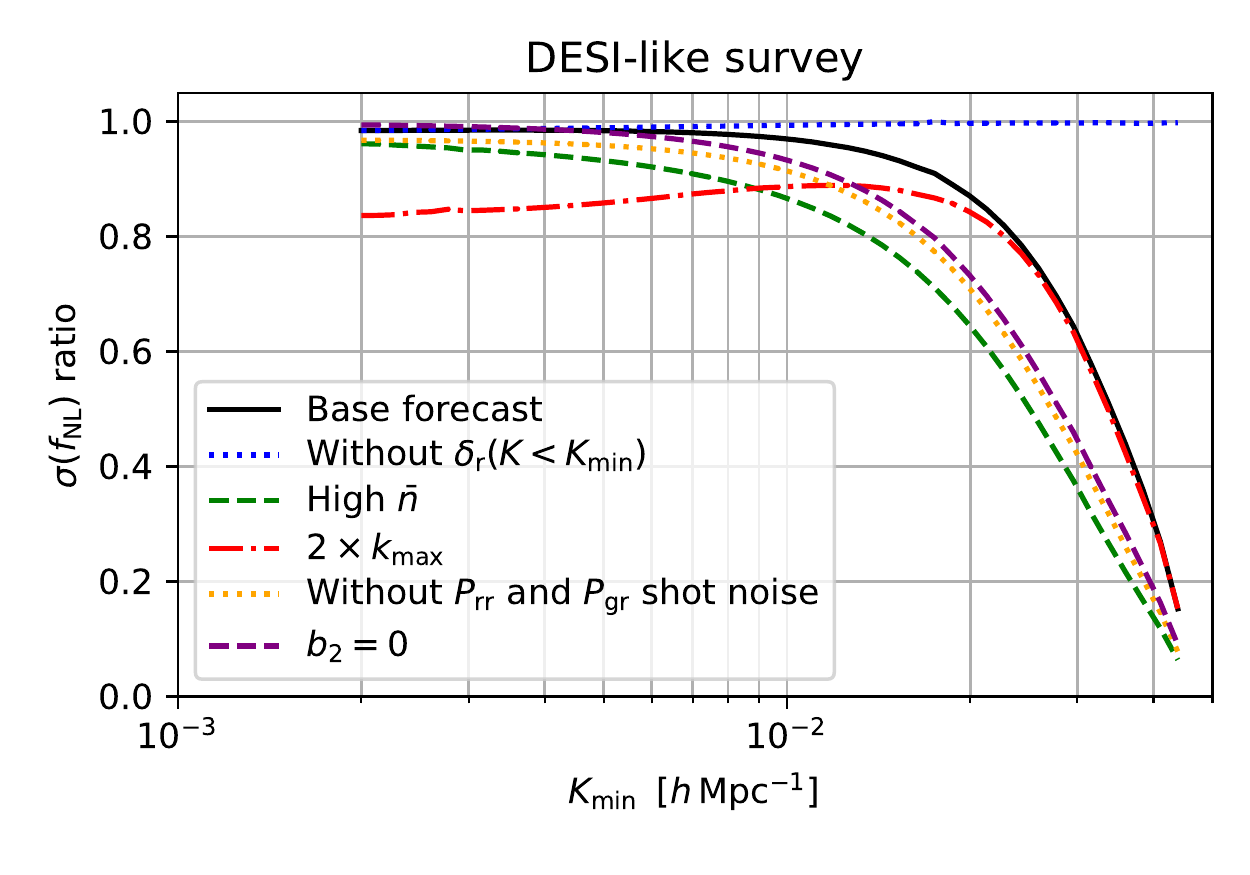} 
\caption{
\label{fig:desi-options}
The analog of the right panel of Fig.~\ref{fig:desi}, with a variety of (mostly artificial) modifications to the forecasts. There is no improvement in $\sigma(\fnl)$ when $\delta_{\rm r}$ is neglected at $K<K_{\rm min}$, indicating that the inclusion of $\delta_{\rm r}$ at $K<K_{\rm min}$ drives the improvement. Greater improvements are achieved for higher galaxy number density or if $k_{\rm max}$ can be increased by a factor of~2, with milder changes if  the fiducial $b_2$ value is set to zero or shot noise on $P_{\rm rr}$ and $P_{\rm gr}$ is neglected.
}
\end{figure}

To determine the origin of this behavior, we show several modifications of this forecast in Fig.~\ref{fig:desi-options}. In particular, when reconstructed modes with $K<K_{\rm min}$ are not included, there is no improvement of $\sigma(\fnl)$, indicating that these modes are entirely responsible for the improvement. Therefore, DESI is not powerful enough to allow for cosmic variance cancellation between $\delta_{\rm g}$ and $\delta_{\rm r}$ at the same scales; rather, the primary use of reconstruction is to access scales ($K<K_{\rm min}$) where $\delta_{\rm g}$ cannot be directly measured. This naturally explains why the improvement of $\sigma(\fnl)$ grows for higher~$K_{\rm min}$. While the absolute values of $\sigma(\fnl)$ are not impressive at such high $K_{\rm min}$ -- at $K_{\rm min}=0.02\invMpc$, for example, $\sigma(\fnl) \approx 50$ without reconstruction and $40$ with reconstruction -- the improvement comes ``for free," without requiring any other datasets.

The other curves in Fig.~\ref{fig:desi-options} illuminate other aspects of this forecast. Increasing $\bar{n}$ to an unrealistically high value of $10^2\,{\rm Mpc}^{-3}$ improves the $\delta_{\rm g}$-only forecast by roughly 10\% (not shown), and also increases the improvement on $\sigma(\fnl)$ from including $\delta_{\rm r}$, indicating that shot noise is a limiting factor in this improvement. Simply neglecting $P_{\rm rr,shot}$ and $P_{\rm gr,shot}$ has a similar effect, clarifying that shot noise in the galaxy power spectrum itself is comparatively less important than in these other spectra.

Also, we see the same type of change if we alter the fiducial value of $b_2$. As mentioned at the end of Sec.~\ref{sec:noisecont}, if $P_{\rm gg} \gg P_{\rm gg,shot}$ (as it is here), then $P_{\rm rr}/N_{\rm GG} \propto (1+b_2/b_1)^2$, so increasing $b_2$ from $-0.3$ to $0$ boosts the signal to noise on the reconstructed modes. This would lead to a larger improvement if not for the large contribution of $P_{\rm rr,shot}$. Boosting $k_{\rm max}$ by a factor of 2 leads to a better $\sigma(\fnl)$ improvement at low $K_{\rm min}$. This change lowers the Gaussian reconstruction noise $N_{\rm GG}$, but also raises $P_{\rm rr,shot}$ and $P_{\rm gr,shot}$ by different amounts, and the combination of these changes ends up slightly boosting the constraining power of $\delta_{\rm r}$.

It may seem counterintuitive that $\delta_{\rm r}$ adds anything at all to our forecasts, since the reconstruction noise and shot noise on $P_{\rm rr}$ are much larger than $P_{\rm rr}$ itself: one would expect such large noise to lead to a low cross-correlation coefficient between $\delta_{\rm r}$ and $\deltag$, and also make it difficult to extract information from the auto spectrum of $\delta_{\rm r}$. However, the presence of a cross shot noise contribution to $P_{\rm gr}$ alters this picture, contributing to the $\delta_{\rm r}$-$\deltag$ cross-correlation coefficient and altering the structure of the covariance matrix. While it is not trivial to see in the Fisher matrix expression in Eq.~\eqref{eq:fisheronepar}, the net effect is to enhance the information content of $\delta_{\rm r}$ with respect to $\fnl$. \cite{Liu:2020izx} reached a similar conclusion when examining cosmic variance cancellation between different line intensity maps, noticing that lowering the cross shot noise contribution led to worsened constraints on $\fnl$.

\subsubsection{MegaMapper}

\begin{figure}[t]
\includegraphics[width=\textwidth, trim = 10 10 0 10 ]{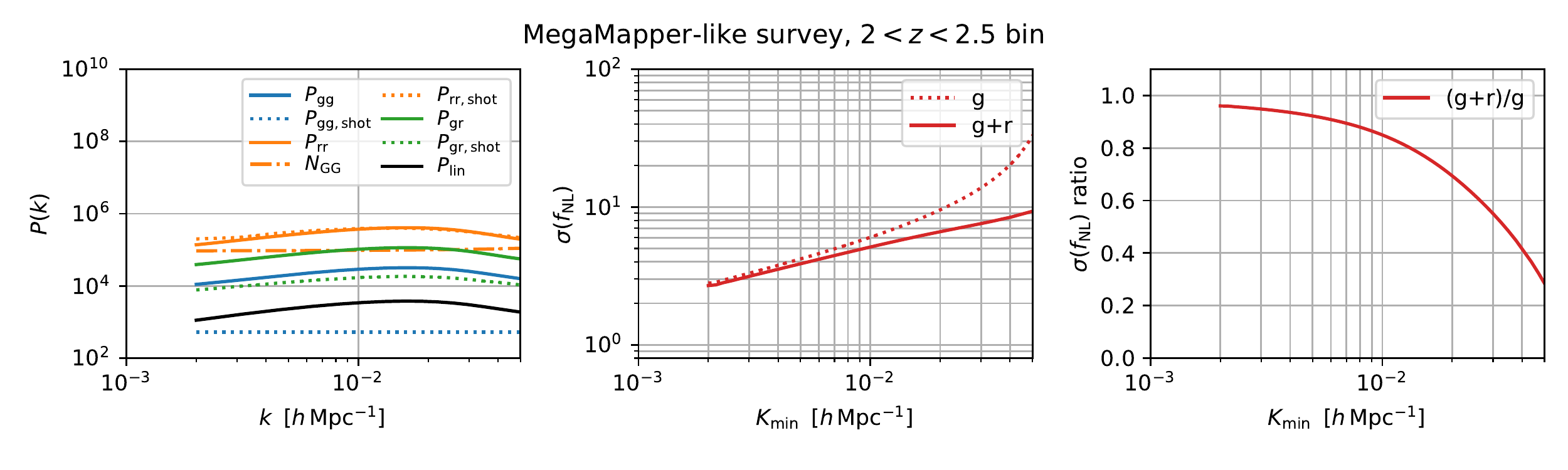} 
\includegraphics[width=\textwidth, trim = 10 10 0 10 ]{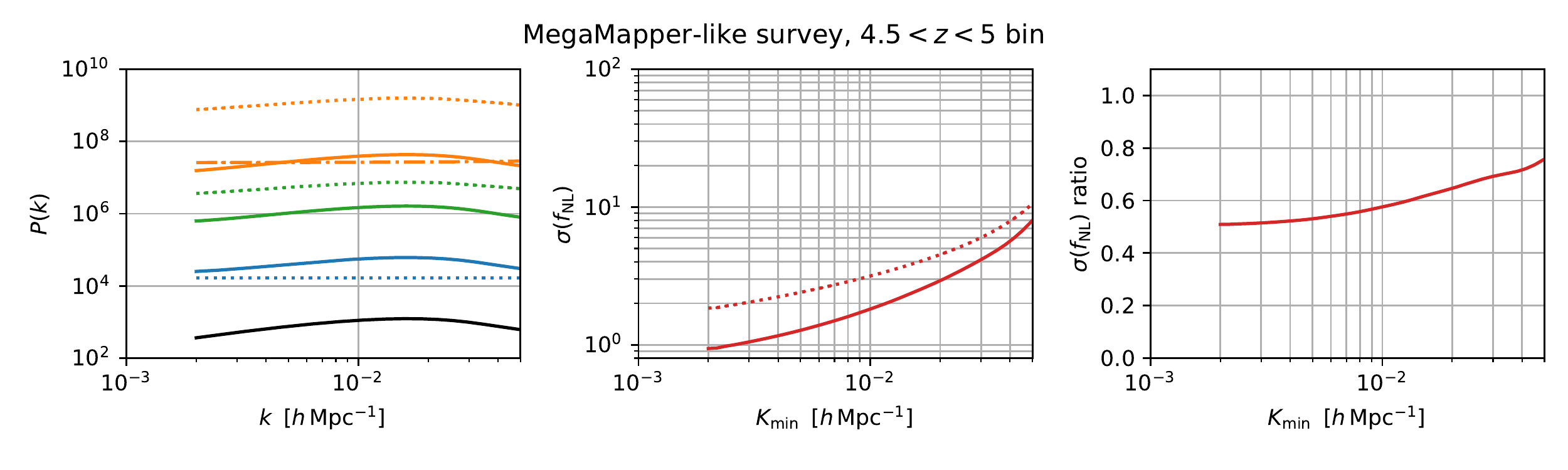} 
\caption{
\label{fig:mm}
As Fig.~\ref{fig:desi}, for low-redshift {\it (top panels)} and high-redshift {\it (bottom panels)} bins of a MegaMapper-like survey. The former has greater signal to noise on reconstructed modes than DESI, leading to a greater improvement in $\sigma(\fnl)$ when these modes are included in the forecast. For the latter, the shot noise contributions to $P_{\rm rr}$ and $P_{\rm gr}$ are comparatively much larger, leading to a different scale-dependence for the improvement in $\sigma(\fnl)$.
}
\end{figure}

\begin{figure}[h]
\includegraphics[width=\textwidth, trim = 10 10 0 10 ]{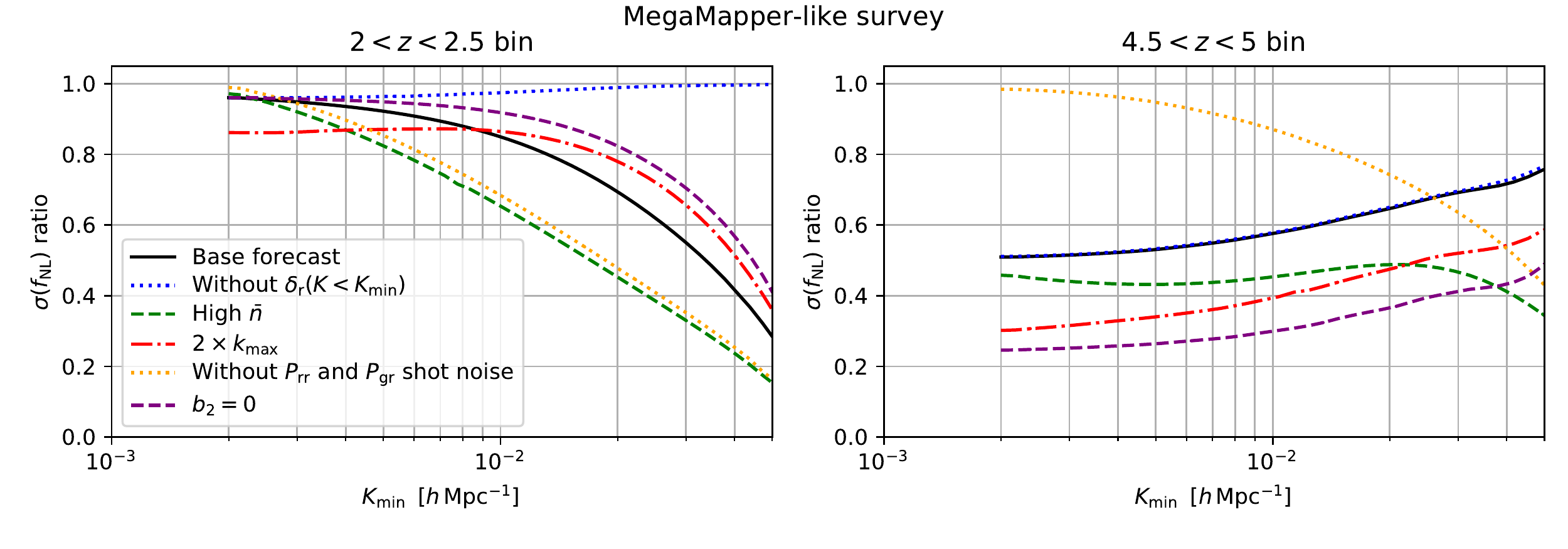} 
\caption{
\label{fig:mm-options}
Modifications to the base MegaMapper forecasts. For the low-redshift bin {\it (left panel)}, as for DESI, the improvement in $\sigma(\fnl)$ is driven mostly by modes of $\delta_{\rm r}$ with $K<K_{\rm min}$, with further improvement possible for higher galaxy number density. For the high-redshift bin {\it (right panel)}, neglecting $\delta_{\rm r}$ at $K<K_{\rm min}$ makes no difference, indicating that for lower $K_{\rm min}$ values, cosmic variance cancellation between $\deltag$ and $\delta_{\rm r}$ at the same $K$ is driving the improvement in $\sigma(\fnl)$. There are several ways to obtain greater improvements, as discussed in the main text.
}
\end{figure}

Our results for the MegaMapper-like survey are shown in Fig.~\ref{fig:mm}. For the low-$z$ bin, the signal to reconstruction noise on the reconstructed modes is higher than for DESI, thanks to a combination of higher $\bar{n}$, higher $k_{\rm max}$, and higher bias, and the signal to shot noise ratio is also correspondingly smaller. This leads to a greater improvement in $\sigma(\fnl)$ when reconstructed modes are included. The left panel of Fig.~\ref{fig:mm-options} shows that, like DESI, this improvement comes not from cosmic variance cancellation, but from reconstructed modes with $K<K_{\rm min}$, where we assume that $\delta_{\rm g}$ cannot be directly measured. We see large changes if $\bar{n}$ is boosted or $P_\text{gr,shot}$ and $P_\text{rr,shot}$ are neglected, indicating that shot noise is a limiting factor in this bin.
Changing the fiducial $b_2$ from $1.1$ to $0$ reduces the usefulness of the reconstructed modes for the same reason that changing~$b_2$ increased their usefulness for DESI.

We see rather different behavior in the high-$z$ bin. There, we find that the reconstruction noise is of the same order as $P_{\rm rr}$ while the shot noise contribution is much greater than the signal, and the shot noise contribution to $P_{\rm gr}$ is also greater than the signal. Despite this, the improvement in $\sigma(\fnl)$ is larger than for the low-$z$ bin, reaching 50\% at $K_{\rm min}=K_{\rm f}$. The right panel of Fig.~\ref{fig:mm-options} shows that the improvement is the same whether or not we include modes of $\delta_{\rm r}$ with $K<K_{\rm min}$, and therefore, cosmic variance cancellation between $\delta_{\rm g}$ and $\delta_{\rm r}$ is solely responsible for the change in $\sigma(\fnl)$.

We also see from Fig.~\ref{fig:mm-options} that the low number density ($\bar{n}=2\times10^{-5}\,{\rm Mpc}^{-3}$) in the high-$z$ bin is not a huge limiting factor, with only a modest change if we use a much larger number density. This is because the reconstruction noise remains comparable to $P_{\rm rr}$ even for a much denser survey, while further improvements are possible for a higher $k_{\rm max}$ but the same number density. If $P_{\rm rr,shot}$ and $P_{\rm gr,shot}$ are ignored, the results revert to the same situation as the low-$z$ bin, with only slight gains in $\sigma(\fnl)$ possible for low $K_{\rm min}$. Finally, if $b_2$ is changed from $17$ to $0$, there is significantly more improvement in $\sigma(\fnl)$: the amplitudes of $P_{\rm rr}$ and $P_{\rm gr}$ are reduced, but the relative uncertainty on $\fnl$ from marginalizing over $b_2$ is also reduced, and the latter effect wins.

\subsubsection{PUMA}
\label{sec:puma}

\begin{figure}[t]
\includegraphics[width=\textwidth, trim = 10 10 0 10 ]{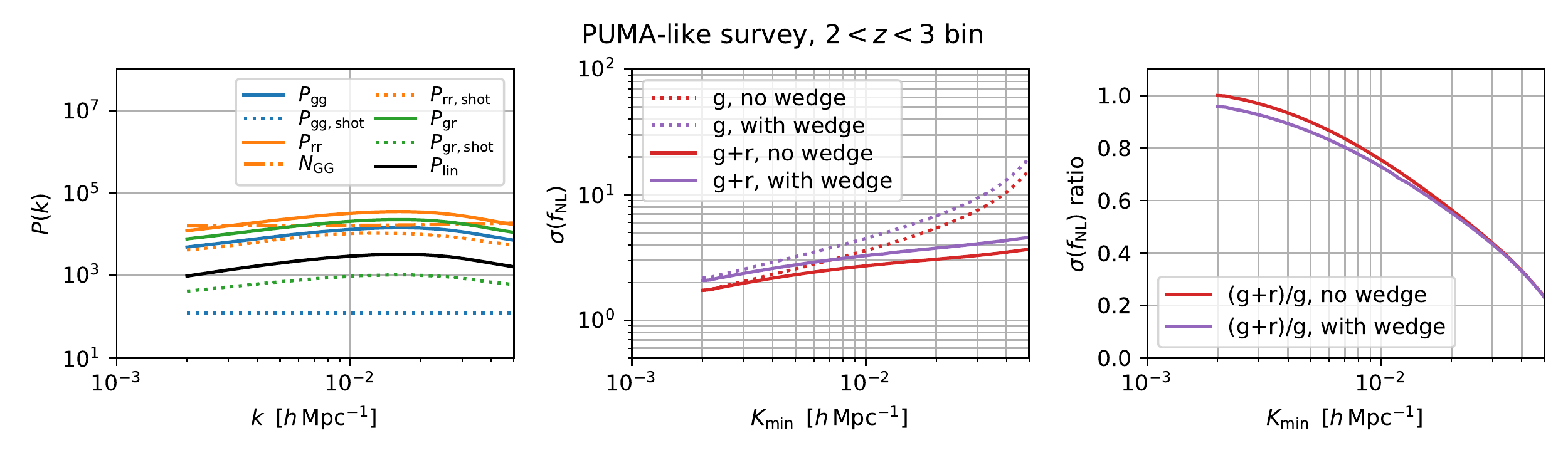} 
\includegraphics[width=\textwidth, trim = 10 10 0 10 ]{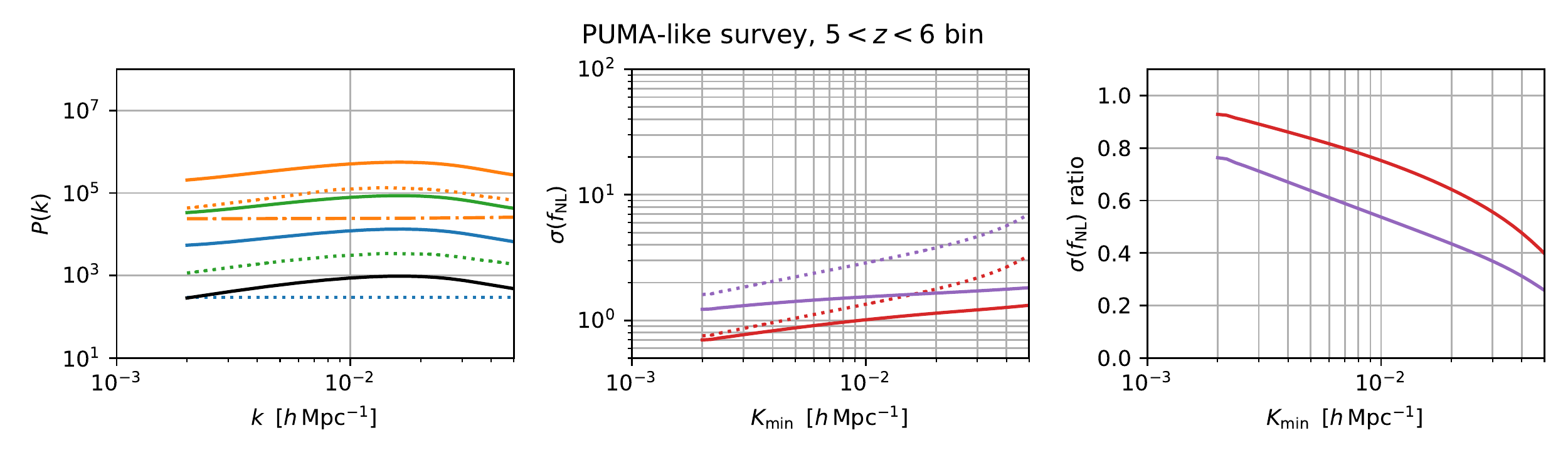} 
\caption{
\label{fig:puma}
As Fig.~\ref{fig:desi}, for low-redshift (\textit{top panels}) and high-redshift (\textit{bottom panels}) bins of a PUMA-like survey, treating the \tcm brightness temperature in the same way as $\delta_{\rm g}$ in our other forecasts, and translating thermal noise on the brightness temperature into an effective tracer number density for computing shot noise. We show $\sigma(\fnl)$ either neglecting or incorporating the effects of the \tcm foreground wedge; at high $z$, the benefit to $\sigma(\fnl)$ from including reconstructed modes is greater in the presence of the wedge, since there are fewer $\delta_{\rm g}$ modes that can be directly measured in that case. The results for the low-redshift bin are similar to those for MegaMapper, while larger improvements in $\sigma(\fnl)$ are possible at higher redshift.}
\end{figure}

We show results for the PUMA-like survey in Fig.~\ref{fig:puma}, either neglecting or including the effects of the foreground wedge. Note that the left panels in Fig.~\ref{fig:puma} only show noise curves corresponding to the no-wedge case. As for the other surveys, we assume an isotropic $K_{\rm min}$ for $\delta_{\rm g}$ in Fig.~\ref{fig:puma}; we show results for a cutoff on $K_{\parallel}$, which are qualitatively similar to those in Fig.~\ref{fig:puma}, in Appendix~\ref{app:fg-pumakparmin}.

\begin{figure}[t]
\includegraphics[width=\textwidth, trim = 10 10 0 10 ]{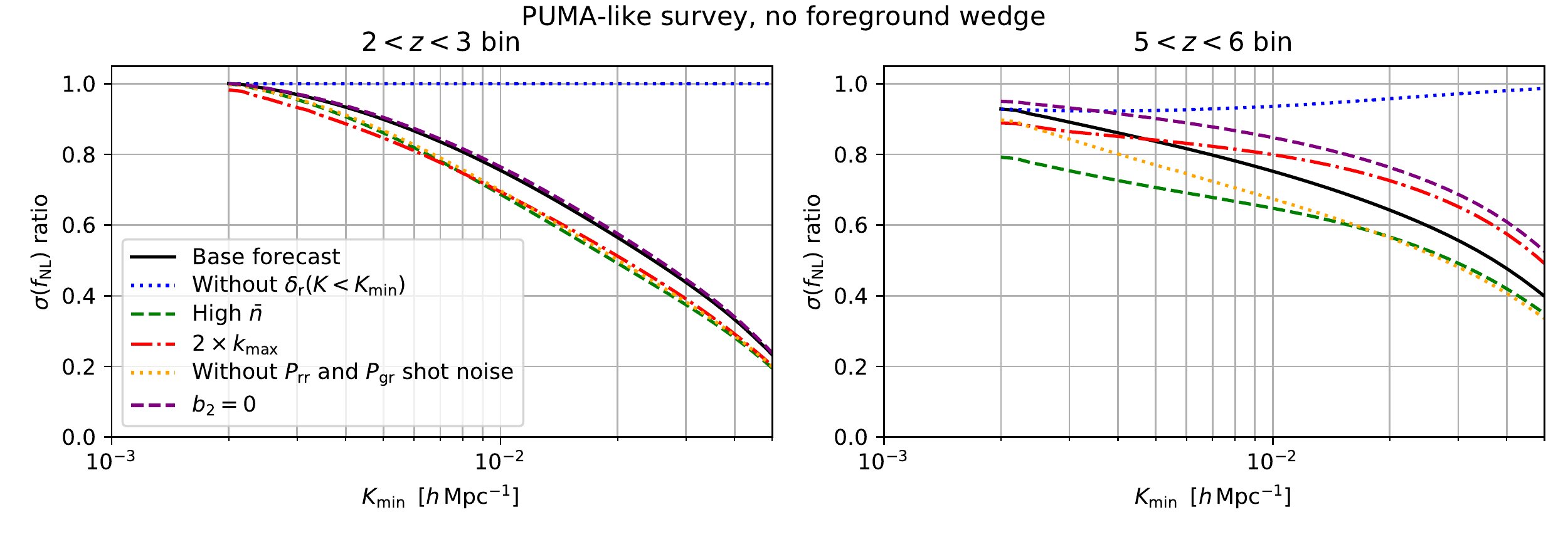} 
\caption{
\label{fig:puma-options}
Modifications to the base PUMA forecasts, neglecting the foreground wedge. In the low-$z$ bin, modes of $\delta_{\rm r}$ with $K<K_{\rm min}$ are entirely responsible for the improvement in $\sigma(\fnl)$, with most other modifications having little effect. In the high-$z$ bin, the blue dotted curve demonstrates that the $\sigma(\fnl)$ improvement comes from a combination of low-$K$ modes of $\delta_{\rm r}$ and cosmic variance cancellation at higher $K$. The improvement would get better if the thermal noise could be reduced (which maps onto a higher $\bar{n}$ in these forecasts).
}
\end{figure}

For both redshift bins, the shot noise in $C^{\rm gg}$, $C^{\rm rr}$, and $C^{\rm gr}$ is below the signal. However, the reconstruction noise is high enough in the low-redshift bin that the effect of reconstructed modes on $\sigma(\fnl)$ is similar to DESI and the low-$z$ MegaMapper bin, with the vast majority of the extra constraining power coming from reconstructed modes with $K<K_{\rm min}$ (see the left panel of Fig.~\ref{fig:puma-options}). The impacts of taking a higher $k_{\rm max}$ or tracer number density (equivalent to thermal noise in the interferometer) would only be mild.

Meanwhile, in the high-$z$ bin, the improvement in $\sigma(\fnl)$ arises from a combination of low-$K$ reconstructed modes and cosmic variance cancellation between $\delta_{\rm g}$ and $\delta_{\rm r}$. There is greater improvement in the presence of the wedge, as reconstruction helps to recover modes that would otherwise be lost. This improvement is around 20\% at the lowest~$K_{\rm min}$, and increases as more $\deltag$ modes are lost, implying that reconstruction will be extremely useful for single-tracer constraints on $\fnl$ from PUMA or other high-$z$ intensity mapping. The right panel of Fig.~\ref{fig:puma-options} shows that lower thermal noise would lead to further improvements, while a lower value of $b_2$ would worsen the results due to a lowering of the signal to noise on $P_{\rm rr}$.

\subsubsection{...and beyond}

To demonstrate how the constraints on $\fnl$ scale for surveys with extremely low shot noise and reconstruction noise, we also examine forecasts for the PUMA high-redshift bin where $k_{\rm max}$ is artificially increased, assuming that our quadratic bias model is valid to arbitrarily high $k$.\footnote{In practice the quadratic bias model will break down at sufficiently high $k$, but a theoretical framework such as the response function formalism (e.g. \citealt{Barreira:2017sqa, Barreira:2017kxd}) may allow the use of higher $k_{\rm max}$, with suitable modifications of the reconstruction procedure. We leave this topic to future work.} We take the galaxy number density to infinity in these forecasts, to prevent shot noise from becoming the limiting factor. In this case, we expect the uncertainty on $\fnl$ to scale like the inverse of the signal to noise on the reconstructed modes (see Sec.~\ref{sec:lowshotlimit}). In turn, in this limit, the signal to noise scales like $k_{\rm max}^{3/2}$ because the reconstruction noise spectrum $N_{\rm GG}$ becomes proportional to the number of modes with $k_{\rm min}<k<k_{\rm max}$ (see Eq.~\ref{eq:NgglowK}).

\begin{figure}[t]
\centering
\includegraphics[width=\textwidth, trim = 10 10 0 10 ]{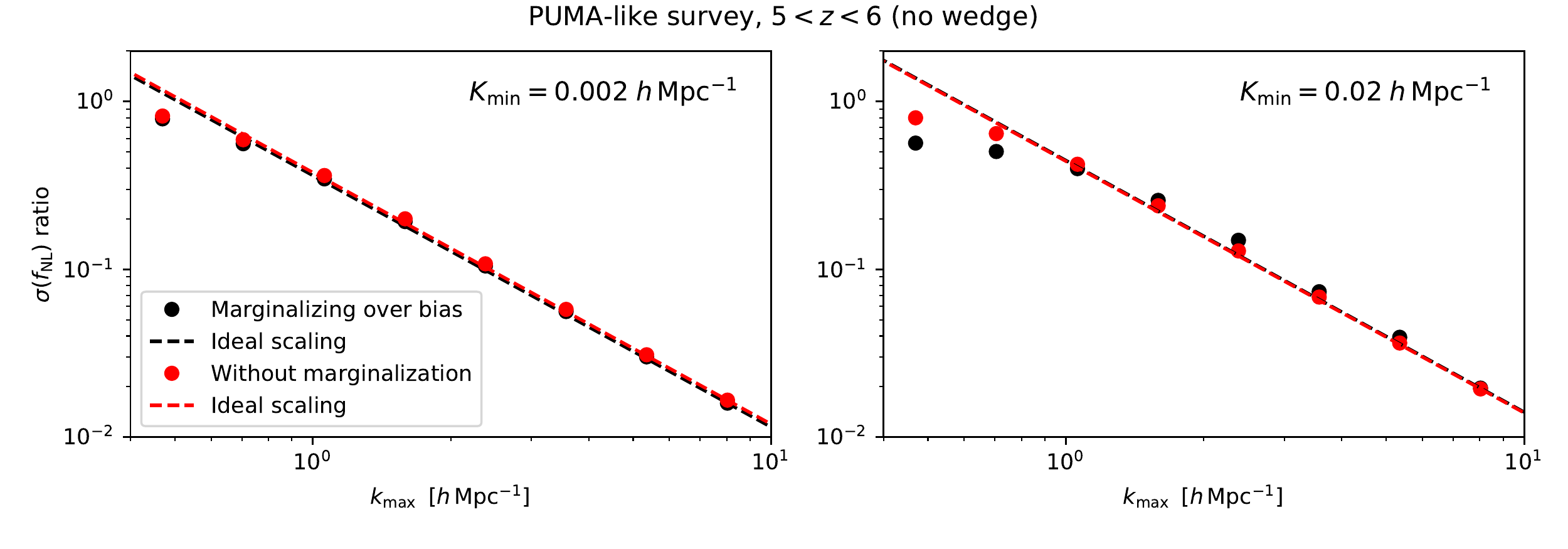} 
\caption{
\label{fig:sigfnl-scaling}
The ratio of $\sigma(\fnl)$ for the $\delta_{\rm g}+\delta_{\rm r}$ and $\delta_{\rm g}$-only forecasts for the high-$z$ PUMA bin, where $\bar{n}$ is taken to infinity and $k_{\rm max}$ is artificially increased, assuming that our quadratic bias model is valid to arbitrarily high $k$. The left and right panels correspond to two representatives values of $K_{\rm min}$. We show forecasts after marginalizing over bias parameters ({\it black points}) and without any marginalization ({\it red points}), along with the expected $k_{\rm max}^{-3/2}$ scaling ({\it dashed lines}, each normalized to the highest-$k_{\rm max}$ point plotted). We find the unmarginalized curves quickly approach the ideal scaling, while the marginalized forecasts show small deviations from this scaling. This shows that large increases in constraining power are possible in principle for surveys with very high number density and a large allowed value of $k_{\rm max}$.
}
\end{figure}

In Fig.~\ref{fig:sigfnl-scaling}, we show the ratio of $\sigma(\fnl)$ from g+r or g-only forecasts as a function of $k_{\rm max}$ for two representative values of $K_{\rm min}$. We indeed find that as the signal to noise on reconstructed modes is increased, the improvement on $\sigma(\fnl)$ also increases, with the unmarginalized forecasts quickly satisfying the expected scaling. (Marginalization over bias parameters causes small deviations from this scaling.) This demonstrates the huge increases in constraining power that are possible in principle for a survey with high galaxy number density and many small-scale modes whose correlations can be used in reconstruction. We have also numerically verified that the $\sigma(\fnl)$ ratio stays flat with increasing $k_{\rm max}$ if the noise in $P_{\rm gg}$ is taken very high, or if zero cross-correlation between $\deltag$ and $\delta_{\rm r}$ is assumed, further demonstrating that the scaling seen in Fig.~\ref{fig:sigfnl-scaling} arises from the joint constraining power between $\deltag$ and $\delta_{\rm r}$ measured from the same volume.

\section{Discussion}
\label{sec:discussion}

The results presented here can be compared to other methods either utilizing reconstruction and/or combining a tower of $n$-point correlation functions. Compared to most methods proposed in the literature, this work presents an optimal quadratic estimator to reconstruct the large scale mode. As explained in Sec.~\ref{sec:prec_on_r}, in principle this reconstructed mode can be used for several (cosmological) applications, and here we only explored $\fnl$ as an application of interest. When comparing this work with previous works, the main question is if the amount of information captured in the statistics of the tracer field is fully exploited. While it will be hard to compare methods directly, here we propose some heuristic arguments where we think our methods overlap and where they differ. 

As mentioned in the introduction, some publications have aimed to simplify the search for primordial non-Gaussianities by proposing more compressed versions of the full bispectrum \citep{Schmittfull:2014tca,Fergusson:2010ia,Byun:2017fkz,Dai:2020adm,MoradinezhadDizgah:2019xun,Chiang:2014oga,dePutter:2018jqk,Gualdi:2018pyw}. Common to these works is the fact that the information accessed is captured by the 2- and 3-point functions. In this work, besides the 3-point function, the 4-point function is also used and is important in obtaining cosmic variance cancellation. In other words, as shown in Fig.~\ref{fig:sigfnl-scaling}, significant improvements are possible when some conditions are met that would not be possible when considering the compressed statistics proposed in these earlier works. 

Even if cosmic variance cancellation is not achieved, we generally observe improvements between $20-50\%$. These numbers are similar to those projected in for example \cite{Dai:2020adm} for compressed statistics, but direct comparisons between our method and others are generally difficult. In the method presented in this paper, the improvement can roughly be attributed to coming from access to larger scales through the reconstruction, or, when both the linear and reconstructed mode are combined, cosmic variance cancellation. The projected improvement on the amplitude $\fnl$ from compressed statistics is the result of adding the bispectrum information on top of the power spectrum. For a detailed comparison we would need to carefully associate every mode with improved signal-to-noise side by side for the two different methods. Although this would be interesting by itself, as it would help us understand to what extent these methods are overlapping or how they complement one another, we will leave this to future work. 

The paper which our work has most in common with is \cite{dePutter:2018jqk}, which discusses the information content of a joint analysis of the two point function and squeezed three- and four-point functions.  This work has several commonalities with our analysis. To perform forecasts, \cite{dePutter:2018jqk} uses the squeezed-limit position-dependent power spectrum as a field, in an approach that is quite similar to our long wavelength mode reconstruction. The author also makes similar arguments for how sample variance cancellation can significantly influence and improve constraints.

However, there are also many important differences to our approach. Most importantly, the specific squeezed-limit power spectrum picture in \cite{dePutter:2018jqk} is discussed as a tool to enable better forecasting of joint 2, 3 and 4-point analyses of local non-Gaussianity, rather than a practical data analysis method. In contrast, our method has been proposed as an analysis method and estimator to rapidly jointly analyze 2, 3 and 4- point functions, that is not only computationally tractable, but has been tested (to some extent) on simulations.

There are also significant differences in the details of the methodology. Our reconstruction quadratic estimator can infer the long wavelength mode from mode-pairs that are not much smaller than the mode to be reconstructed; in contrast, \cite{dePutter:2018jqk} always operates in the squeezed limit when analyzing the position-dependent power spectrum. While it is expected that the majority of information about local non-Gaussianity in the 3 and 4-point functions is contained in very squeezed shapes, it is not clear that non-squeezed shapes do not contribute to long-wavelength mode reconstruction and hence sample variance cancellation. On the other hand, we note that in our analysis method we combine all quadratic estimator mode pairs into one long-wavelength mode estimate; in contrast, \cite{dePutter:2018jqk} shows that additional sample variance cancellation can be obtained when treating each mode pair (or position-dependent power spectrum bin) as a separate tracer. Although this suggests that further improvements to our method might be possible, the results of \cite{dePutter:2018jqk} suggest that this would only give significant improvements for very high $k$ and very low noise, beyond the capabilities of next-generation surveys.

Finally, shortly before the completion of this work, in a follow-up to \cite{Li:2020uug},  \cite{Li:2020luq} presented results relating to reconstruction of large-scale density modes using biased tracers, although without discussing the application to constraining non-Gaussianity. While the core of this work is similar (using a quadratic estimator as proposed by \citealt{Foreman:2018gnv}), here we explicitly account for the mode-coupling from higher-order biasing (which is non-negligible) in our estimator and compare theoretical estimates of the reconstruction noise, including bi- and trispectrum shot noise with additional contributions from primordial non-Gaussianity, to simulations. \cite{Li:2020luq} include observations on the light-cone in their formalism, and also include the effect of redshift space distortions up to second order in the linear density, which we neglect in this work (although see Sec.~\ref{sec:rsd} for a discussion of the impact of the Kaiser term).

\section{Conclusions}
\label{sec:conclusions}

In this paper, we have further developed a method for reconstructing modes of the cosmic density field using a quadratic estimator. This estimator extracts information about (typically) large-scale modes from correlations between smaller-scale modes, similar to standard methods for CMB lensing reconstruction. We have improved upon the estimator introduced in \cite{Foreman:2018gnv} by incorporating nonlinear biasing and local-type primordial non-Gaussianity, up to second order in the linear density field. At this order, there are several distinct sources of couplings between small-scale modes of the tracer density field, with amplitudes (i.e.\ bias coefficients) that are unknown a priori. We have found that an estimator based on the mode-coupling due to isotropic growth of the perturbations results in the lowest noise on reconstructed modes, and have enumerated the various multiplicative biases that will accompany the output of this estimator. We have also applied this estimator (along with those based on large-scale bulk flows and tidal interactions) to halos in $N$-body simulations, verifying that the results agree with analytical predictions.

In the course of this study, we have identified that it is crucial to include the shot noise contribution to the covariance between directly-observed tracer modes and reconstructed modes when performing an analysis. The shot noise not only adds a white noise contribution to the tracer power spectrum itself (in the case that the tracers Poisson-sample the density field, which we assume here), but also adds noise to the reconstructed-mode power spectrum and the cross-spectrum with the tracer modes. For sufficiently low tracer number density, this contribution can actually overwhelm the reconstruction noise from the quadratic estimator, and the cross spectrum alters the correlation coefficient between the tracer and reconstructed modes. We self-consistently include these features in our forecasts.

We have carried out forecasts that apply this formalism to several upcoming large-scale structure surveys: the emission-line galaxy survey from DESI \citep{Aghamousa:2016zmz}, the high-$z$ dropout survey envisioned in the MegaMapper proposal \citep{Ferraro:2019uce,Schlegel:2019eqc}, and the \tcm line-intensity survey from the PUMA proposal \citep{Ansari:2018ury,Bandura:2019uvb}, treated like a galaxy survey with effective number density derived from PUMA's thermal noise model. Examining the expected errorbars on the power spectrum of the reconstructed modes for $K<0.02\invMpc$, we find that these errorbars are several times larger than the signal for DESI and a high-redshift bin of MegaMapper. The latter is limited by the low number density of tracers, leading to a high shot noise contribution to the reconstruction noise, while the former's high reconstruction noise is sourced both by shot noise and a low number of modes used in the reconstruction (i.e.\ low $k_{\rm max}$). In the other forecasts, we find that high-S/N reconstructions of the large-scale density power spectrum can be obtained, with the caveat that this spectrum comes with multiplicative biases with known shapes but unknown amplitudes.

We have also computed the expected improvement in constraints on the amplitude of local-type primordial non-Gaussianity, $\fnl$, arising from analyzing reconstructed modes along with directly-observed tracer modes. For DESI and low-$z$ bins of MegaMapper and PUMA, the improvement arises solely from being able to access reconstructed modes with $K<K_{\rm min}$, where $K_{\rm min}$ denotes the minimum wavenumber at which we assume tracer modes can be measured (for systematics obscuring tracer modes with $K<K_{\rm min}$ but not affecting the modes used for reconstruction). On the other hand, for a high-$z$ bin of MegaMapper, the improvement in $\fnl$ constraints arises solely from cosmic variance cancellation between tracer and reconstructed modes at the same wavenumbers, similar to what can happen with different tracer populations or tracer-lensing cross-correlations \citep{Seljak:2008xr,McDonald:2008sh,Schmittfull:2017ffw,Liu:2020izx}. For a high-$z$ bin of PUMA, the $\sigma(\fnl)$ improvement comes from a combination of cosmic variance cancellation and reconstructed modes alone. Generally, cosmic variance cancellation depends on having a sufficiently high cross-correlation coefficient between the tracer and reconstructed modes, but this depends on shot noise in a somewhat complicated way, due to the aforementioned cross shot noise contribution.

The improvement in $\sigma(\fnl)$ also depends on the assumed value of $K_{\rm min}$, so we have plotted the expected constraints as a function of $K_{\rm min}$. In general, reconstructed modes improve $\sigma(\fnl)$ by tens of percents: for example, at $K_{\rm min}=0.01\invMpc$, $\sigma(\fnl)$ improves by a few percent for DESI, 15\% and 40\% for the low-$z$ and high-$z$ MegaMapper bins we consider, and at least 20\% for both $z$-bins of PUMA, depending on what is assumed for the \tcm foreground wedge. We have also shown that in the limit of zero shot noise, and if our quadratic bias model were valid to arbitrarily high~$k$, $\sigma(\fnl)$ scales like $k_{\rm max}^{-3/2}$, reflective of the number of small-scale modes used for reconstruction.

There are several possible ways that this work could be extended. For example, we have neglected redshift-space distortions, but they should clearly be incorporated in advance of applying this technique to data. One could also consider applying reconstruction to photometric surveys, after an assessment of the impact of photometric redshift errors on the results. It would be interesting to see how things change if one were to consider the bias model from \cite{Schmittfull:2018yuk}, based on shifted versions of bias operators designed to more fully incorporate large-scale displacements. Finally, one could consider investigating nonlinear response functions \citep{Barreira:2017sqa, Barreira:2017kxd} as a way to increase the number of small-scale modes that could be used in the quadratic estimator. Overall, we expect there to be many applications for reconstructed modes beyond constraints on local-type non-Gaussianity, and we therefore advocate for this reconstruction procedure as a useful tool  to increase the scientific returns of upcoming large-scale structure surveys.

\acknowledgments

We thank Emanuele Castorina, Azadeh Moradinezhad Dizgah, Simone Ferraro, Mat Madhavacheril, Moritz M{\"u}nchmeyer, Will Percival, Matias Zaldarriaga, and Hong-Ming Zhu for useful conversations. We also thank Emanuele Castorina, Azadeh Moradinezhad Dizgah, Emmanuel Schaan and Marcel Schmittfull for thoughtful comments on a draft of this paper.
Research at Perimeter Institute is supported in part by the Government of Canada through the Department of Innovation, Science and Economic Development Canada and by the Province of Ontario through the Ministry of Colleges and Universities. S.F.\ thanks the Kavli Institute for Cosmology, Cambridge for hospitality while part of this work was carried out. The numerical part of this work was performed using the DiRAC COSMOS supercomputer and greatly
benefited from the support of K. Kornet. M.A. acknowledges support from the Cambridge Commonwealth Trust, the Higher Education Commission, Pakistan, and the Cambridge Centre for Theoretical Cosmology. T.B. acknowledges support from the Cambridge Center for Theoretical Cosmology through the Stephen Hawking Advanced Fellowship. P.D.M.\  acknowledges support of the Netherlands organization for scientific research (NWO) VIDI grant (dossier 639.042.730). B.D.S. acknowledges support from an Isaac Newton Trust Early Career Grant, from a European Research Council (ERC) Starting Grant under the European Union’s Horizon 2020 research and innovation programme (Grant agreement No. 851274), and from an STFC Ernest Rutherford Fellowship. OD is funded by the STFC CDT in Data Intensive Science.

\appendix

\section{Derivation of density reconstruction from the bispectrum}
\label{app:rec-from-b}

We will here consider how we can reconstruct an unknown field $X$ given the knowledge of i) its bispectrum $B$ with two other fields $Y$, $Z$,  and ii) a measurement of these two other fields. While we write our argument exploiting the connection between bispectra and quadratic estimators in a form that is generally valid, for this paper we will assume that $X=\delta_1$, the linear density field, and both $Y, Z$ are the observed non-linear density field $Y, Z = \deltag$; given that we can calculate the $\delta_1 \deltag \deltag$ bispectrum, we can easily write an estimator for $\delta_1$ given an observed $\deltag$. We will assume statistical homogeneity and isotropy of the fields and the bispectrum, which is a good approximation for large-scale structure surveys (although it may be broken for other applications).

We begin by defining the bispectrum of the unknown field with two observed fields:
\beq
\langle X(\vk_1) Y(\vk_2) Z(\vk_3) \rangle = (2\pi)^3 \dirac(\vk_1+\vk_2+\vk_3) B_{XYZ}(k_1,k_2,k_3)\ .
\eeq
We will now write an ansatz for recovering the unknown field $X$ from a quadratic estimator involving $Y, Z$:
\beq
\label{eq:phiEstimator}
\hat X(\vK) =  \int_{\vq}  g(\vq,\vK-\vq)  Y(\vq) Z(\vK -\vq)\ ,
\eeq
where we have introduced a function $g$ which weights these pairs of modes. As we will see, the arguments of $Y, Z$ assumed here are a consequence of the delta function momentum constraint in the bispectrum.

We now derive this function $g$. The function must obviously give an unbiased estimator. In a situation where one may not wish to average over $Y, Z$ at fixed $X$ (e.g.\ because they are the same fields), we will define unbiasedness by the condition that
\beqn
\langle X(\vK') \hat{X}(\vK)\rangle = (2 \pi)^3 \dirac(\vK'+\vK)P_{XX}(K)\ .
\eeqn
For the estimator, this implies that since
\beqn
\langle X(\vK') \hat{X}(\vK)\rangle &=& \int_{\vq}  g(\vq,\vK-\vq) \langle X(\vK') Y(\vq) Z(\vK -\vq)\rangle \nonumber \\
&=& (2 \pi)^3 \dirac(\vK'+\vK)\int_{\vq}g(\vq,\vK-\vq) B_{XYZ}(K,q,|\vK-\vq|) \ ,
\eeqn
we have a normalization condition on $g$
\beq
I[g] \equiv \int_{\vq}  g(\vq,\vK-\vq) \frac{B_{XYZ}(K,q,|\vK-\vq|)}{P_{XX}(K)} =1\ .
\eeq

We would also like the estimator $\hat{X}$ to have as little variance per mode as possible. We will assume that, for the purposes of variance calculation, the fields can be approximated as Gaussian and statistically isotropic. Under these assumptions, the variance $V[f](\vK)$ is given by:
\beqn
\cmbav{\hat X (\vK) \hat X(\vK') } &=& (2\pi)^3 ~ V[g](\vK) ~ \dirac(\vK + \vK') \\
\nn &=&   \int_{\vq,\vq'} g(\vq,\vK-\vq) g(\vq',\vK'-\vq) ~  \cmbav{Y(\vq) Z(\vK -\vq)   Y(\vq') Z(\vK' -\vq')}\\
\nn &=& 
\int_{\vq,\vq'} g(\vq,\vK-\vq) g(\vq',\vK'-\vq') [\cmbav{Y(\vq) Y(\vq')}  \cmbav{Z(\vK - \vq) Z(\vK' -\vq')} \\
\nn &~&+ \cmbav{Y(\vq) Z(\vK' - \vq')}  \cmbav{Z(\vK - \vq) Y(\vq')} ] \\
\nn &=& \int_{\vq,\vq'}
	g(\vq,\vK-\vq) g(\vq',\vK'-\vq') 
	[ (2\pi)^3 P_{YY}(q) \dirac (\vq + \vq') (2\pi)^3 P_{ZZ}(\modu{\vK -\vq}) \dirac(\vK + \vK'-\vq-\vq') \\
	&~&+ (2\pi)^3 P_{YZ}(q) \dirac(\vq+\vK' - \vq') (2\pi)^3 P_{YZ}(\modu{\vK -\vq}) 
	\dirac (\vK-\vq + \vq' )]\\
\nn &=&  (2\pi)^3 \dirac(\vK + \vK') \int_{\vq}\left[  g(\vq,\vK-\vq)(-\vq,-\vK+\vq) P_{YY}(q) P_{ZZ}(\modu{\vK -\vq} )\right. \\
&~&\hspace{3.2cm}+\left. g(\vq,\vK-\vq) g(\vq-\vK,-\vq) P_{YZ}(q)  P_{YZ}(\modu{\vK -\vq})\right] \ ,
\eeqn
where we have used Wick's Theorem. In the following we will specialize to the case of $Y=Z$ (which is the relevant case for our application) and can thus consider $g$ to be symmetric under exchange of its arguments.
With the requirement for the reconstructed field to be real, this implies the following expression for the variance as a functional of $g$:
\beqn
V[g](\vK)&=&  2\int_{\vq}  {g^2(\vq,\vK-\vq)} P_{YY}(q) P_{YY}(\modu{\vK -\vq})  \ .
\eeqn
We can thus solve for $g$ by minimizing the variance $V[g](\vK)$ subject to the constraint $I[g]=1$. We can do this by introducing a Lagrange multiplier $N_{\alpha \alpha}$ and minimizing
\beq
V[g] - N_{\alpha \alpha} \times I[g] \ ,
\eeq
with respect to $g$. Minimizing this expression, we obtain
\beq
g(\vq,\vK-\vq) = N_{\alpha \alpha}(K)\frac{1}{4 P_{YY}(q) P_{YY}(\modu{\vK -\vq})}  \frac{B_{XYY}(K,q,|\vK-\vq|)}{P_{XX}(K)} \ ,
\eeq 
where from the constraint equation $I=1$ we find 
\beq
N_{\alpha \alpha}(K) = \left[   \int_{\vq}  \frac{1}{4 P_{YY}(q) P_{YY}(\modu{\vK -\vq}) } \left(\frac{B_{XYY}(K,q,|\vK-\vq|)}{P_{XX}(K)}\right)^2   \right]^{-1} \ .
\label{eq:weightfrombi}
\eeq

Applying this to our choice of fields, i.e., evaluating the $\langle \delta_{1} \deltag \deltag \rangle$ bispectrum, we note that we recover a function $g$ which gives the same expression for the quadratic estimator as used in the main part of our paper.
With Eq.~\eqref{eq:galpha} we have for the bispectrum of a linear mode $X=\delta_1$ and two galaxy modes (ignoring biases) $Y=Z=\delta_\text{g}$
\beq
B_{\delta_1,\delta_\text{g},\delta_\text{g}}(k_1,k_2,k_3)= 2\left[F_\alpha(\vk_1,\vk_3)P_\text{lin}(k_1)P_\text{lin}(k_3)+F_\alpha(\vk_1,\vk_2)P_\text{lin}(k_1)P_\text{lin}(k_2)\right]\, .
\eeq
Plugging this into Eq.~\eqref{eq:weightfrombi} we get
\beq
g(\vq,\vK-\vq) = N_{\alpha \alpha}(K) \frac{F_\alpha(\vK,-\vq)P_\text{lin}(q)+F_\alpha(\vK,-\vK+\vq)P_\text{lin}(|\vK-\vq|)}{2 P_\text{tot}(q) P_\text{tot}(\modu{\vK -\vq})}\, ,
\eeq
which agrees with Eq.~\eqref{eq:galpha} in the main text.

\section{Noise expressions for quadratic estimator}
\label{app:shot}

In this appendix, we derive expressions for the noise power spectrum corresponding to the auto-correlation of the reconstructed field $\hat{\Delta}_{\alpha}(\vec{K})$ and its cross-correlation with the original input tracer field $\deltag(\vec{K})$. We will see that the noise comes from a combination of shot noise, due to discrete sampling of the underlying matter field, and cosmic variance.

\subsection{Noise for the auto correlation of the reconstructed field}
In this appendix, we calculate the covariance of our quadratic estimators, which is defined as
\begin{multline}
    \big\langle \hat{\Delta}_{\alpha}(\vec{K})\hat{\Delta}_{\beta}(\vec{K'})\big\rangle-\big\langle \hat{\Delta}_{\alpha}(\vec{K})\big\rangle\big\langle \hat{\Delta}_{\beta}(\vec{K'})\big\rangle=\\
    \int_{\vec{q}}\int_{\vec{q'}}g_{\alpha}(\vec{q}, \vec{K}-\vec{q})g_{\beta}(\vec{q'}, \vec{K'}-\vec{q'})\Big(\big\langle\delta_{\text{g}}(\vec{q})\delta_{\text{g}}(\vec{K}-\vec{q})\delta_{\text{g}}(\vec{q'})\delta_{\text{g}}(\vec{K'}-\vec{q'})\big\rangle-\big\langle\delta_{\text{g}}(\vec{q})\delta_{\text{g}}(\vec{K}-\vec{q})\big\rangle \big\langle \delta_{\text{g}}(\vec{q'})\delta_{\text{g}}(\vec{K'}-\vec{q'})\big\rangle\Big) \ .
\end{multline}
To compute this expression we have to first derive the four point function for the input tracer field. It is possible to derive the shot noise formulae directly in Fourier space, with a discretized version of the tracer field. 
Alternative derivations of bispectrum shot noise can be found in \citep{DJeong:2010PhD} following \citep{1994Feldman}. We cross-checked our results with \cite{sugiyama2019perturbation, Chan_2017}.

Let us start by rederiving the stochasticity contributions to the power spectrum, bispectrum and trispectrum.
Let us consider a finite number $N$ of point-like tracers, such as galaxies, at positions $\vec x_i$ in a finite volume $V$.\footnote{It is useful to recall that all wave vectors in a finite volume are integer multiples of the fundamental wavenumber. The Dirac delta distribution thus becomes a Kronecker delta $(2\pi)^2\delta_\text{D}(\vec k_1+\vec k_2)\to V \delta_\text{K}(\vec k_1,\vec k_2)$.}
Their Fourier space density field is then given as a sum of plane waves
\begin{equation}
\delta_\text{g}(\vec k)=\frac{1}{\bar n}\sum_i \exp\left[i \vec k \vec x_i\right]\ ,
\end{equation}
where $\bar n=N/V$.
The power spectrum of the discrete tracers in the finite volume can then be computed as 
\begin{equation}
\begin{split}
P_\text{g}(k)=&\frac{1}{V}\left\langle \delta_\text{g}(\vec k) \delta_\text{g}(-\vec k)\right\rangle\; \\
=&\frac{V}{N^2}\sum_{i=j}\left\langle\exp\left[i \vec k (\vec x_i-\vec x_j)\right]\right\rangle+\frac{V}{N^2}\sum_{i\neq j}\left\langle\exp\left[i \vec k (\vec x_i-\vec x_j)\right]\right\rangle\; \\
=&\frac{1}{\bar n}+P_\text{g,cont}(k)\; .
\end{split}
\end{equation}
Here, the constant $1/\bar{n}$ is denoted the shot noise term and we have identified the non-zero lag expectation value with the continuous part of the discrete tracer power spectrum $P_\text{g,cont}(k)$. In the local bias model at linear order we have $P_\text{g,cont}(k)=b_1^2 P_\text{lin}(k)$, which becomes $P_\text{g,cont}(k)=[b_{10}+b_{01}/M(k)]^2 P_\text{lin}(k)$ in the presence of primordial non-Gaussianities of the local kind. 
Let us now consider the bispectrum. Following the same steps that led to the power spectrum above, we have to consider the case where all three positions coincide, the case where two positions coincide but are different from the third, and finally the case where all three positions are distinct:
\begin{equation}
\begin{split}
B_\text{g}(\vec k_1,\vec k_2)=&\frac{1}{V}\left\langle \delta_\text{g}(\vec k_1) \delta_\text{g}(\vec k_2)\delta_\text{g}(-\vec k_1-\vec k_2)\right\rangle\; \\
=&\frac{V^2}{N^3}\sum_{i=j=l}\left\langle\exp\left[i \vec k_1 (\vec x_i-\vec x_l)+i \vec k_2 (\vec x_j-\vec x_l)\right]\right\rangle\\
&+\frac{V^2}{N^3}\sum_{i=l\neq j}\left\langle\exp\left[i \vec k_1 (\vec x_i-\vec x_l)+i \vec k_2 (\vec x_j-\vec x_l)\right]\right\rangle+\text{2 perm.}\\
&+\frac{V^2}{N^3}\sum_{i\neq j,j \neq l,i\neq l}\left\langle\exp\left[i \vec k_1 (\vec x_i-\vec x_l)+i \vec k_2 (\vec x_j-\vec x_l)\right]\right\rangle\; \\
=&\frac{1}{{\bar n}^2}+\frac{1}{\bar n}\left[P_\text{g,cont}(k_1)+\text{2 perm.}\right]+B_\text{g,cont}(\vec k_1,\vec k_2)\; .
\label{eq:bispectnoisedelta}
\end{split}
\end{equation}
Again, the non-zero lag correlators are identified with the continuous power spectrum and bispectrum of the tracer field. We see that two different stochasticity corrections arise: a $1/{\bar n}^2$ constant shot noise term and a product of the shot noise and the continuous power spectrum. As above for the power spectrum, in the presence of primoridal non-Gaussianity, both of these continuous statistics contain the respective non-Gaussian bias corrections. Note that there is now a coupling between stochasticity and clustering which is enhanced with respect to the pure noise term on large scales.

Let us connect this result to the noise terms introduced in Eq.~\eqref{eq:nongausbiasexpansion}, where the relevant contributions are given by
\begin{equation}
    \delta_\text{g}\supset \epsilon+\epsilon_\delta \delta +\epsilon_\varphi \varphi\, .
\end{equation}
The three-point correlator of the noise fields $\epsilon$ can be associated with the white-noise term in Eq.~\eqref{eq:bispectnoisedelta}
\begin{equation}
  \left\langle \epsilon(\vk_1) \epsilon(\vk_2) \epsilon(\vk_3) \right \rangle=(2\pi)^3 \delta_\text{D}(\vec k_1+\vec k_2+\vec k_3)\frac{1}{\bar n^2}\, .
\end{equation}
The contributions from $\epsilon_\delta \delta$ and $\epsilon_\varphi \varphi$ arise by correlating with the linear $\delta$ and $\varphi$ fields and the linear noise term $\epsilon$
\begin{equation}
\begin{split}
    \left\langle \epsilon(\vec k_1) ([\epsilon_\delta \star\delta]+ [\epsilon_\varphi \star\varphi])(\vec k_2) (b_{10}\delta+b_{01}\varphi)(\vec k_3) \right \rangle=&
    \int_{\vec q}\Bigl[\left\langle \epsilon(\vec k_1) \epsilon_\delta(\vec q)\right\rangle \bigl(b_{10}\left\langle \delta(\vec k_2-\vec q) \delta(\vec k_3)\right \rangle+b_{01}\left\langle \varphi(\vec k_2-\vec q) \delta(\vec k_3)\right \rangle \bigr)\\
    &+\left\langle \epsilon(\vec k_1) \epsilon_\varphi(\vec q)\right\rangle \bigl(b_{10}\left\langle \varphi(\vec k_2-\vec q) \delta(\vec k_3)\right \rangle+b_{01}\left\langle \varphi(\vec k_2-\vec q) \varphi(\vec k_3)\right \rangle \bigr)\Bigr] \ .%
\end{split}
\end{equation}
Using \citep{Schmidt:2015gwz,Desjacques:2016bnm} we have for the noise correlators
\begin{align}
    \left\langle \epsilon \epsilon_\delta\right\rangle=(2\pi)^3\delta_\text{D}(\vec k+\vec k')\frac{b_{10}}{\bar n}\, , &&
    \left\langle \epsilon \epsilon_\varphi\right\rangle=\frac{b_{01}}{b_{10}}\left\langle \epsilon \epsilon_\delta\right\rangle=(2\pi)^3\delta_\text{D}(\vec k+\vec k')\frac{b_{01}}{\bar n}\, .
\end{align}
We finally obtain for the mixed contribution to the three-point correlator
\begin{equation}
\begin{split}
    \left\langle \epsilon (\epsilon_\delta \delta+ \epsilon_\varphi \varphi) (b_{10}\delta+b_{01}\varphi) \right \rangle
    =&(2\pi)^3 \delta_\text{D}(\vec k_1+\vec k_2+\vec k_3)\frac{P_\text{lin}(k_3)}{\bar n}\left[b_{10} \left(b_{10}+\frac{b_{01}}{M(k_3)} \right)+\frac{b_{01}}{M(k_3)} \left(b_{10}+\frac{b_{01}}{M(k_3)} \right)\right]\, .
    \end{split}
\end{equation}
In summary, we have
\begin{equation}
    B_\text{g}(k_1,k_2,k_3)=\frac{1}{\bar n^2}+\frac{1}{\bar n}\left[\left(b_{10}+ \frac{b_{01}}{M(k_1)}\right)^2P_\text{lin}(k_1)+\text{2 perm.}\right]+B_\text{g,cont}(k_1,k_2,k_3) \ .
\end{equation}
This is equivalent to Eq.~\eqref{eq:bispectnoisedelta}, as long as the galaxy power spectrum $P_\text{g,cont}$ in that equation is taken to be the one with the  scale dependent non-Gaussian bias $b_{10}+b_{01}/M(k)$.

For the connected trispectrum, we have four positions, which allow for five different configurations: all positions equal, three positions equal but different from the fourth one, two pairs of positions equal but different from the other pair, one pair of positions equal but different from all other positions and finally, all four positions distinct. The trispectrum can then be written as
\begin{equation}
\begin{split}
T_\text{g,conn}(\vec k_1,\vec k_2,\vec k_3)=&\frac{1}{V}\left\langle \delta_\text{g}(\vec k_1) \delta_\text{g}(\vec k_2)\delta_\text{g}(\vec k_3)\delta_\text{g}(-\vec k_1-\vec k_2-\vec k_3)\right\rangle\; \\
=&\frac{V^3}{N^3}\frac{1}{N}\sum_{i,j,s,t}\left\langle\exp\left[i \vec k_1 (\vec x_i-\vec x_t)+i \vec k_2 (\vec x_j-\vec x_t)+i \vec k_3 (\vec x_s-\vec x_t)\right]\right\rangle\\
=&\frac{V^3}{N^3}\frac{1}{N}\sum_{i=j=s=t}\left\langle\exp\left[i \vec k_1 (\vec x_i-\vec x_t)+i \vec k_2 (\vec x_j-\vec x_t)+i \vec k_3 (\vec x_s-\vec x_t)\right]\right\rangle\\
&+\frac{V^2}{N^2}\frac{V}{N^2}\sum_{i=j=s\neq t}\left\langle\exp\left[-i \vec k_4 (\vec x_i-\vec x_t)\right]\right\rangle+\text{3 perm.}\\
&+\frac{V^2}{N^2}\frac{V}{N^2}\sum_{i=j\neq s= t}\left\langle\exp\left[i (\vec k_1+\vec k_2) (\vec x_i-\vec x_t)\right]\right\rangle+\text{2 perm.}\\
&+\frac{V}{N}\frac{V^2}{N^3}\sum_{i= j\neq s\neq t, j\neq t}\left\langle\exp\left[i (\vec k_1+\vec k_2) (\vec x_i-\vec x_t)+i \vec k_3 (\vec x_s-\vec x_t)\right]\right\rangle+\text{5 perm.}\\
&+\frac{V^3}{N^4}\sum_{i\neq j\neq s\neq t,i\neq t,i\neq s,j\neq t}\left\langle\exp\left[i \vec k_1 (\vec x_i-\vec x_t)+i \vec k_2 (\vec x_j-\vec x_t)+i \vec k_3 (\vec x_s-\vec x_t)\right]\right\rangle\\
=&\frac{1}{\bar n^3}+\frac{1}{\bar n^2}[P_\text{g,cont}(\vec k_4)+3\text{ perm.}]+\frac{1}{\bar n^2}[P_\text{g,cont}(\vec k_1+\vec k_2)+2\text{ perm.}]\\
&+\frac{1}{\bar n}\left[B_\text{g,cont}(\vec k_1+\vec k_2,\vec k_3)+\text{5 perm.}\right]+T_\text{g,cont}(\vec k_1,\vec k_2,\vec k_3).
\end{split}
\end{equation}

Furthermore, there is a disconnected cosmic variance contribution for counter-aligned pairs of momenta
\begin{equation}
\begin{split}
T_\text{g,disconn}(\vec k_1,-\vec k_1,\vec k_3)=&\frac{1}{V}\left\langle \delta_\text{g}(\vec k_1) \delta_\text{g}(-\vec k_1)\delta_\text{g}(\vec k_3)\delta_\text{g}(-\vec k_3)\right\rangle\; \\
=&\frac{V^3}{N^3}\frac{1}{N}\sum_{i,j,s,t}\left\langle\exp\left[i \vec k_1 (\vec x_i-\vec x_j)+i \vec k_3 (\vec x_s-\vec x_t)\right]\right\rangle\\
=&V\frac{V}{N^2}\sum_{i,j}\left\langle\exp\left[i \vec k_1 (\vec x_i-\vec x_j)\right]\right\rangle
\frac{V}{N^2}\sum_{s,t}\left\langle\exp\left[i \vec k_3 (\vec x_s-\vec x_t)\right]\right\rangle.\\
\end{split}
\end{equation}
In the continuous case this disconnected contribution to the four-point function becomes
\begin{equation}
    \left\langle \delta_\text{g}(\vec k_1) \delta_\text{g}(\vec k_2)\delta_\text{g}(\vec k_3)\delta_\text{g}(\vec k_4)\right\rangle\supset(2\pi)^6\dirac\left(\vec k_1+\vec k_2\right)\dirac\left(\vec k_3+\vec k_4\right)\left[P_\text{g,cont}(k_1)+\frac{1}{\bar n}\right]\left[P_\text{g,cont}(k_3)+\frac{1}{\bar n}\right]+\text{2 perm.}
\end{equation}

This will give the Gaussian contribution to the covariance of the estimator.

Thus, the total covariance for the auto-spectrum of the reconstructed field is
\begin{align*}
\big\langle \hat{\Delta}_{\alpha}(\vec{K})\hat{\Delta}_{\beta}(\vec{K'})\big\rangle
    -\big\langle \hat{\Delta}_{\alpha}(\vec{K})\big\rangle\big\langle \hat{\Delta}_{\beta}(\vec{K'})\big\rangle
=&\int_{\vec{q}}\int_{\vec{q'}}
	g_{\alpha}(\vec{q}, \vec{K}-\vec{q})g_{\beta}(\vec{q'}, \vec{K'}-\vec{q'})\\	&\hspace{1cm}\times\Big[\big\langle\delta_{\text{g}}(\vec{q})\delta_{\text{g}}(\vec{K}-\vec{q})\delta_{\text{g}}(\vec{q'})\delta_{\text{g}}(\vec{K'}-\vec{q'})\big\rangle\\
    &\hspace{1cm}-\big\langle\delta_{\text{g}}(\vec{q})\delta_{\text{g}}(\vec{K}-\vec{q})\big\rangle \big\langle \delta_{\text{g}}(\vec{q'})\delta_{\text{g}}(\vec{K'}-\vec{q'})\big\rangle\Big] \\
=&(2\pi)^3\dirac(\vec{K}+\vec{K'}) \left[
    \frac{N_{\alpha\alpha}(\vec{K})N_{\beta\beta}(\vec{K})}{N_{\alpha\beta}(\vec{K})}+N_{\alpha\beta,\rm{shot}}(\vec{K})\right] \ ,
    \numberthis
\end{align*}
where
\begin{align*}
N^{\text{T}}_{\alpha\beta,\rm{shot}}(\vec{K})
	&\equiv  \int_{\vec{q}}\int_{\vec{q'}}g_{\alpha}(\vec{q}, \vec{K}-\vec{q})
	g_{\beta}(\vec{q'}, -\vec{K}-\vec{q'})
	T_{\rm{g,conn}}(\vec{q}, \vec{K}-\vec{q}, \vec{q'}, -\vec{K}-\vec{q'})\\
&=\int_{\vec{q}}\int_{\vec{q'}}g_{\alpha}(\vec{q}, \vec{K}-\vec{q})g_{\beta}(\vec{q'}, -\vec{K}-\vec{q'})\\
&\qquad\quad\times \Bigg\{ \frac{1}{\bar{n}^3}
	+\frac{1}{\bar{n}^2}\bigg[ P_{\rm{g},\rm{cont}}(\vec{q})
	+P_{\rm{g},\rm{cont}}(\vec{K}-\vec{q})+P_{\rm{g},\rm{cont}}(\vec{q}')
	+P_{\rm{g},\rm{cont}}(\vec{K}+\vec{q'}) \bigg]\\
&\qquad\qquad
	+\frac{1}{\bar{n}^2}\bigg[P_{\rm{g},\rm{cont}}(\vec{K})
	+P_{\rm{g},\rm{cont}}(\vec{q}-\vec{K}-\vec{q'})
	+P_{\rm{g},\rm{cont}}(\vec{q}+\vec{q'})\bigg]\\
&\qquad\qquad
	+\frac{1}{\bar{n}} \bigg[B_{\rm{g},\rm{cont}}(\vec{K},\vec{q'},-\vec{K}-\vec{q'})
	+B_{\rm{g},\rm{cont}}(\vec{q}+\vec{q'}, \vec{K}-\vec{q},-\vec{K}-\vec{q'})  \\
&\qquad\qquad\qquad
	+B_{\rm{g},\rm{cont}}(\vec{q}-\vec{K}-\vec{q'},\vec{K}-\vec{q},\vec{q'} )
	+B_{\rm{g},\rm{cont}}(\vec{K}-\vec{q}+\vec{q'}, \vec{q},-\vec{K}-\vec{q'}) \\
&\qquad\qquad\qquad
	+B_{\rm{g},\rm{cont}}(\vec{K}-\vec{q}-\vec{K}-\vec{q'},\vec{q},\vec{q'} )
	+B_{\rm{g},\rm{cont}}(\vec{q'}-\vec{K}-\vec{q'},\vec{q},\vec{K}-\vec{q})\bigg] \Bigg\}\ .
\numberthis
\label{eq:nrrshot}
\end{align*}
For our forecasts, we use the ``growth" estimator, and therefore the conversion to the notation of the main text is
\beq
P_\text{rr,shot}(K) = N_{\rm GG,shot}^{\rm T}(K)\ .
\label{eq:prrshot-appendix}
\eeq
Note an important feature of Eq.~\eqref{eq:nrrshot}: the shot noise contribution to the quadratic estimator's noise power spectrum depends on the tracer power spectrum and bispectrum at the same scale $K$ as the mode being reconstructed. This is to be contrasted with the Gaussian estimator noise in Eq.~\eqref{eq:nab}, which is mainly determined by the tracer power spectrum at the smallest scale $k_{\rm max}$ used in the estimator. Since the tracer power spectrum and bispectrum both increase at smaller wavenumbers (down to the matter-radiation equality scale), Eq.~\eqref{eq:nrrshot}'s sensitivity to large scales can cause it to dominate over the Gaussian estimator noise if $\bar{n}$ is sufficiently low. In our forecasts in the main text, this condition is met for DESI (Fig.~\ref{fig:desi}) and MegaMapper (Fig.~\ref{fig:mm}).

We can simplify the above expression by noticing that changes of variables can make some terms of the integrand equivalent. For example, we can simplify the final 3 lines of Eq.~\eqref{eq:nrrshot} into
\begin{equation}
\int_{\vec{q}}\int_{\vec{q'}} g_{\alpha}(\vec{q}, \vec{K}-\vec{q})g_{\beta}(\vec{q'}, -\vec{K}-\vec{q'})  
\frac{1}{\bar{n}}
\Big( B_{\rm{g},\rm{cont}}(\vec{K}, \vec{q'}, -\vec{K}-\vec{q'})
+4B_{\rm{g},\rm{cont}}(-\vec{q}-\vec{q'}, \vec{q}, \vec{q'})
+B_{\rm{g},\rm{cont}}(-\vec{K}, \vec{q}, \vec{K}-\vec{q}) \Big)   \ .
\end{equation}
In our calculations, we take the tree-level expression for the bispectrum, obtainable from Eq.~\eqref{eq:deltag-condensed} as (see also \citealt{Baldauf:2010vn,Tellarini:2015faa})
\begin{equation}
    B_{\rm{g},\rm{cont}}(\vec{k}_1, \vec{k}_2, \vec{k}_3) = 2\left(b_{10}+f_{\rm{NL}}\frac{c_{01}}{M(\vec{k}_1)}\right)\left(b_{10}+f_{\rm{NL}}\frac{c_{01}}{M(\vec{k}_2)}\right)\sum_{\alpha}c_{\alpha}F_{\alpha}(\vec{k}_1, \vec{k}_2)P_{\text{lin}}(\vec{k}_1)P_{\text{lin}}(\vec{k}_2)+\text{ 2 perms.}
\end{equation}

\subsection{Noise for the cross correlation of the reconstructed field with the tracer field}

The noise calculation for the cross-correlation of the reconstructed field with the tracer field is very similar to the one above. The variance of the cross-correlation of the reconstructed field with the tracer field depends on the bispectrum of the tracer field derived above in
Eq.~\eqref{eq:bispectnoisedelta} (see also \citealt{DJeong:2010PhD}):
\begin{multline}
\Big\langle \hat{\Delta}_{\alpha}(\Vec{K})\deltag(\Vec{K'}) \Big\rangle-\Big\langle\hat{\Delta}_{\alpha}(\Vec{K})\Big\rangle\Big\langle\deltag(\Vec{K'}) \Big\rangle = \Big\langle \hat{\Delta}_{\alpha}(\Vec{K})\deltag(\Vec{K'}) \Big\rangle = \int_{\Vec{q}} g_{\alpha}(\Vec{q}, \Vec{K}-\Vec{q}) \Big\langle \deltag(\Vec{q})\deltag(\Vec{K}-\Vec{q})\deltag(\Vec{K'}) \Big\rangle 
\\
= (2\pi)^3\dirac(\Vec{K}+\Vec{K'})\int_{\Vec{q}}g_{\alpha}(\Vec{q},\Vec{K}-\Vec{q}) \Bigg[B_{\rm{g},\rm{cont}}(\Vec{q},\Vec{K}-\Vec{q},-\Vec{K})+\frac{1}{\Bar{n}}\Big(P_{\rm{g},\rm{cont}}(\Vec{q})+P_{\rm{g},\rm{cont}}(\Vec{K}-\Vec{q})+P_{\rm{g},\rm{cont}}(-\Vec{K})\Big)+\frac{1}{\Bar{n}^2}\Bigg]\ .
\end{multline}
Then, the shot noise power is given by
\begin{equation}
    N^{\text{B}}_{\alpha,\rm{shot}}(\vec{K})\equiv \frac{1}{\Bar{n}} \int_{\Vec{q}}g_{\alpha}(\Vec{q},\Vec{K}-\Vec{q}) \Bigg[P_{\rm{g},\rm{cont}}(\Vec{q})+P_{\rm{g},\rm{cont}}(\Vec{K}-\Vec{q})+P_{\rm{g},\rm{cont}}(-\Vec{K})+\frac{1}{\Bar{n}}\Bigg]\ .
    \label{eq:nrtshot}
\end{equation}
For the growth estimator we use in our forecasts we thus have
\beq
P_\text{gr,shot}(K) = N_\text{G,shot}^{\rm B}(K)\ .
\label{eq:pgrshot-appendix}
\eeq

\section{Bias-hardening}
\label{app:biashardening}

In Sec.~\ref{sec:qe}, we saw that a quadratic estimator designed to have unit response to a specific form of mode-coupling will generically acquire a mean-field contamination from other forms of mode coupling that are not incorporated in the estimator's weights (see Eq.~\ref{eq:meanfield}). Our main approach in this paper is to include that contamination in our model for the estimator's output, marginalizing over the associated free (bias) parameters where necessary. Alternatively, one can attempt to define an estimator that is orthogonal to those extra mode-couplings; such a ``bias-hardening" procedure has been applied to weak lensing of the CMB (e.g.\ \citealt{Namikawa:2012pe,Osborne:2013nna}) and line intensity maps \citep{Foreman:2018gnv}. In this appendix, we explore this approach and explain why we did not find it to be useful for this study.

\subsection{General derivation}

First, we derive a form of bias-hardening that is a light generalization of the standard form (e.g. \citealt{Namikawa:2012pe}). Recall that a quadratic estimator with weights $g_{\alpha}$ is given by
\beq
    \hat{\Delta}_{\alpha}(\vec{K})
    =\int_{\vec{q}}g_{\alpha}(\vec{q}, \vec{K}-\vec{q})\deltag(\vec{q})\deltag(\vec{K}-\vec{q})\ ,
\eeq
with expectation value
\beq
\left\langle \hat{\Delta}_{\alpha}(\vec{K}) \right\rangle_{\delta_1(\vec{K})\text{ fixed}}  
      = \sum_{\beta} \lb c_\beta \int_{\vq} g_{\alpha}(\vq, \vK-\vq) f_{\beta}(\vq, \vK-\vq) \rb 
      b_1\delta_1(\vec{K})\ .
      \label{eq:meanfieldinapp}
\eeq
The estimator will be unbiased with respect to the $\alpha$ mode-coupling if
\begin{equation}
    \int_{\vec{q}}g_{\alpha}(\vec{q}, \vec{K}-\vec{q}) f_{\alpha}(\vec{q}, \vec{K}-\vec{q})  = 1\ .
    \label{eq:gab-constraint}
\end{equation}
The $\beta\neq\alpha$ terms in the sum in Eq.~\eqref{eq:meanfieldinapp} could be subtracted if we knew the values of the $c_\beta$ coefficients ahead of time, but this will generally not be true. Instead, we can attempt to set the weights $g_\alpha$ such that
\begin{equation}
    \int_{\vec{q}} g_{\alpha}(\vec{q}, \vec{K}-\vec{q})f_{\beta}(\vec{q}, \vec{K}-\vec{q}) = 0 \ ,
\end{equation}
for all $\beta\neq\alpha$.

We begin by assuming that there is only one additional mode-coupling $\beta$ that we are concerned with, and requiring that the estimator's response to it is not necessarily zero, but a chosen constant $c$ instead:
\begin{equation}
    \int_{\vec{q}} g_{\alpha}(\vec{q}, \vec{K}-\vec{q})
    f_{\beta}(\vec{q}, \vec{K}-\vec{q}) =c\ ,
    \label{eq:constraint0}
\end{equation}
while also imposing Eq.~\eqref{eq:gab-constraint} and minimizing the Gaussian contribution to the variance of $\hat{\Delta}_{\alpha}(\vec{K})$, given by
\begin{equation}
    {\rm Var}_{\rm G}\! \lb \hat{\Delta}_{\alpha}(\vec{K}) \rb 
    = 2\int_{\vec{q}}g_{\alpha}(\vec{q}, \vec{K}-\vec{q})
    g^{*}_{\alpha}(\vec{q}, \vec{K}-\vec{q})
    P_{\rm tot}(q)P_{\rm tot}(|\vec{K}-\vec{q}|) \ .
    \label{eq:covmatcalc}
\end{equation}
We find $g_\alpha$ that satisfies these conditions by the method of Lagrange multipliers, starting with the following function:
\begin{align*}
    L[g_{\alpha}, \lambda, \lambda_*]
    &= 2\int_{\vec{q}} g_{\alpha}(\vec{q}, \vec{K}-\vec{q})
    g^{*}_{\alpha}(\vec{q}, \vec{K}-\vec{q})
    P_{\rm tot}(\vec{q}) P_{\rm tot}(\vec{K}-\vec{q}) \\
&\quad - \lambda \lb \int_{\vec{q}} g_{\alpha}(\vec{q}, \vec{K}-\vec{q})f_{\alpha}(\vec{q}, \vec{K}-\vec{q})-1 \rb 
    - \lambda_{*} \lb \int_{\vec{q}} g_{\alpha}(\vec{q}, \vec{K}-\vec{q})f_{\beta_*}(\vec{q}, \vec{K}-\vec{q})
    -c \rb\ .
    \numberthis
\end{align*}
If we wanted to control the effect of other mode-couplings in Eq.~\eqref{eq:meanfieldinapp}, we would simply add other terms with similar constraints to this equation. We demand that the functional derivative of $L$ with respect to $g_{\alpha}(\vq', \vK-\vq')$ vanishes:
\begin{align*}
0 \stackrel{!}{=} \frac{\delta L[g_{\alpha}, \lambda, \lambda_*]}{\delta g_{\alpha}(\vec{q'}, \vec{K}-\vec{q'})}
	&= 2\int_{\vq} \lb \dirac(\vq-\vq') g^{*}_{\alpha}(\vq, \vK-\vq) 
	+ g_{\alpha}(\vq, \vK-\vq) \dirac(\vq-\vq') \rb P_{\rm tot}(\vq) P_{\rm tot}(\vK-\vq) \\
&\quad 
	- \lambda  \int_{\vq} \dirac(\vq-\vq') f_{\alpha}(\vq, \vK-\vq)  
	- \lambda_{*}  \int_{\vq} \dirac(\vq-\vq')f_{\beta}(\vq, \vK-\vq)  \\
&= 4g_{\alpha}(\vq', \vK-\vq')P_{\rm tot}(\vq')P_{\rm tot}(\vK-\vq')-\lambda f_{\alpha}(\vq', \vK-\vq') -\lambda_{*}f_{\beta}(\vq', \vK-\vq')\ ,
	\numberthis
\end{align*}
where in the last line, we took $g_\alpha$ to be real. This implies that (relabelling $\vq'\to\vq$)
\begin{equation}
    g_{\alpha}(\vec{q}, \vec{K}-\vec{q})
    = \frac{\lambda f_{\alpha}(\vec{q}, \vec{K}-\vec{q}) +\lambda_{*}f_{\beta}(\vec{q}, \vec{K}-\vec{q})}
    {4P_{\rm tot}(\vec{q})P_{\rm tot}(\vec{K}-\vec{q})} \ ,
    \label{eq:galpha-intermediate}
\end{equation}
and plugging this into Eq.~\eqref{eq:gab-constraint} gives
\begin{equation}
    \lambda = 2N_{\alpha\alpha}-\lambda_*\frac{N_{\alpha\alpha}}{N_{\alpha\beta}}\ ,
    \label{eq:lambda}
\end{equation}
using the definition of $N_{\alpha\beta}$ from Eq.~\eqref{eq:nab}. (Note that we obtain the original filter if $\lambda_*=0$.) Inserting Eqs.~\eqref{eq:galpha-intermediate} and~\eqref{eq:lambda} into~\eqref{eq:constraint0}, we get\footnote{To generalise to several mode-couplings, the coefficients $\lambda_{i*}$ are obtained from $\vec{\lambda}_*=-2\mathbf{A}^{-1}\vec{c}$, where $\mathbf{A}$ has elements $A_{ij}=(N_{\beta_i\beta_j}^{-1}-N_{\alpha\alpha}N_{\alpha\beta_j}^{-1}N_{\alpha\beta_i}^{-1})$ and $\vec{c}$ has elements $N_{\alpha\alpha}N_{\alpha\beta_i}^{-1}$.}
\begin{equation}
    \lambda_* = 2\frac{1}{1-r^2_{\alpha\beta}}N_{\beta\beta}(\vec{K})
    \lb c(\vec{K})-\frac{N_{\alpha\alpha}(\vec{K})}{N_{\alpha\beta}(\vec{K})} \rb \ ,
\end{equation}
where
\beq
r^2_{\alpha\beta}\equiv \frac{N_{\alpha\alpha} N_{\beta\beta}}{N_{\alpha\beta}^{2}}\ .
\label{eq:rab-appendix}
\eeq

Thus, the final form of the weight function is
\begin{align*}
g_{\alpha}(\vec{q}, \vec{K}-\vec{q})
	&= \frac{N_{\alpha\alpha}(\vK)f_{\alpha}(\vq, \vK-\vq)} {2P_{\rm tot}(\vq) P_{\rm tot}(\vK-\vq)} \\
&\quad
	+\frac{f_{\beta_*}(\vq, \vK-\vq) N_{\beta\beta}(\vK)
	- \frac{N_{\alpha\alpha}(\vK)N_{\beta\beta}(\vK)}{N_{\alpha\beta}(\vK)} f_{\alpha}(\vq, \vK-\vq)}
	{2P_{\rm tot}(\vq)P_{\rm tot}(\vK-\vq)}
	\frac{1}{1-r^2_{\alpha\beta}} \lb c(\vK)-\frac{N_{\alpha\alpha}(\vK)}{N_{\alpha\beta_*}(\vK)} \rb \ . 
	\numberthis
	\label{eq:filterg}
\end{align*}
This is the standard filter for mode-coupling $\alpha$, plus some additional terms related to the response to mode-coupling~$\beta$. For the case $c=N_{\alpha\alpha} N_{\alpha\beta}^{-1}$, one obtains the standard estimator and contamination term.

\subsection{Application to long-mode reconstruction}

\begin{figure}[t]
\centering
\includegraphics[width=0.5\textwidth, trim = 10 10 10 10 ]{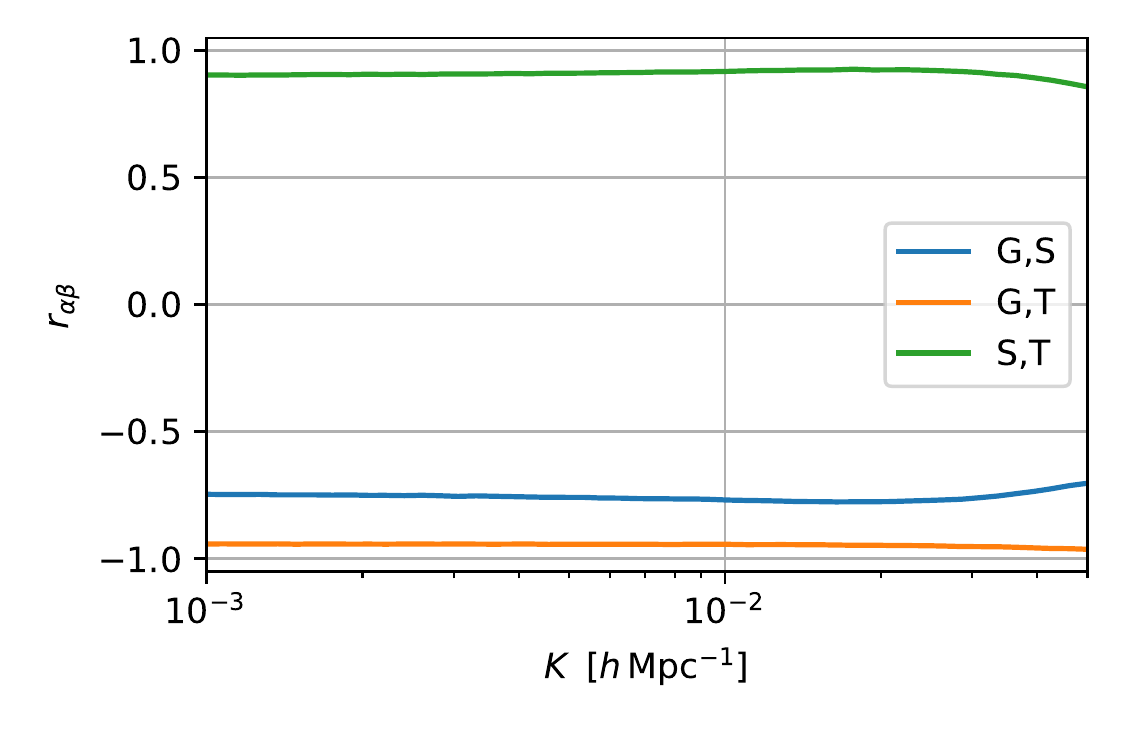} 
\caption{
\label{fig:rab}
Cross-correlation coefficients between G, S, and T estimators, as given by Eq.~\eqref{eq:rab-appendix}, for our DESI-like survey forecast (the other surveys give similar results) We see that all three estimators are highly correlated, implying, through Eq.~\eqref{eq:bhsimple}, that applying bias-hardening will strongly increase the noise of the resulting estimator.
}
\end{figure}

To relate the bias-hardened estimator derived above to the specific application we consider in this paper, let us examine the variance of the estimator:
\begin{equation}
    {\rm Var}\! \lb \hat{\Delta}_{\alpha}^{\rm H}(\vec{K}) \rb 
    = N_{\alpha\alpha}(\vec{K})+N_{\beta\beta}(\vec{K})
    \frac{ \lb c(\vec{K})-\frac{N_{\alpha\alpha}(\vec{K})}{N_{\alpha\beta}(\vec{K})} \rb^2}
    {1-r^2_{\alpha\beta}(\vec{K})} \ .
    \label{eq:bhsimple}
\end{equation}
If $|c(\vec{K})|<|N_{\alpha\alpha}(\vec{K}) N_{\alpha\beta_*}(\vec{K})^{-1}|$, then the increase in the estimator's variance scales with $(1-r_{\alpha\beta}^2)^{-1}$, where $r_{\alpha\beta}$ is the correlation coefficient between the un-hardened estimator $\hat{\Delta}_{\alpha}$ and the analogous estimator for the other mode-coupling, $\hat{\Delta}_{\beta}$. In Fig.~\ref{fig:rab}, we show $r_{\alpha\beta}(K)$ for the growth, shift, and tidal mode-couplings introduced in Sec.~\ref{sec:qe}. It is clear that the corresponding estimators are highly correlated, so that any bias-hardened estimator will have a much larger variance than without bias-hardening. In our numerical tests (with $c=0$), when one of these three estimators was bias-hardened with respect to the other two, we found that the variance increased enough to eliminate any advantages of removing the mean-field contamination, and therefore we did not implement any bias-hardening in our final forecasts.

In the course of this investigation, we derived a compact form for a bias-hardened quadratic estimator in the case of three mode-couplings, and we reproduce this derivation here in case it may be useful in other contexts. Considering only G, S, and T, the expectation values of the corresponding quadratic estimators (see Eq.~\ref{eq:meanfield}) can be written in matrix form:
\beq
\lb \begin{array}{c} 
	\langle \hat{\Delta}_{\rm G} \rangle \\
	\langle \hat{\Delta}_{\rm S} \rangle \\ 
	\langle \hat{\Delta}_{\rm T} \rangle 
	\end{array} \rb
	=
	b_1 \lb \begin{array}{ccc} 
	1 & N_{\rm GG} N_{\rm GS}^{-1} & N_{\rm GG} N_{\rm GT}^{-1} \\
	N_{\rm SS} N_{\rm SG}^{-1}& 1 & N_{\rm SS} N_{\rm TS}^{-1} \\
	N_{\rm TT} N_{\rm TG}^{-1} &  N_{\rm TT} N_{\rm TS}^{-1} & 1
	\end{array} \rb
	\lb \begin{array}{c} c_{\rm G} \\ c_{\rm S} \\ c_{\rm T}
	\end{array}\rb
	\delta_1\ .
	\label{eq:Deltasystem}
\eeq
We can derive bias-hardened estimators (in the $c(\vK)=0$ case) by inverting this system, solving for each $b_1 c_\alpha\delta_1$. After some lengthy algebra, the results can be written in terms of the original variances plus certain combinations of the original cross-correlation coefficients:
\begin{align*}
{\rm Var}\!\lb \hat{\Delta}^{\rm H}_{\rm G} \rb
	&= N_{\rm GG} \times  \frac{1-r_{\rm ST}^2}{\det M}\ , \\
{\rm Var}\!\lb \hat{\Delta}^{\rm H}_{\rm S} \rb
	&= N_{\rm SS} \times  \frac{1-r_{\rm GT}^2}{\det M}\ , \\
{\rm Var}\!\lb \hat{\Delta}^{\rm H}_{\rm T} \rb
	&= N_{\rm TT} \times  \frac{1-r_{\rm GS}^2}{\det M}\ , 	
	\numberthis
	\label{eq:bhvar-r}
\end{align*}
where $M$ denotes the matrix in Eq.~\eqref{eq:Deltasystem}, and
\beq
\label{eq:detM-r}
\det M = 1-r_{\rm GS}^2 - r_{\rm GT}^2 - r_{\rm ST}^2 +2 r_{\rm GS} r_{\rm GT} r_{\rm ST}\ .
\eeq
There can be nontrivial cancellations within the above determinant, and in our case, these lead to large increases in the variances of the bias-hardened estimators.

Finally, we mention a few other possible solutions to the problem of mean-field contamination. Instead of fixing the contamination to some value, as in Eq.~\eqref{eq:constraint0}, one could require it to be smaller than some fixed value, or one could minimise some total function that depends on the contamination and the (Gaussian) variance. For example it is possible to define a ``bouncing estimator" by solving the following minimization problem:
\begin{align*}
    L[g_{\alpha},\lambda, \vec{K}] 
    &= 2\int_{\vec{q}}|g_{\alpha}(\vec{q}, \vec{K}-\vec{q})|^2P_{\rm tot}(\vec{q}) P_{\rm tot}(\vec{K}-\vec{q})
    + A \lb \int_{\vec{q}}g_{\alpha}(\vec{q}, \vec{K}-\vec{q}) f_{\beta}(\vec{q}, \vec{K}-\vec{q}) \rb^2 \\
&\quad 
	-\lambda \lb \int_{\vq} g_{\alpha}(\vq, \vK-\vq) f_{\alpha}(\vq, \vK-\vq)-1 \rb\ .
	\numberthis
\end{align*}
The intuition behind this is that we want to minimize the variance of the $\alpha$ estimator, trying also to take into account the contamination from the other mode-coupling. The square is to ensure that the modulus of the contamination is minimized in the combination that makes $L$ the smallest. If we go to the standard minimum variance solution, the solution here will ``bounce" from it, because it would increase $L$, if we take into account the square of the contamination. We also want to decrease the contamination, but without it taking a large negative value. The last term enforces the standard unbiasedness condition.

We reiterate, however, that for some applications, such as the $\fnl$ constraints we consider in this paper, a mean-field contamination can actually be advantageous. We leave the problem of finding a fully optimal estimator for long-wavelength reconstruction to future work.

\section{Contamination of quadratic estimator by $f_{\rm NL}$ terms: analytical expressions}
\label{app:meanfield}

The overdensity of a biased tracer has second order contributions in the linear field coming from the presence of primordial non-Gaussianity. In this appendix, we show that when reconstructing the new field on large scales, we get an $f_{\rm{NL}}$ term proportional to $\frac{1}{M(\vec{K})}$. We will show this for the low-$|\vec{K}|$ limit, which is the relevant regime for the reconstructed modes we are concerned with.

\subsection{Expansion of basic quantities}

We need to expand a few quantities first. Expanding the linear power spectrum around $\Vec{q}$
gives 
\begin{equation}
\ba
    \lim_{|\vec{K}|\rightarrow 0} P_{\rm{lin}}(|\vec{K}-\vec{q}|)&= P_{\rm{lin}}(|\vec{q}|)-\vec{\nabla}_{\vec{q}}P_{\rm{lin}}\cdot \vec{K}+O(|\vec{K}|^2)\\
    &=P_{\rm{lin}}(|\vec{q}|)-\frac{\partial P_{\rm{lin}}}{\partial |\vec{q}|}\vec{\nabla}_{\vec{q}}|\vec{q}|\cdot \vec{K}+O(|\vec{K}|^2)
    = P_{\rm{lin}}(\vec{q})\left( 1-\frac{|\vec{K}|\mu}{|\vec{q}|}\frac{\partial{ \rm{ln}}P_{\rm{lin}}}{\partial{ \rm{ln}}|\vec{q}|} \right)+O(|\vec{K}|^2) \\ &\equiv P_{\rm{lin}}(\vec{q})\left( 1+\Delta_P \right)+O(|\vec{K}|^2) \ ,
\ea
\end{equation}
where the last expression is useful because we can see the expansion in powers of $\frac{|\vec{K}|}{|\vec{q}|}$, and where $\mu$ is the cosine of the angle between $\vec{K}$ and $\vec{q}$. In the same way, we can also expand for
\begin{equation}
    \lim_{|\vec{K}|\rightarrow 0} P_{\rm{NL}}(|\vec{K}-\vec{q}|)=  P_{\rm{NL}}(\vec{q})\left( 1-\frac{|\vec{K}|\mu}{|\vec{q}|}\frac{\partial{\rm{ln}}P_{\rm{NL}}}{\partial{\rm{ln}}|\vec{q}|} \right)+O(|\vec{K}|^2) \equiv P_{\rm{NL}}(\vec{q})\left( 1+\Delta_{NL} \right)+O(|\vec{K}|^2) \ ,
\end{equation}{}
and 
\begin{equation}
    \lim_{|\vec{K}|\rightarrow 0} M|\vec{K}-\vec{q}|)= 
 M(\vec{q})\left( 1-\frac{|\vec{K}|\mu}{|\vec{q}|}\frac{\partial{\rm{ln}}M}{\partial {\rm{ln}}|\vec{q}|} \right)+O(|\vec{K}|^2)\ \equiv M(\vec{q})\left( 1+\Delta_{M} \right) .
\end{equation}{}

We can also write
\begin{equation}
    \lim_{|\vec{K}|\rightarrow 0} |\vec{K}-\vec{q}| =  \lim_{|\vec{K}|\rightarrow 0} \sqrt{|\vec{K}|^2+|\vec{q}|^2-2|\vec{K}||\vec{q}|\mu} \approx |\vec{q}|\lp 1-\frac{|\vec{K}|}{|\vec{q}|}\mu \rp \ ,
\end{equation}{}
and the expression for the cosine of the angle between $\vec{K}$ and $\vec{K}-\vec{q}$ as
\begin{equation}
    \mu'= \lim_{|\vec{K}|\rightarrow 0} \approx -\mu+(1-\mu^2)\frac{|\vec{K}|}{|\vec{q}|}+\frac{|\vec{K}|^2}{|\vec{q}|^2}\mu \approx -\mu+(1-\mu^2)\frac{|\vec{K}|}{|\vec{q}|}\ .
\end{equation}{}

\subsection{Expansion of mode coupling expressions}

As $|\vec{K}|\rightarrow 0$ for the next calculations we will assume $\frac{|\vec{K}|}{|\vec{q}|}\ll1$  so that we keep terms linear in this variable. If the linear terms completely cancel, we include terms at the next order. 

Recall the definition of $f_{\beta}(\vec{q}, \vec{K}-\vec{q})$ from Eq.~\eqref{eq:falpha}:
\begin{equation}
   f_{\beta}(\vec{q}, \vec{K}-\vec{q})=2\Big[F_{\alpha}\big(\vec{K}, -\vec{q}\big)P_{\rm{lin}}(|\vec{q}|)+F_{\alpha}\big(\vec{K}, -\vec{K}+\vec{q}\big)P_{\rm{lin}}\big(|\vec{K}-\vec{q}|\big)\Big] \ . \label{eq:explicitfalpha}
\end{equation}{}
Starting from Eq.~\eqref{eq:explicitfalpha}, we have that for the $\varphi\varphi$ term 

\begin{equation}
\ba
   f_{\varphi\varphi}(\vec{q}, \vec{K}-\vec{q}) &= \frac{2}{M(\vec{K})}\left[\frac{M(\vec{K}-\vec{q})}{M(-\vec{q})}P_{\rm{lin}}(\vec{q})+\vec{q}\rightarrow \vec{K}-\vec{q}\right]\\
   &= \frac{2P_{\rm{lin}}(\vec{q})}{M(\vec{K})}\left [ \frac{M(\vec{q})(1+\Delta_M(\vec{q}))}{M(\vec{q})}+\frac{M(\vec{q})}{M(\vec{q})(1+\Delta_M(\vec{q}))}(1+\Delta_P(\vec{q})) \right] \approx
    \frac{2P_{\rm{lin}}(\vec{q})}{M(\vec{K})}\left [ 2+\Delta_P(\vec{q}) \right] \ ,
\ea    
\end{equation}
and for the $b_{01}$ term
\begin{equation}
\ba
   f_{01}(\vec{q}, \vec{K}-\vec{q}) &= \frac{2}{M(\vec{K})}\frac{1}{2}\left[(\vec{K}\cdot (-\vec{q}))(\frac{1}{|-\vec{q}|^2}+\frac{M(\vec{K})}{|\vec{K}|^2M(-\vec{q})})P_{\rm{lin}}(-\vec{q})+\vec{q}\rightarrow \vec{K}-\vec{q}\right] \\
    &=\frac{1}{2} \frac{2P_{\rm{lin}}(\vec{q})}{M(\vec{K})}\Big[ \vec{K}\cdot(-\vec{q})\left( \frac{1}{|\vec{q}|^2}+\frac{M(\vec{K})}{|\vec{K}|^2M(\vec{q})} \right) + \vec{K}\cdot (-1)(\vec{K}-\vec{q}) \big(\frac{1}{|\vec{q}|^2} (1+\frac{|\vec{K}|}{|\vec{q}|}\mu)+\frac{M(\vec{K})}{K^2M(\vec{q})}(1-\Delta_M(\vec{q})) \big)\\
    &\times \big(1+\Delta_P(\vec{q})\big) \Big]\\
    &\approx \frac{1}{2}\frac{2P_{\rm{lin}}(\vec{q})}{M(\vec{K})}\left[ -\frac{|\vec{K}|^2}{|\vec{q}|^2} - \frac{|\vec{K}|^3}{|\vec{q}|^3}\mu - \frac{M(\vec{K})}{M(\vec{q})} (1+\Delta_P(\vec{q})-\Delta_M(\vec{q}))+\frac{|\vec{K}|^2}{|\vec{q}|^2}\mu^2+\frac{q\mu M(\vec{K})}{|\vec{K}|M(\vec{q})} (\Delta_P(\vec{q})-\Delta_M(\vec{q})) \right]\\
    &\approx \frac{1}{2}\frac{2P_{\rm{lin}}(\vec{q})}{M(\vec{q})}\left[ -1 - \Delta_P(\vec{q})-\Delta_M(\vec{q}) - (1-\mu^2)\frac{|\Vec{K}|^2}{|\Vec{q}|^2}\right] \ ,
\ea    
\end{equation}
where we remember that $M(\vec{q})\propto |\vec{q}|^{-2}$. For the $b_{11}$ term,
\begin{equation}
\ba
   f_{11}(\vec{q}, \vec{K}-\vec{q})& = \frac{2}{M(\vec{K})}\frac{1}{2}\left[
   \left(1+\frac{M(\vec{K})}{M(-\vec{q})}\right)P_{\rm{lin}}(-\vec{q})+\vec{q}\rightarrow \vec{K}-\vec{q}\right] \\
   &=\frac{P_{\rm{lin}}(\vec{q})}{M(\vec{K})}\Big[ 1+\frac{M(\vec{K})}{M(\vec{q})}+1
   +\frac{M(\vec{K})}{M(\vec{q})}(1-\Delta_M(\vec{q}))(1+\Delta_P(\vec{q}))  \Big]\\
   &\approx \frac{P_{\rm{lin}}(\vec{q})}{M(\vec{K})}\left[ 2+ \frac{M(\vec{K})}{M(\vec{q})} (1+\Delta_P(\vec{q})-\Delta_M(\vec{q})) \right],
 \ea   
\end{equation}
and finally for the $b_{02}$ term
\begin{equation}
\ba
   f_{02}(\vec{q}, \vec{K}-\vec{q}) &= \frac{2}{M(\vec{K})}\left[
   \frac{P_{\rm{lin}}(-\vec{q})}{M(-\vec{q})}+\vec{q}\rightarrow \vec{K}-\vec{q}\right]\\
   &=\frac{2P_{\rm{lin}}(\vec{q})}{M(\vec{K})}\left[\frac{1}{M(\vec{q})} + \frac{1}{M(\vec{q})} (1-\Delta_M(\vec{q})+\Delta_P(\vec{q})) \right] \\
   &\approx  \frac{2P_{\rm{lin}}(\vec{q})}{M(\vec{K})}\frac{1}{M(\vec{q})}\Big[ 2-\Delta_M(\vec{q}) +\Delta_P(\vec{q}) \Big]\ .
\ea    
\end{equation}

We wrote all of them in such a way that, when possible, we can factor out a $\frac{1}{M(\vec{K})}$. Finally the growth term can be written as
\begin{equation}
    f_{\text{G}}(\vec{q}, \vec{K}-\vec{q}) = 2\left[ \frac{17}{21}P(\vec{
    q})+\frac{17}{21}P_{\rm{lin}}(\Vec{K}-\Vec{q})\right]=2\frac{17}{21}P(\vec{
    q})\Big[ 2+ \Delta_P(\Vec{q}) \Big]\ .
\end{equation}

\subsection{Writing the large scale contamination terms}

Recall that the terms that contaminate the expectation value of the quadratic estimator are of the form $c_{\alpha}\frac{N_{\text{GG}}}{N_{\text{G}\alpha}}$ (see Eq.~\ref{eq:meanfield}). Therefore, for the large scale limit, we need to consider
\begin{equation}
\ba
    \lim_{|\Vec{K}|\rightarrow 0} N_{\text{G}\alpha}^{-1}(\Vec{K})&= \lim_{|\Vec{K}|\rightarrow 0} \frac{2\pi}{(2\pi)^3}\int_{-1}^{1}d\mu\int_{q_{\rm{min}}}^{q_{\rm{max}}}dq q^2 \frac{f_{\alpha}(\Vec{q}, \Vec{K}-\Vec{q})}{2P_{\rm{tot}}(\Vec{q})P_{\rm{tot}}(\Vec{K}-\Vec{q})}f_g(\Vec{q}, \Vec{K}-\Vec{q})\\
    &= \lim_{|\Vec{K}|\rightarrow 0}\frac{1}{b_1^4} \frac{2\pi}{(2\pi)^3}\int_{-1}^{1}d\mu\int_{q_{\rm{min}}}^{q_{\rm{max}}}dq q^2 \frac{f_{\alpha}(\Vec{q}, \Vec{K}-\Vec{q})}{2P^2_{\rm{NL}}(\Vec{q})}2\frac{17}{21}P_{\rm{lin}}(\vec{
    q})\left[2+ \Delta_{\rm{P}}(\Vec{q})\right](1-\Delta_{\mathrm{PN}}) \ , 
\ea    
\end{equation}
where we assume no shot noise in the total galaxy power spectrum. Dropping any non-zero power of $\frac{|\Vec{K}|}{|\Vec{q}|}$, for small $\Vec{K}$ (with respect to reconstruction modes $\vec{q}$), we obtain
\begin{equation}
\ba
\label{eq:NgglowK}
    N_{\text{GG}}(\Vec{K}) &\approx \left[\frac{1}{b_1^4}\frac{2\pi}{(2\pi)^3}\left(2\frac{17}{21}\right)^2 4\int_{q_{\rm{min}}}^{q_{\mathrm{max}}} dq q^2 \frac{P_{\mathrm{lin}}^2(\Vec{q})}{P^2_{\mathrm{NL}}(\Vec{q})}[1+\Delta_{\rm{P}}(\Vec{q})-\Delta_{\rm{PN}}(\Vec{q})]\right]^{-1}\\
    &\approx \left[\frac{1}{b_1^4}\frac{4\pi}{(2\pi)^3}8\left(\frac{17}{21}\right)^2\int_{q_{\rm{min}}}^{q_{\mathrm{max}}} dq q^2 \frac{P_{\mathrm{lin}}^2(\Vec{q})}{P^2_{\mathrm{NL}}(\Vec{q})}\right]^{-1} \ ,
\ea
\end{equation}
and we can see that if we approximate $P_{\mathrm{lin}} \approx P_{\mathrm{NL}}$ for the small scales of reconstruction, then the Gaussian reconstruction noise is roughly proportional to the volume shell between $q_{\rm{min}}$ and $q_{\rm{max}}$. Thus, we can take the noise as roughly proportional to $q_{\rm{max}}^3$, although in practice this relation is not exactly correct.

At this point, we can start listing the $N_{{\rm G}\alpha}$ terms where $\alpha$ is a mode-coupling involving $\fnl$. We begin with  $\alpha=\varphi\varphi$:
\begin{equation}
    N_{\text{G}\varphi\varphi}^{-1}(\Vec{K}) \approx \frac{1}{b_1^4}\frac{1}{2\pi^2}\left(\frac{136}{21}\right)\frac{1}{M(\Vec{K})}\int_{q_{\rm{min}}}^{q_{\mathrm{max}}} dq q^2 \frac{P_{\mathrm{lin}}^2(\Vec{q})}{P^2_{\mathrm{NL}}(\Vec{q})} \ ,
\end{equation}
such that multiplying by Eq.~\eqref{eq:NgglowK}, we have a scaling for the $\alpha=\varphi\varphi$ term in the bias of the reconstructed field:
\begin{equation}
    N_{\text{GG}}(\Vec{K}) N_{\text{G}\varphi\varphi}^{-1}(\Vec{K}) \approx \left(\frac{21}{17}\right)\frac{1}{M(\Vec{K})}\ .
\end{equation}
Similarly we can calculate an approximate expression for the $\alpha=01$ term:
\begin{equation}
    N_{\text{G} 01}^{-1}(\Vec{K}) \approx (-1)\frac{1}{b_1^4}\frac{1}{2\pi^2}\frac{1}{2}\frac{68}{21}\int_{q_{\rm{min}}}^{q_{\mathrm{max}}} dq q^2 \frac{P_{\mathrm{lin}}^2(\Vec{q})}{P^2_{\mathrm{NL}}(\Vec{q})}\frac{1}{M(\Vec{q})} \ .
\end{equation}
Multiplying this by the approximate $N_{\text{G}\text{G}}$  expression, we obtain
\begin{equation}
    N_{\text{GG}}(\Vec{K}) N_{\text{G} 01}^{-1}(\Vec{K}) \propto -\frac{42}{17}\ ,
\end{equation}
implying that on large scales for the $\alpha=01$ term we do not have a $\frac{1}{K^2}$ behaviour, but a negative constant bias.

Turning to the $\alpha=02$ and $11$ terms, we find
\begin{equation}
    N^{-1}_{\text{G}02}(\Vec{K}) \approx \frac{1}{b^4}\frac{1}{2\pi^2}\left(\frac{136}{21}\right)\frac{1}{M(\Vec{K})}\int_{q_{\rm{min}}}^{q_{\mathrm{max}}} dq q^2 \frac{P_{\mathrm{lin}}^2(\Vec{q})}{P^2_{\mathrm{NL}}(\Vec{q})}\frac{1}{M(\Vec{q})} \ ,
\end{equation}
and
\begin{equation}
    N^{-1}_{\text{G}11}(\Vec{K}) \approx \frac{1}{b^4}\frac{1}{2\pi^2}\left(\frac{68}{21}\right)\frac{1}{M(\Vec{K})}\int_{q_{\rm{min}}}^{q_{\mathrm{max}}} dq q^2 \frac{P_{\mathrm{lin}}^2(\Vec{q})}{P^2_{\mathrm{NL}}(\Vec{q})}\left( 1+\frac{M(\Vec{K})}{2M(\Vec{q})} \right) \ .
\end{equation}
In both cases we end up with a $1/K^2$ behavior:
\begin{equation}
    N_{\text{GG}}(\Vec{K}) N_{\text{G} 02}^{-1}(\Vec{K}) = 21/17 \frac{1}{M(\Vec{K})} \Bigg(\int_{q_{\rm{min}}}^{q_{\mathrm{max}}} dq q^2 \frac{P_{\mathrm{lin}}^2(\Vec{q})}{P^2_{\mathrm{NL}}(\Vec{q})}\frac{1}{M(\Vec{q})}/\int_{q_{\rm{min}}}^{q_{\mathrm{max}}} dq q^2 \frac{P_{\mathrm{lin}}^2(\Vec{q})}{P^2_{\mathrm{NL}}(\Vec{q})} \Bigg)\ ,
\end{equation}
\begin{equation}
    N_{\text{GG}}(\Vec{K}) N_{\text{G} 11}^{-1}(\Vec{K}) = \frac{1}{2}\frac{21}{17} \frac{1}{M(\Vec{K})}\ .
\end{equation}

\begin{figure}[t]
\centering
\includegraphics[width=0.5\textwidth, trim = 10 10 10 10 ]{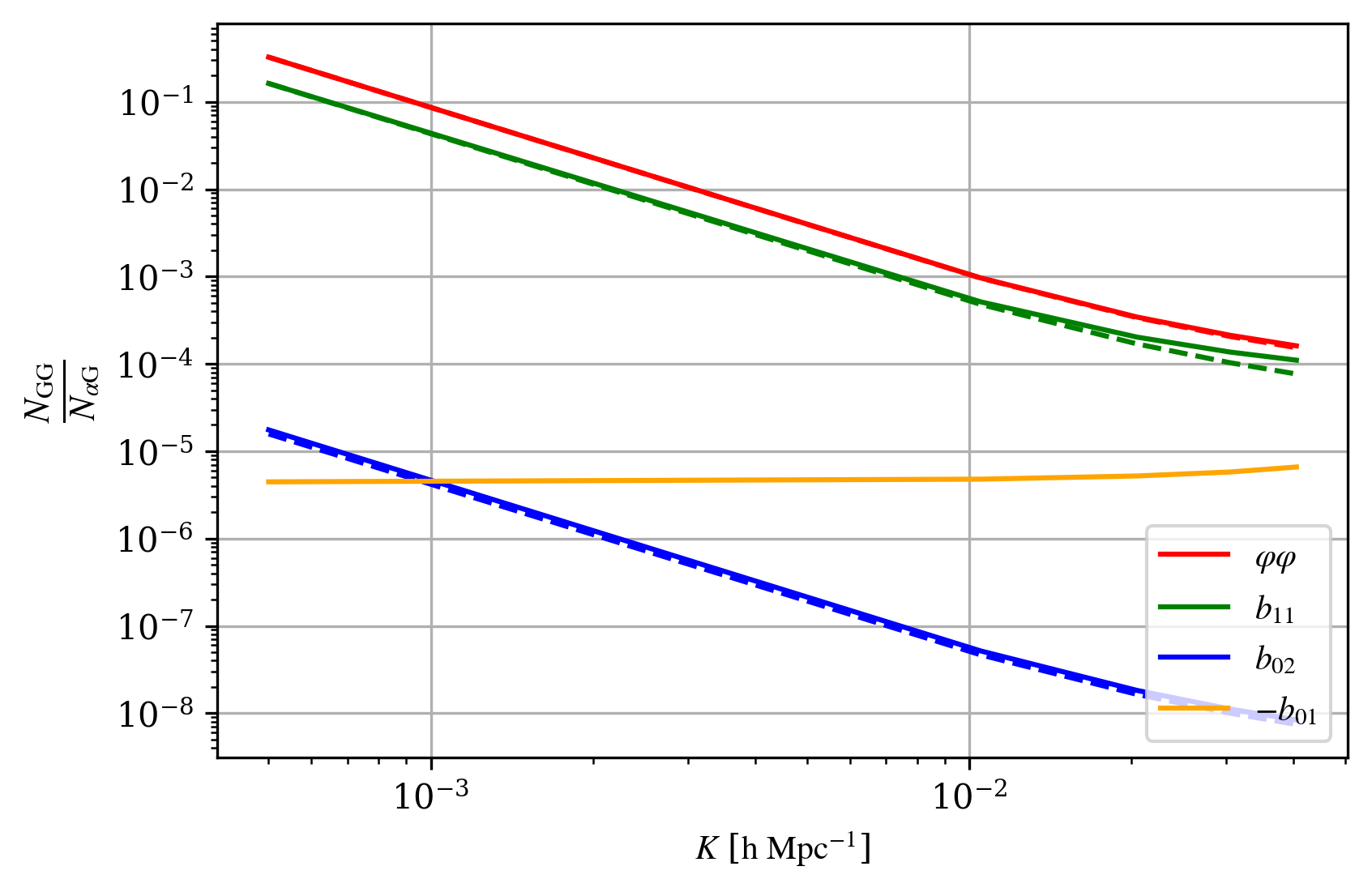} 
\caption{
\label{fig:contaminationapprox}
We plot the contamination curves from numerical and analytic approximation. It can be seen that for the $\varphi\varphi$, $b_{11}$ and $b_{02}$ terms we have a $\frac{1}{K^2}$ behaviour. For the analytical calculation of the $b_{02}$ term we use an approximation $ \Bigg(\int_{q_{\rm{min}}}^{q_{\mathrm{max}}} dq q^2 \frac{P_{\mathrm{lin}}^2(\Vec{q})}{P^2_{\mathrm{NL}}(\Vec{q})}\frac{1}{M(\Vec{q})}/\int_{q_{\rm{min}}}^{q_{\mathrm{max}}} dq q^2 \frac{P_{\mathrm{lin}}^2(\Vec{q})}{P^2_{\mathrm{NL}}(\Vec{q})} \Bigg) \approx \frac{1}{2M(q_{\rm{min}})}$.
We also show the absolute value of the numerical $b_{01}$ curve, with an approximate constant value on large scales.
}
\end{figure}

To summarise, we have found an induced contamination on the G estimator of the following form:
 \begin{equation}
    \sum_{\alpha \in \{\varphi\varphi, 01, 11, 02\}}\int_{\vec{q}}c_{\alpha}g_{\rm G}(\vec{q}, \vec{K}-\vec{q})f_{\alpha}(\vec{q}, \vec{K}-\vec{q}) = \frac{1}{M(|\vec{K}|)}\Big[f_{\text{NL}}A(\vec{K})+f_{\text{NL}}^2B(\vec{K})\Big] \ ,
\end{equation}
where $A, B$ are some functions that can be calculated from the definitions or numerically. In Fig.~\ref{fig:contaminationapprox}, we show that the analytical approximations are in excellent agreement with the full numerical computations for the contamination curves.

\section{Foregrounds for \tcm intensity mapping}
\label{app:fg}

\subsection{Implementation in forecasts}
\label{app:fg-implementation}

As discussed in Sec.~\ref{sec:config}, the presence of foregrounds in \tcm intensity mapping limits the modes of $\delta_{\rm g}$ that can be directly observed. Specifically, foregrounds impose a minimum $k_\parallel$ value for these modes, and also obscure modes within a wedge-shaped region in the $k_\parallel-k_\perp$ plane. The modes within this wedge satisfy (e.g.\ \citealt{Ansari:2018ury})
\beq
k_\parallel < \beta(z) k_\perp\ ,
\eeq
where
\beq
\beta(z) \equiv \frac{\chi(z) H(z)}{c(1+z)} \sin(\theta_{\rm w})\ ,
\eeq
$\chi(z)$ is the comoving distance to redshift $z$, and $\theta_{\rm w}$ is the maximum angle from the beam center at which the power of a spectrally-smooth source will leak into other regions of Fourier space. The angle $\theta_{\rm w}$ is typically related to the width of the primary beam; following \cite{Ansari:2018ury}, we take it to be 3 times the primary beam width of PUMA, or $\theta_{\rm w} \approx 3 \times 1.22\lambda(z)/D_{\rm eff}$, where $\lambda(z) = 21(1+z)\,{\rm cm}$ and $D_{\rm eff} \approx  5\,{\rm m}$ is the effective dish diameter ($\eta_{\rm a}^{1/2} \times 6\,{\rm m}$ with aperture efficiency factor $\eta_{\rm a}=0.7$). Using the mean redshifts of the low-$z$ and high-$z$ bins we use in our forecasts, this yields $\beta\approx 0.38$ and $1.3$ for each bin respectively. 

The wedge will restrict the small-scale modes that can be used for reconstructing the longer modes via the quadratic estimator in Eq.~\eqref{eq:quadest}, and we can account for this by restricting the reconstruction noise integral in Eq.~\eqref{eq:nab} to modes outside the wedge. This means that, when reconstructing a mode with wavevector~$\vK$, the integration variable~$\vq$ must satisfy
\beq
\label{eq:q-wedge-conditions}
|q_\parallel| > \beta q_\perp\ , \quad
	|K_\parallel - q_\parallel| > \beta \lp \vK-\vq \rp_\perp\ .
\eeq
Rather than implementing these restrictions directly in the integral for $N_{\alpha\beta}$, which would cause the result to depend on the full vector $\vK$ instead of just the norm $K$, we use an approximation based on the fact that in the $q\gg K$ limit, $N_{\alpha\beta}$ scales like the inverse of the number of modes that contribute to the reconstruction.\footnote{This is identical to what happens to the noise on the standard quadratic estimator for CMB lensing in the $\ell\gg L$ limit (e.g.~\citealt{Hanson:2010rp}).} Thus, the effect of the wedge is mostly to rescale $N_{\alpha\beta}$ by the inverse of the fraction of modes that are outside of the wedge, i.e.\ the fraction of the integration domain that satisfies Eq.~\eqref{eq:q-wedge-conditions}. This fraction will depend slightly on the direction of~$\vK$, and we account for this dependence by averaging the fraction over $\mu_K \equiv K_\parallel/K$, although the dependence is only mind.

\begin{figure}[t]
\centering
\includegraphics[width=0.5\textwidth, trim = 10 10 10 10 ]{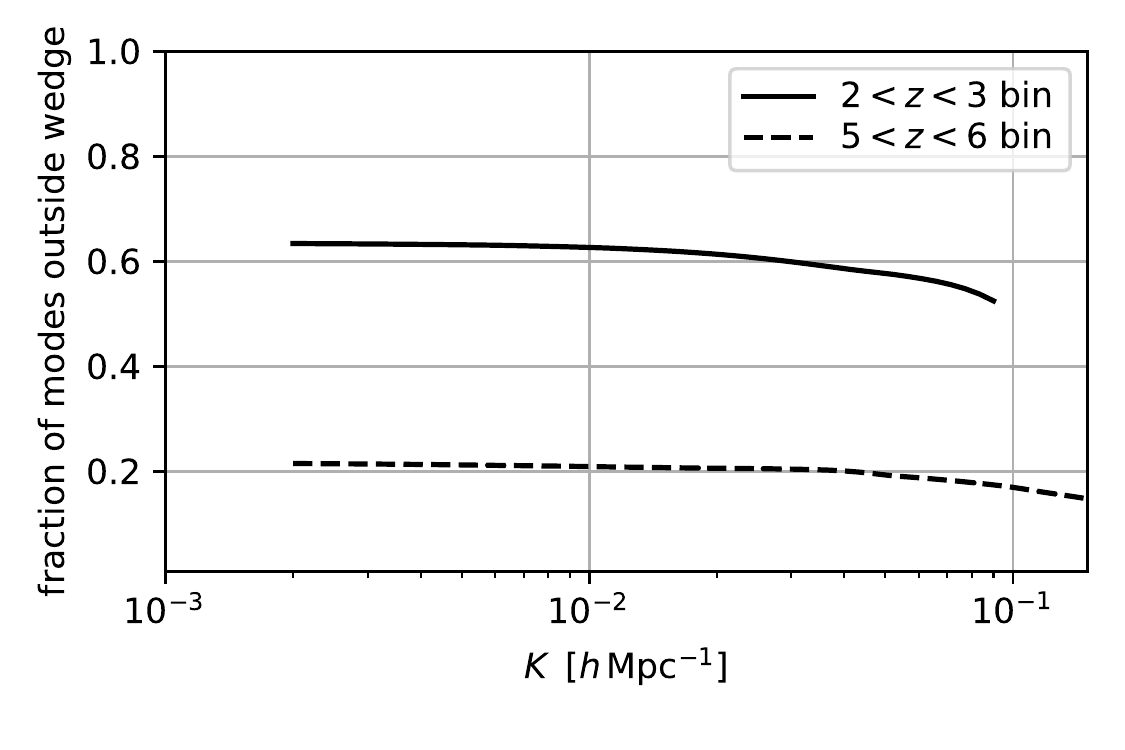} 
\caption{
\label{fig:wedgefractions}
The fraction of small-scale modes lying outside the foreground wedge, for the low-$z$ {\em (solid)} and high-$z$ {\em (dashed)} PUMA redshift bins we use in our forecasts. To a good approximation, the reconstruction noise integrals $N_{\alpha\beta}$ will simply be scaled by the inverse of these fractions in the presence of the wedge.
}
\end{figure}

We plot this angle-averaged fraction in Fig.~\ref{fig:wedgefractions} for both redshift bins we use for our PUMA forecast. In the $2<z<3$ bin, we find that around $60\%$ of the small-scale modes are untouched by the wedge, while for the $5<z<6$ bin, only $20\%$ of the modes remain, corresponding to a factor of $5$ increase in the reconstruction noise compared to the no-wedge case.

For the shot noise contributions to the $\delta_{\rm r}$ auto spectrum and $\delta_{\rm r}$-$\deltag$ cross spectrum (Eqs.~\ref{eq:nrrshot}-\ref{eq:prrshot-appendix} and~\ref{eq:nrtshot}-\ref{eq:pgrshot-appendix}, respectively), we directly implement the wedge in the angular limits of the integrals, but we find that it has a negligible effect, since these integrals are normalized with $N_{\alpha\beta}$ and the fractional change in the integrals and $N_{\alpha\beta}$ is very similar.

We also need to incorporate the wedge when we integrate the Fisher matrix in Eq.~\eqref{eq:fisherpermodegeneral} over long-wavelength modes used for the $\fnl$ constraint (Eq.~\ref{eq:integratedfishermatrix}). Within the wedge, we will not have access to $\delta_{\rm g}$, but we will have access to modes $\delta_{\rm r}$ reconstructed with the quadratic estimator. Thus, as in our baseline forecasts with an isotropic $K_{\rm min}$, we sum the outside-wedge and inside-wedge Fisher matrices, each restricted to the appropriate integration domain, with the latter Fisher matrix determined solely by the covariance of the reconstructed modes.

In the next subsection, where we consider a $K_{\parallel,{\rm min}}$ instead of an isotropic $K_{\rm min}$, we likewise implement the restriction $K>K_{\parallel,{\rm min}}$ in Eq.~\eqref{eq:integratedfishermatrix}, and add the contribution from reconstructed modes with $K<K_{\parallel,{\rm min}}$.

\subsection{PUMA forecasts with $K_{\parallel,{\rm min}}$}
\label{app:fg-pumakparmin}

\begin{figure}[t]
\includegraphics[width=\textwidth, trim = 10 10 0 10 ]{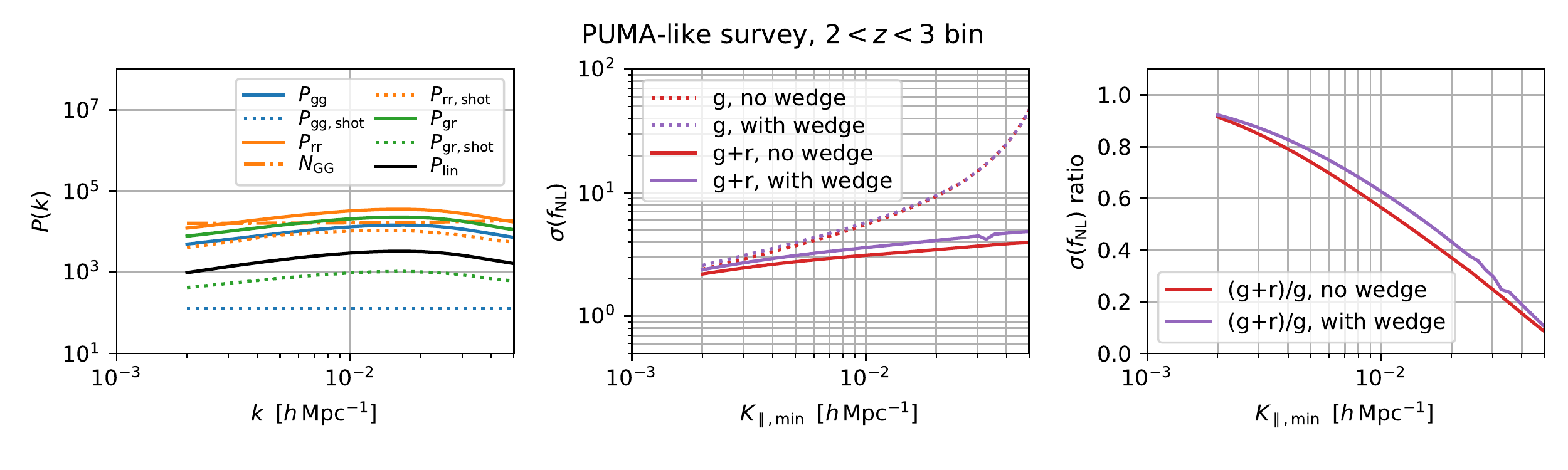} 
\includegraphics[width=\textwidth, trim = 10 10 0 10 ]{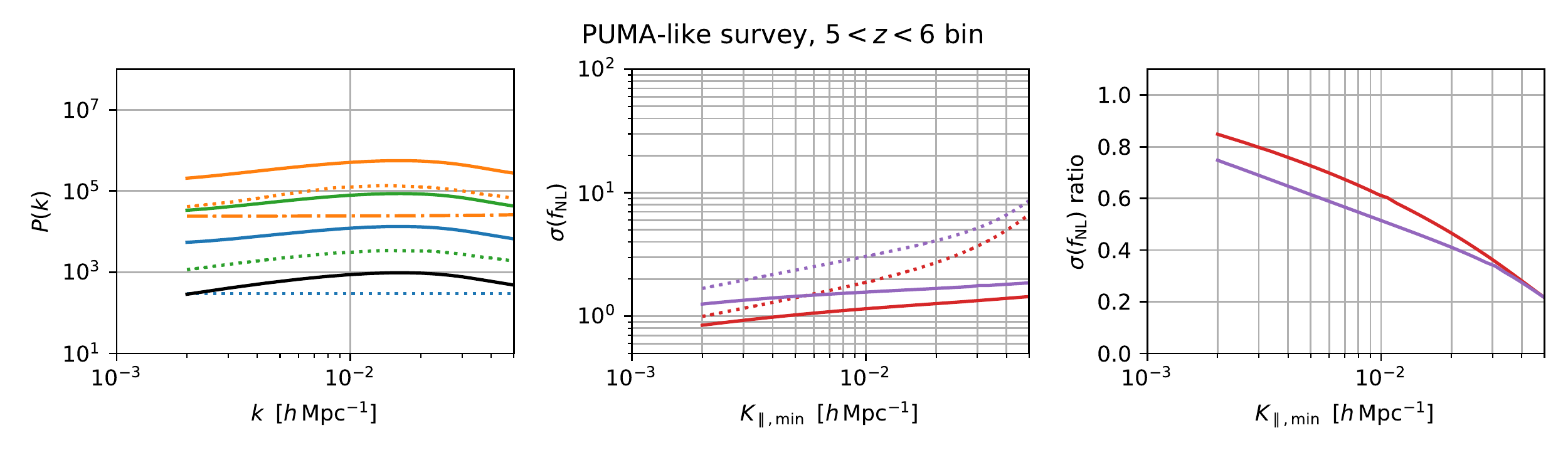} 
\caption{
\label{fig:puma-kparmin}
Forecasts for a PUMA-like survey, analogous to Fig.~\ref{fig:puma} except for using a cutoff on the line-of-sight component of accessible $\delta_{\rm g}$ modes instead of an isotropic $K$ cutoff. This is motivated by the fact that \tcm foregrounds will preferentially obscure modes with low wavenumber components along the line of sight. The absolute values of $\sigma(\fnl)$ are slightly higher when using $K_{\parallel,{\rm min}}$ instead of $K_{\rm min}$, since more modes are eliminated with a $K_\parallel$ cut, but the improvement in $\sigma(\fnl)$ from including reconstructed modes is qualitatively similar to the case with $K_{\rm min}$.
}
\end{figure}

In Sec.~\ref{sec:puma}, we showed forecasts assuming an isotropic $K_{\rm min}$ for $\delta_{\rm g}$. In Fig.~\ref{fig:puma-kparmin}, we repeat those forecasts, but with a $K_{\parallel,{\rm min}}$, assuming that all values of $K_{\perp}$ within the survey volume can be accessed. The absolute values we find for $\sigma(\fnl)$ are slightly higher, due to the number of inaccessible modes being larger with a $K_{\parallel,{\rm min}}$ cutoff, but the results for the improvement in $\sigma(\fnl)$ due to the inclusion of reconstructed modes are qualitatively similar to those in Sec.~\ref{sec:puma}.

\bibliography{biblio}

\end{document}